\documentclass[fleqn,usenatbib]{mnras}

\usepackage{newtxtext,newtxmath}

\usepackage[T1]{fontenc}

\DeclareRobustCommand{\VAN}[3]{#2}
\let\VANthebibliography\thebibliography
\def\thebibliography{\DeclareRobustCommand{\VAN}[3]{##3}\VANthebibliography}

\usepackage{graphicx}	
\usepackage{amsmath}	
\usepackage{euclid}
\usepackage[flushleft]{threeparttable}
\usepackage[dvipsnames]{xcolor}
\usepackage{ulem}

\newcommand{\orcid}[1]{} 

\title[\Euclid galaxy property estimate with machine learning]{\Euclid preparation: XXIII. Derivation of galaxy physical properties with deep machine learning using mock fluxes and H-band images}

\author[L.Bisigello et al.]{Euclid Collaboration: L.~Bisigello\orcid{0000-0003-0492-4924}$^{1,2,3}$\thanks{laura.bisigello@inaf.it}, C.J.~Conselice$^{4}$, M.~Baes\orcid{0000-0002-3930-2757}$^{5}$, M.~Bolzonella\orcid{0000-0003-3278-4607}$^{2}$, M.~Brescia\orcid{0000-0001-9506-5680}$^{6}$, S.~Cavuoti\orcid{0000-0002-3787-4196}$^{6,7,8}$, \newauthor O.~Cucciati\orcid{0000-0002-9336-7551}$^{2}$, A.~Humphrey$^{9}$, L.~K. Hunt\orcid{0000-0001-9162-2371}$^{10}$
% ...
\\
    (Affiliations can be found after the references)}

\date{Accepted XXX. Received YYY; in original form ZZZ}

\pubyear{2022}

\begin{document}
\label{firstpage}
\pagerange{\pageref{firstpage}--\pageref{lastpage}}
\maketitle

\clearpage
\begin{abstract}
Next generation telescopes, like \Euclid, \textit{Rubin}/LSST, and \textit{Roman}, will open new windows on the Universe, allowing us to infer physical properties for tens of millions of galaxies. Machine learning methods are increasingly becoming the most efficient tools to handle this enormous amount of data, because they are often faster and more accurate than traditional methods.
We investigate how well redshifts, stellar masses, and star-formation rates (SFR) can be measured with deep learning algorithms for observed galaxies within data mimicking the \Euclid and \textit{Rubin}/LSST surveys. We find that Deep Learning Neural Networks and Convolutional Neutral Networks (CNN), which are dependent on the parameter space of the training sample, perform well in measuring the properties of these galaxies and have a better accuracy than methods based on spectral energy distribution fitting. 
CNNs allow the processing of multi-band magnitudes together with $H_{\scriptscriptstyle\rm E}$-band images. We find that the estimates of stellar masses improve with the use of an image, but those of redshift and SFR do not. Our best results are deriving i) the redshift within a normalised error of less than 0.15 for 99.9$\%$ of the galaxies with S/N>3 in the $H_{\scriptscriptstyle\rm E}$-band; ii) the stellar mass within a factor of two ($\sim0.3 \rm dex$) for 99.5$\%$ of the considered galaxies; iii) the SFR within a factor of two ($\sim0.3 \rm dex$) for $\sim$70$\%$ of the sample. We discuss the implications of our work for application to surveys as well as how measurements of these galaxy parameters can be improved with deep learning.

\end{abstract}




\section{Introduction}
Understanding the physical processes driving galaxy evolution is one of the most outstanding issues in astronomy today.  It is clear that galaxies are assembling through star formation and mergers through cosmic time \citep[e.g.,][]{Madau2014,Conselice2014}, and morphologically evolve from irregular/peculiar galaxies at $z > 1$ to more normal regular systems at lower redshifts \citep[e.g.,][]{Mortlock2013}. Whilst these basic features are now well understood in a generalised way within an evolving galaxy population, the exact details of this process are still unknown but there will be significant improvements in their understanding in the next decade with large telescope projects such as the \Euclid Space Telescope \citep{Laureijs2011}, the \textit{Vera C. Rubin} Observatory \citep[\textit{Rubin}/LSST][]{Ivezic2008}, and the \textit{Nancy Roman} Space Telescope \citep{WFIRST}.

One of the most important ways for carrying out the analysis of galaxy evolution is through measuring properties of galaxies, like stellar mass and star-formation rates (SFRs), at different distances (or redshifts). These properties of galaxies can be difficult to measure accurately even through standard methodologies \citep[e.g.][]{Ciesla2017,Bisigello2016,Bisigello2017}, which are generally based on fitting the galaxy spectral energy distribution (SED) with theoretical or empirical models. In the epoch of large data projects such as \Euclid and LSST the use of these standard techniques will require huge computing power in order to measure these properties for the hundreds of millions of galaxies that will be observed within these data. Therefore, these large surevys will require the use of methods that go beyond the traditional ones for an efficient and accurate data analysis (i.e. statistical methods, including those based on machine learning). This shows the necessity of improving machine learning algorithms in the near future.\par
However, except for some pioneering works \citep[e.g.][]{Tagliaferri2003,Hoyle2016,StensboSmidt2016,DIsanto2018,DelliVeneri2019,Surana2020,Mucesh2021,Razim2021}, the measurements of redshift, stellar mass and SFR in an automatic way with machine learning is still largely under development. 
Other measurements, such as the galaxy morphology and structure, e.g. CAS parameters \citep[i.e. concentration, asymmetry, clumpiness,][]{Conselice2003}, have been more extensively tested and can indeed be retrieved through deep learning methods \cite[e.g.,][]{Cheng2020, tohill2021}. \par

In a recent work by the \citet{Euclidz2020}, a careful comparison was performed between the photometric redshift obtained with different standard and machine learning techniques, showing the strengths and weaknesses of both methods. In particular, the photometric redshift measurement obtained with machine learning is challenging where the colour space regions of the training sample are not well covered, while traditional methods have issues at very low-\emph{z} (i.e., $z<0.5$), at least when considering optical and near-IR filters, perhaps because of a lack of a valid set of templates or priors. Thus, various methods should be investigated to determine the optimal ways to measure these properties.\par
In general, galaxy images can contain more information than integrated magnitudes, as the morphology, size, and the presence of companions hold information about their nature. SED fitting methods and machine learning networks used to derive physical properties are mainly based on integrated quantities. However, recent works \citep{Hoyle2016,Pasquet2019} have shown the power of using images to derive photometric redshifts and the improvement caused by adding morphological information when estimating stellar masses \citep{Dobbles2019}. A similar analysis on the direct use of images to estimate stellar masses and SFR is however still missing. \par
In this paper we discuss if we can retrieve the most basic galaxy properties from deep learning neural networks (DLNNs\footnote{A DLNN is similar to a Multilayer Perceptron, but it has more than one fully-connected layer, as it happens in our case.}) and from convolutional neural networks (CNNs), which indeed can make use of galaxy images. We make use of the Cosmos Evolution Survey \citep[COSMOS;][]{Scoville2007} field and the catalogue from \citet{Laigle2016} as well as imaging from the COSMOS-Drift And SHift \citep[COSMOS-DASH;][]{Mowla2019} survey with the \HST (HST) Wide Field Camera 3 (WFC3). Thanks to these data, we verify if we are able to retrieve with \Euclid \citep[$I_{\scriptscriptstyle\rm E}$ previously called VIS, $Y_{\scriptscriptstyle\rm E}$, $J_{\scriptscriptstyle\rm E}$, $H_{\scriptscriptstyle\rm E}$;][]{Schirmer2022} and \textit{Rubin}/LSST filters ($u$, $g$, $r$, $i$, $z$) the same SFRs and stellar masses derived through SED fitting, but based on a larger number of filters (i.e. 30 from ultra-violet to near-infrared). This would mean that the machine learning networks are able to correctly interpolate between filters to retrieve the same output quantities using less input information.
In this work we perform a first step by deriving the point estimates of redshift, stellar mass, and SFR, while we leave to a future work the complex analysis of the uncertainties and the probability distribution functions associate to each of them. \par
This paper is organised as follows. In Section \ref{sec:mock} we introduce the mock Euclid catalogue and the simulated $H_{\scriptscriptstyle\rm E}$-band images. In Section \ref{sec:ML} we describe the machine learning algorithms considered, while Section \ref{sec:propderiv} contains the redshift, stellar mass, and SFR estimates. The main findings are summarised in Section \ref{sec:end}.
In this paper we consider a $\Lambda$CDM cosmology with $H_0=70\,\kmsMpc $, $\Omega_{\rm m}=0.27$, $\Omega_\Lambda=0.73$, a \citet{Chabrier2003} initial mass function (IMF), and all magnitudes are in the AB system \citep{Oke1983}.

\section{mock observations}\label{sec:mock}
\begin{figure}
    \centering
    \includegraphics[width=0.9\linewidth, keepaspectratio]{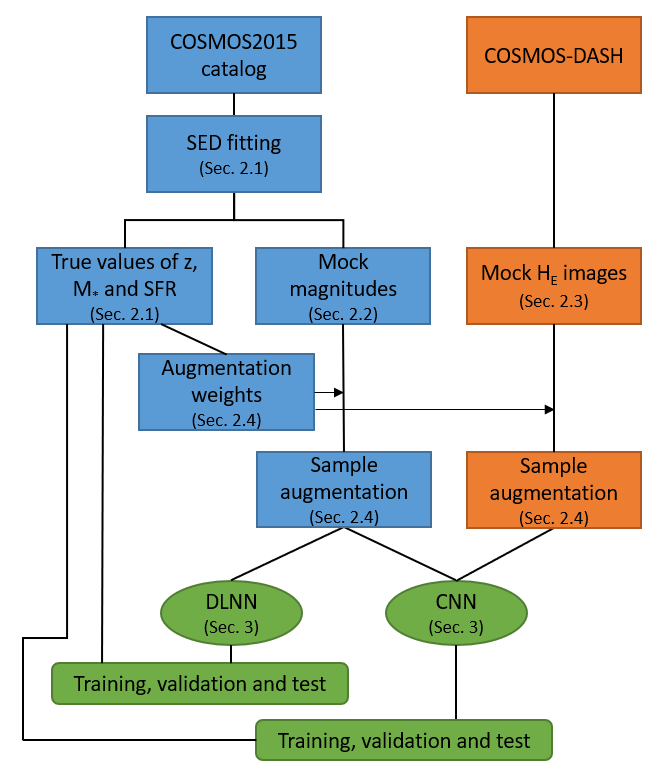}
    \caption{Workflow illustrating the different steps considered in this work. We highlight in blue the parts derived from the COSMOS2015 catalogue and in orange the part derived from the COSMOS-DASH survey. The green part indicates the two machine learning networks considered (see Sec. \ref{sec:ML}).}
    \label{fig:workflow}
\end{figure}

In the next sections we report the procedure considered to derive mock \Euclid magnitudes and Euclidized $H_{\scriptscriptstyle\rm E}$-band images starting from observed galaxies. These are the inputs required by the two neural networks analysed in this work, which are described in details in Section \ref{sec:ML}. In Figure \ref{fig:workflow} we report the full workflow to guide the reader.

\subsection{Mock catalogue and SED fitting procedure}\label{sec:cat}
In this work we made use of an updated version of the mock catalogue of the Euclid Wide Survey presented by \citet{Bisigello2020}. This mock catalogue was created starting from multi-wavelength observations of real galaxies in the 2 $\deg^{2}$ survey of the COSMOS field. These multi-wavelength observations consist of 30 filters, ranging from the ultra-violet to near-infrared wavelengths, and they are part of the public COSMOS2015 catalogue \citep{Laigle2016}. We removed from the original COSMOS2015 catalogue stars and X-ray sources, the latter corresponding to less than 1$\%$ of the galaxy sample.\par
Each observed galaxy was fitted comparing theoretical templates with the fluxes available in the 30 COSMOS15 filters, performing a $\chi^2$ fitting procedure using the \textit{LePhare} code \citep{Arnouts1999,Ilbert2006}. For these fits we considered a broad set of SED templates from \citet{Bruzual2003} with exponentially declining star formation histories with e-folding timescale $\tau$ between 0.1 to 10 Gyr, Solar and sub-solar metallicity (Z$_{\odot}$, 0.04 Z$_{\odot}$), ages from 0.1 to 12 Gyr, 12 values of colour excess from ${E(B-V)}=0$ to 1, and the \citet{Calzetti2000} reddening law. The redshift was fixed to the value reported in the COSMOS2015 catalogue.\par
This fitting procedure allowed us to derive, for each observed galaxy, the best theoretical template and the galaxy physical properties associated to it. In particular, for this work we are interested in the redshift, the stellar mass, and the SFR, while the best fit templates were used to derive mock \Euclid magnitudes (see next sect.). We considered the previously mentioned associated physical properties to be the ground truth. \par
We highlight that the considered physical properties have an associate uncertainty derived from the SED fitting procedure. In particular, as mentioned by \citet{Laigle2016}, the normalised median absolute deviation (NMAD) of the redshift, derived comparing the photometric redshift with available spectroscopic ones, varies from 0.007 to 0.057 moving from bright galaxies ($16<i<21$) to faint galaxies ($25<i<26$). At the same time, the median error on the stellar mass and SFR, which are derived from the output probability distribution of each object, is 0.07 and 0.16 dex, respectively. These errors are stated here, but are not considered when showing, in the next sections, the results for the different networks, as they depend on the number of filters and the method used to derived the physical properties used as the ground truth.

\subsection{Mock magnitudes}\label{sec:fluxes}
The original catalogue from \citet{Bisigello2020} was derived to mimic the Euclid Wide Survey \citep{Scaramella2021} and consists of five filters, i.e. $I_{\scriptscriptstyle\rm E}$, $Y_{\scriptscriptstyle\rm E}$, $J_{\scriptscriptstyle\rm E}$, $H_{\scriptscriptstyle\rm E}$, and the Canada-France Imaging Survey (CFSI) $u$ band. To these filters we add complementary magnitudes in the Sloan Digital Sky Survey \citep[SDSS;][]{Gunn1998}, the $g$, $r$, $i$, and $z$ filters. Observations in similar filters, such as the the ones that will be used by \textit{Rubin}/LSST, will be available to complement \Euclid observations \citep{Scaramella2021}.  \par

\begin{table}
	\caption{Observational depth (point source, $10\sigma$) in AB magnitude, central wavelength, and full width half maximum (FWHM) of the filters considered in our Euclid Wide mock catalogue.}
	\label{tab:magdepth} 
	\centering 
	\begin{tabular}{c c c c}
		\hline\hline 
		\\
		band & $10\sigma$ depth & $\lambda_{\rm cen}$ [\AA] & FWHM [\AA]\\
		\hline
		$I_{\scriptscriptstyle\rm E}$	&  24.50 & 7150 & 3640\\
		NISP/$Y_{\scriptscriptstyle\rm E}$	&  23.24 & 10\,850$^{a}$ & 2660$^{a}$\\
		NISP/$J_{\scriptscriptstyle\rm E}$	&  23.24 & 13\,750$^{a}$ & 4040$^{a}$\\
		NISP/$H_{\scriptscriptstyle\rm E}$	&  23.24 & 17\,725$^{a}$ & 5020$^{a}$\\
		CFSI/$u$  & 24.20 & 3715 & 510\\
		SDSS/$g$  & 24.50 & 4700 & 1263\\
		SDSS/$r$  & 23.90 & 6174 & 1149\\
		SDSS/$i$  & 23.60 & 7534 & 1239\\
		SDSS/$z$  & 23.40 & 8782 & 994\\
		\hline
	\end{tabular}\\
	$a$ The central wavelengths and FWHMs of the NISP filters are slightly ($<0.6\%$ and $<1.3\%$, respectively) different from the more recent values reported in \citet{Schirmer2022}. 
\end{table}
The inclusion of filters blue-ward of the \Euclid bands, such as the previously mentioned $u$, $g$, $r$, $i$, and $z$ filters, may improve the reconstruction of the overall SED template by broadening the wavelength coverage. In particular, such filters are expected to improve the derivation of the SFR, by tracing ultra-violet wavelengths \citep{Pforr2012,Pforr2013}, as well as the Lyman break (i.e. 912\AA), which is one of the most prominent feature in a galaxy spectra.   \par
We included photometric errors by scattering each magnitude around its true value, derived from its best fit SED, and considering the respective survey noise. We did not include any other source of error. In Table \ref{tab:magdepth} we report the observational depths considered, as expected for the Euclid Wide Survey, to perturb the original COSMOS2015 photometry and to perform the SED fitting procedure using \Euclid and \textit{Rubin}/LSST filters. These depths in the \Euclid bands correspond to the values presented in the \Euclid definition study report \citep{Laureijs2011} and are different from the more recent values presented in \citet{Scaramella2021}\footnote{Their median 10$\sigma$ values are 
25.45, 23.55, 23.74, and 23.65 for $I_{\scriptscriptstyle\rm E}$, $Y_{\scriptscriptstyle\rm E}$, $J_{\scriptscriptstyle\rm E}$, and $H_{\scriptscriptstyle\rm E}$, respectively.}. However, as photometric errors are not included in the machine learning networks, we do not expect these differences to impact the results presented in this paper. Photometric errors are not included as inputs because the feature analysis on a similar machine learning algorithm have shown they provide little or no information compared to just using magnitudes (Humphrey et al. in prep). At this stage, we did not apply any magnitude cut, as this will be performed later on the images, but we assigned a value of $-$1 to all magnitudes below $\rm S/N<3$.

\subsection{$H_{\scriptscriptstyle\rm E}$-band Euclidized images}

\begin{figure*}
    \centering
    \includegraphics[width=0.85\linewidth, keepaspectratio]{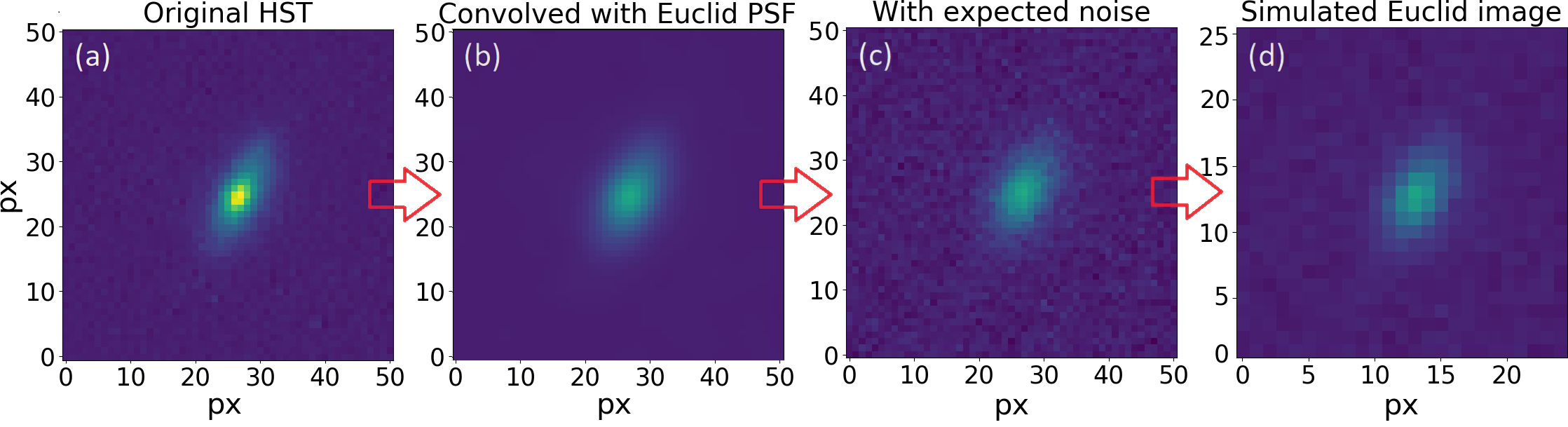}
    \caption{Example of the transformation from an HST/$F160W$ image (a) to a \Euclid $H_{\scriptscriptstyle\rm E}$ band simulated image (d) of a galaxy at $z=0.2$. In the first step we include the \Euclid PSF (b), we then include the expected photometric noise (c), and finally we apply the \Euclid angular resolution (d). The colour scale is linear and it is the same for all panels. }
    \label{fig:imagetrans}
\end{figure*}

\begin{figure*}
	\centering
	\includegraphics[width=0.3\linewidth, keepaspectratio]{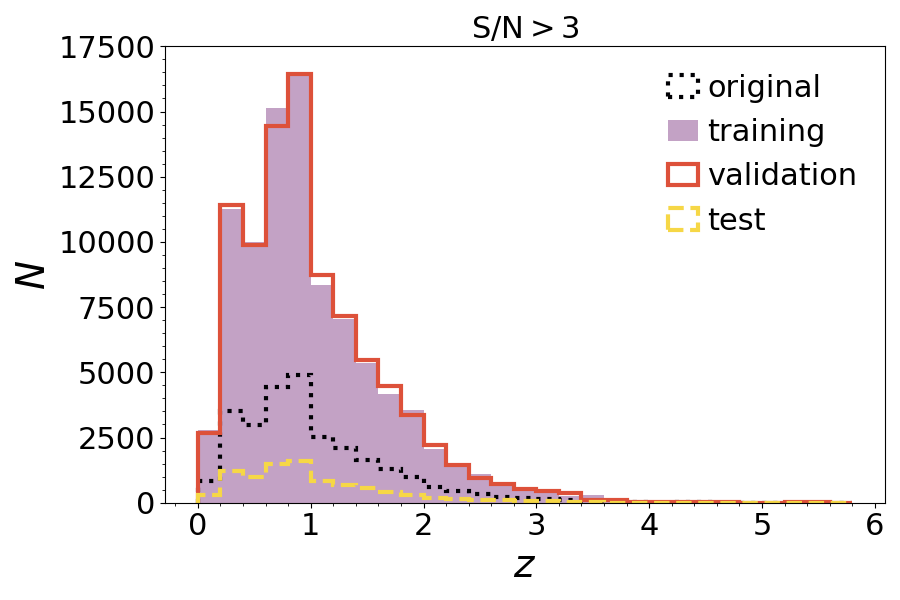}
	\includegraphics[width=0.3\linewidth, keepaspectratio]{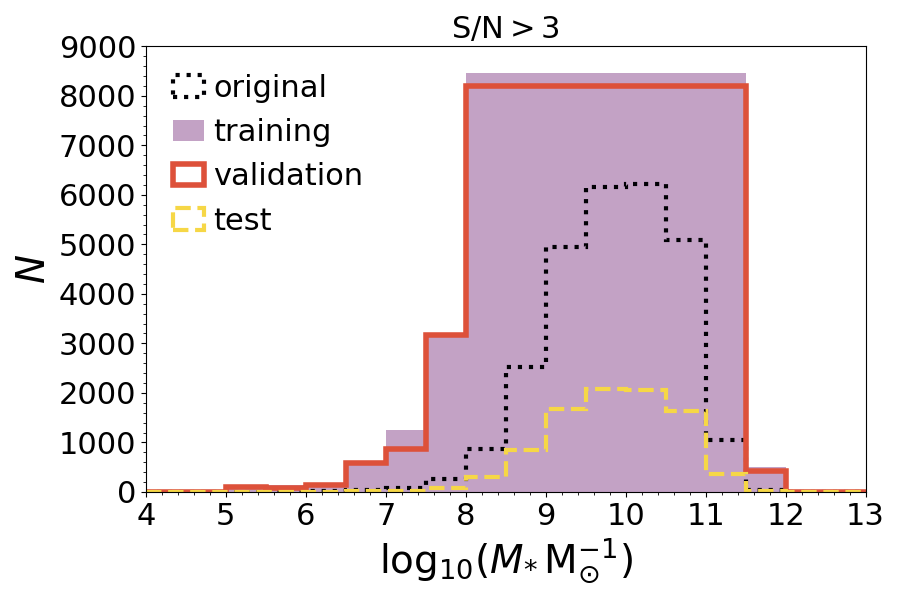}
	\includegraphics[width=0.3\linewidth, keepaspectratio]{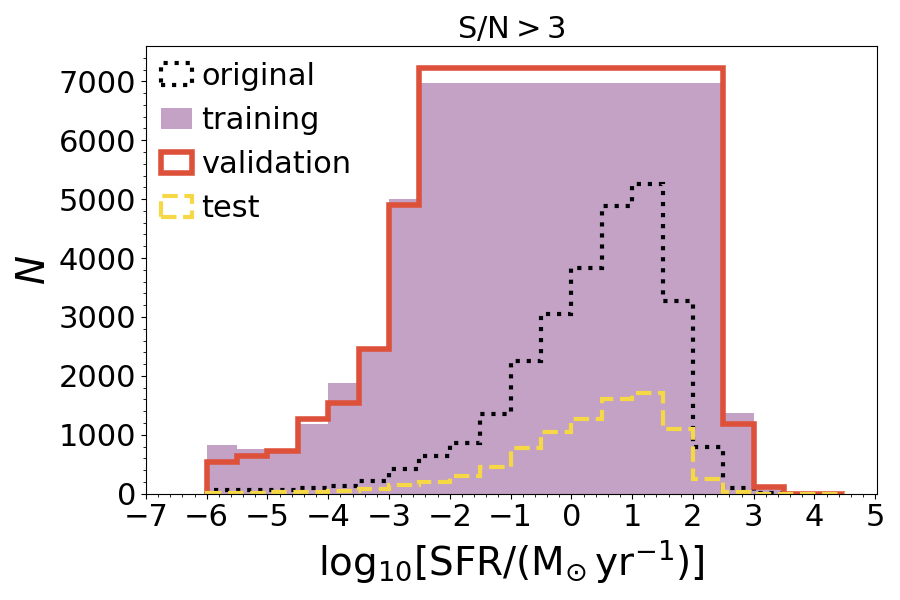}
	\includegraphics[width=0.3\linewidth, keepaspectratio]{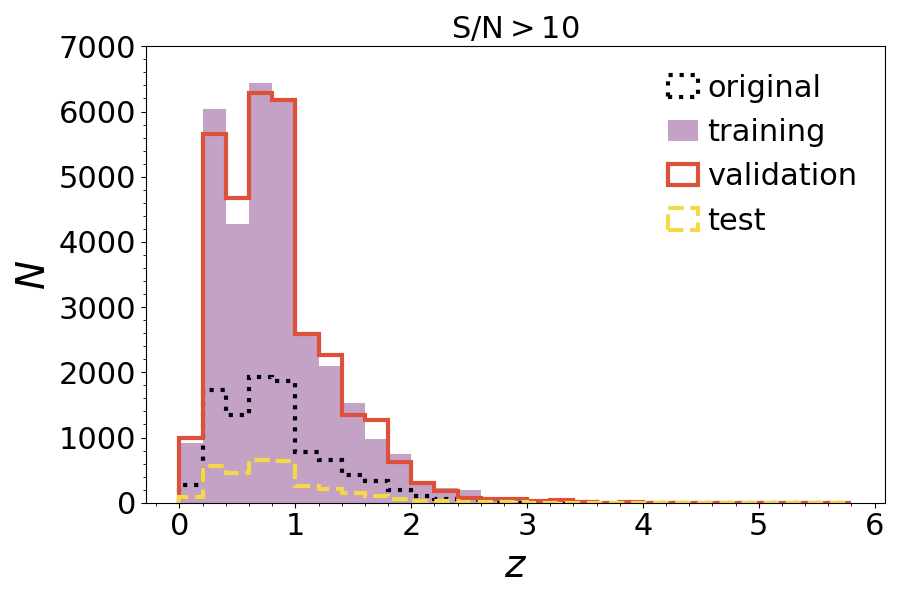}
	\includegraphics[width=0.3\linewidth, keepaspectratio]{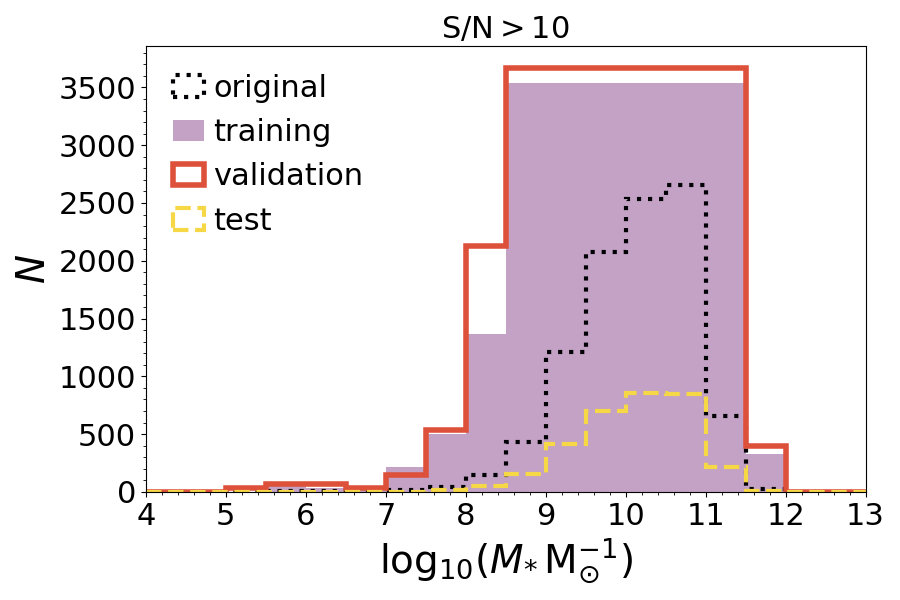}
	\includegraphics[width=0.3\linewidth, keepaspectratio]{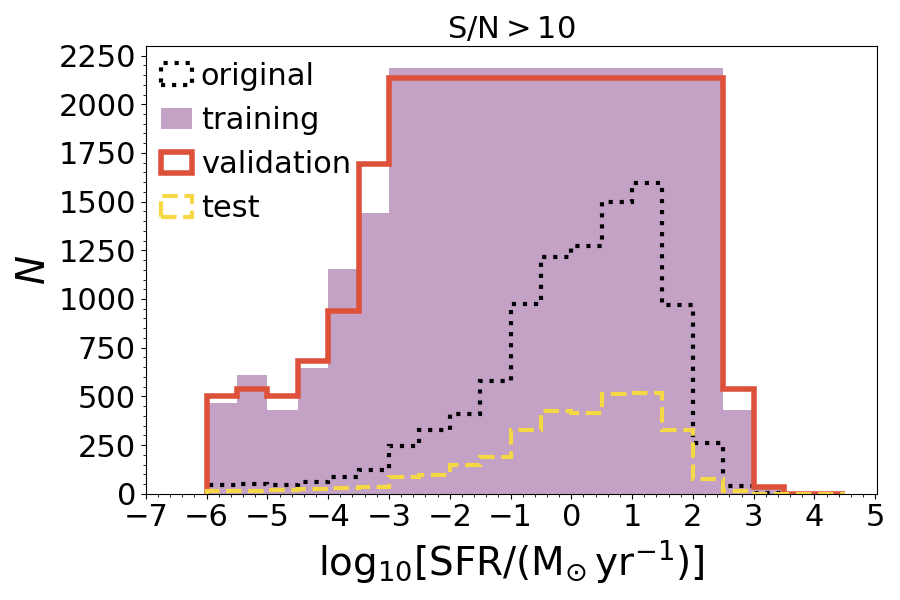}
	\caption{Distribution of the galaxies in the COSMOS field with imaging  ${\rm S/N}>3$ (\textit{top}) and  ${\rm S/N}>10$ (\textit{bottom}) in redshift (\textit{left}), stellar mass (\textit{centre}), and SFR (\textit{right}). The distributions are for the training sample (\textit{filled purple}) and the validation sample (\textit{solid red}) after applying sample augmentation, and for the test sample (\textit{dashed yellow}) and the original sample (\textit{dotted black}) without augmentation. Sample augmentation for SFR and stellar mass is performed to obtain a distribution which is as flat as possible, to avoid biases when training. }
	\label{fig:distributions}
\end{figure*}

In this work, we consider machine learning networks that have as inputs not only magnitudes, but also images. In particular, we make use of simulated images in the $H_{\scriptscriptstyle\rm E}$-filter instead of images in the $I_{\scriptscriptstyle\rm E}$ filter. On one hand, the $I_{\scriptscriptstyle\rm E}$ filter is more sensitive, as it covers a wide 530-920 nm wavelength range (equivalent to $r$, $i$, and $z$ together), and it has three times higher angular resolution, together implying more complex modelling. On the other hand, observations in filters similar to the \Euclid $H_{\scriptscriptstyle\rm E}$-band are already available from HST. \par
We derived our simulated $H_{\scriptscriptstyle\rm E}$-band images from the HST-WFC3 Imaging Survey in the COSMOS Field \citep[COSMOS-DASH;][]{Mowla2019}, which covers a large fraction of the COSMOS field. We consider the HST/$F160W$ images because they correspond to a filter close in wavelength to the  $H_{\scriptscriptstyle\rm E}$ filter. We did not apply any $k$-correction to convert from the HST/$F160W$ filter to the  $H_{\scriptscriptstyle\rm E}$ band. This would imply the use of a SED model and would change the flux values of each pixel but not the morphological features, which are the ones relevant for the CNN analysis. \par

We created $H_{\scriptscriptstyle\rm E}$-band images starting from HST/$F160W$ thumbnails of 51$\times$51 pixels centred around each galaxy. We derive the  S/N of each image by retrieving the background flux and the noise from the median and standard deviation of the fluxes on an area of 51$\times$51 pixels, with the central 18$\times$18 pixels masked to remove the source. For the noise, we derive the value present in the image by calculating the standard deviation after applying a 3$\sigma$-clipping procedure. We then add an additional noise, applying  a scatter from a Gaussian distribution, in order to reproduce the expected \Euclid noise. The signal on the source was then roughly derived from the central 6$\times$6 pixels, which correspond to a square of $\sim\ang{;;0.7}\times\ang{;;0.7}$ \footnote{This procedure, applied to the original HST images, is expected to underestimate the flux of local extended galaxies, for which a specific analysis on the estimation of physical properties will be considered in a future work.}. The  S/N was then derived by subtracting the background from the signal and dividing the result by the retrieved noise. We restricted our analysis to galaxies with a  ${\rm S/N}>3$, to avoid training the networks with images dominated by noise. \par

As a second step, we apply the \Euclid NISP $H_{\scriptscriptstyle\rm E}$-band point-spread-function (PSF)\footnote{In order to have a final image consistent with the \Euclid PSF, we apply a PSF with a standard deviation $\sigma^{2}=\sigma^{2}_{\Euclid}-\sigma^{2}_{\rm HST}$.}, photometric noise, and spatial resolution. Both the \Euclid ($\sim$\ang{;;0.7}) and HST PSFs were approximated by two-dimensional Gaussian functions. In reality, both PSFs are highly non-Gaussian in the wings, we do not expect this to impact extremely our results, as the central small spatial resolution should dominate the training, but it is recommended to train the network on real images in the future. In this way, we start from an HST/$F160W$ image of 51$\times$51 pixels centred on each observed galaxy and we obtain a $H_{\scriptscriptstyle\rm E}$-band simulated image of 25$\times$25 pixels. These sizes generally include the entire galaxy in each image and allow for a sample augmentation through rotation without image loss. In Figure \ref{fig:imagetrans} we give an example of the transformation from an observed HST/$F160W$ image to a $H_{\scriptscriptstyle\rm E}$ band simulated image.\par

In the last step, we matched the COSMOS-DASH catalogue with the COSMOS2015 catalogue, considering a matching radius of 1\arcsecond. In this way we linked each $H_{\scriptscriptstyle\rm E}$-band simulated image to the set of mock magnitudes described in the previous section. We then separate the catalogue into two sub-samples with different signal-to-noise cuts, i.e.  ${\rm S/N}>3$ and  ${\rm S/N}>10$, with the  S/N derived, as mentioned before, on the $H_{\scriptscriptstyle\rm E}$-band simulated images. These sub-samples correspond to 27\,340 and 9799 COSMOS-DASH galaxies with a  ${\rm S/N}>3$ and 10, respectively. The redshift, stellar mass, and SFR distributions of both samples are reported in Figure \ref{fig:distributions}.

\subsection{Sample augmentation}\label{sec:aug}

Sample augmentation is necessary to increase the number of objects and improve the training of the different algorithms. This was done for the magnitudes in the training and validation samples (see Section \ref{sec:ML}) randomly extracting their values from a Gaussian distribution centred on the true flux values and with a dispersion equal to the photometric noise, as expected for the Euclid Wide Survey. We then convert fluxes to magnitudes. On the other hand, to increase the number of images available, we rotated each of them by 10\degree{} for a maximum of 35 times, this change is sufficient to make the network recognise each image as a new one. This method has been often applied in the literature \citep[e.g.,][]{Dieleman2015,HuertasCompany2015} and it has been demonstrated to improve machine learning classifications \citep{Cheng2020}. \par

For the redshift derivation, we increased the number of sources in the catalogue by a factor of 10, while for the stellar mass and SFR, whose measurements are more challenging, we increased the number of objects up to a factor of 35 in order to obtain a flat distribution (Figure \ref{fig:distributions}). This is performed to avoid biases on the training (i.e. most present galaxy having the best estimation) and to obtain an estimation that is similar over a range in mass or SFR as large as possible. However, the number of galaxies with  ${\rm S/N}>3$ (10) and $\logten[{\rm SFR}/(\Msolar {\rm yr}^{-1})]<-2.5$ ($-3$) or $\logten(M_{*}\,\Msolar^{-1})<8$ (8.5) are very low and the resulting SFR and stellar mass distributions are flat only above these values. In addition, we applied a rough conversion of the 3$\sigma$ $I_{\scriptscriptstyle\rm E}$ photometric depth (i.e. $I_{\scriptscriptstyle\rm E}=25.81$, see Table \ref{tab:magdepth}) to SFR by using the relation by \citet{Kennicutt1998a}, which links the ultraviolet luminosity to the un-obscured component of the SFR, and converting it to \citet{Chabrier2003} IMF. This conversion is not possible at $z<1$, as the $I_{\scriptscriptstyle\rm E}$ does not trace the UV light, but it is useful to give a rough estimate of the limits imposed by the photometric noise to the SFR estimation. Indeed, we expect the photometric noise to limit the SFR estimates to $\logten[{\rm SFR}/(\Msolar {\rm yr}^{-1})]>-0.9$ (0.3) at $z=1$ (3). In the following analysis we take into account these effects, showing the results for the full sample and for the sub-sample with a completely flat distribution in stellar mass. For the SFR we instead show the results for the full sample and for the sample with $\logten[{\rm  SFR}/(\Msolar {\rm yr}^{-1})]>0$.  \par
After augmentation, the training samples used for the redshift measurements have 91\,130 galaxies with  ${\rm S/N}>3$ and 32\,660 with  ${\rm S/N}>10$. The augmented samples for the stellar mass (SFR) have instead 63\,295 (82\,899) galaxies with  ${\rm S/N}>3$ and 24\,296 (29\,065) objects with  ${\rm S/N}>10$. Each  S/N cut refers to the $H_{\scriptscriptstyle\rm E}$-band images, as mentioned in the previous Section. The sizes of the samples correspond to the maximum size possible given the input dataset and the augmentation procedure explained. 

\section{Machine learning algorithms}\label{sec:ML}

In this work we considered two different machine learning algorithms, a DLNN, which has as input exclusively tabular data, and a CNN, which includes also images. We investigate these two different machine learning algorithms to determine the performance of both for measuring different physical parameters of galaxies and how they compare with each other. In the future, this work may be extended considering a one-dimensional CNN instead of a DLNN, which may better capture features in the SED, and a more complex CNN with multiple input images. \par 
\subsubsection*{Deep Learning Neural Network}
In the DLNN, we used two different sets of inputs: a) the four \Euclid magnitudes and b) the four \Euclid magnitudes complemented with the magnitudes in the $u$ CHFT and $g, r, i, z$ SDSS ground-based filters. These ground-based filters will be available through ancillary surveys with different facilities \citep{Scaramella2021}. In both cases, we considered the sample with $H_{\scriptscriptstyle\rm E}$-band images with  ${\rm S/N}>3$ and the sub-sample with  ${\rm S/N}>10$, training the network separately for each  S/N cut. \par

\begin{table}
    \centering 
	\caption{The DLNN architecture used in this paper. Linear is a fully-connected layer that applies a linear transformation. We also include a Rectified Linear Unit function between each linear layer.}\label{tab:NN} 
	\begin{tabular}{c c c}
		\hline\hline 
		Layer & $N_{\rm input}$ & $N_{\rm output}$\\
		\hline
		linear	&  $N^{a}$ & 2000\\
		linear	&  2000 & 1000\\
		linear	&  1000 & 500\\
		linear	&  500 & 1 \\
		\hline
	\end{tabular}\\
    $^a$ $N$ is equal to the number of input filters (i.e., 4 or 9) for the single runs, and it is equal to the number of runs, i.e. ten, for the Meta-learner
\end{table}

The architecture of the DLNN is summarised in Table \ref{tab:NN} and consists of four linear layers with the number of neurons ranging from 500 to 2000. Each neuron of each layer is fully connected with the neurons of the previous and the following layer and these connections are updated during the training in order to derive the optimal way to map the input integrated magnitudes into the desired physical properties. Among each linear layer there is a Rectified Linear Unit \citep[ReLu; ][]{Nair2010}, such that $f(x)=0$ if $x<0$ and $f(x)=x$ if $x\geq0$. A similar architecture has been previously used, even if with different inputs, number of neurons and hidden layers, to derive photometric redshift \citep[e.g.][]{Firth2003,Collister2004}. \par
As a simple and direct test, in this work we kept the same architecture when changing the set of inputs (e.g. \Euclid only vs. \Euclid + LSST filters). However, we have not fully explored all the possible combinations of number of neurons and hidden layers, so there may be some architectures than optimise the use of the different set of inputs separately. This would have an impact on the absolute precision of the networks, but this is expected to leave the qualitative statements of the paper untouched. A full optimization will be perform in the future, once real data becomes available. \par

\subsubsection*{Convolutional Neural Network}
\begin{table}
	\caption{The CNN architecture used in this paper. Conv2d indicates a two-dimensional convolutional layer, Max pool corresponds to a pooling layer using the maximum value to down-sample each image, Linear is a fully-connected layer that applies a linear transformation. We included a Rectified Linear Unit function between each convolutional or linear layer. Galaxy images are introduced in the first Conv2d layer, while fluxes are included in the first Linear layer.}\label{tab:CNN} 
	\centering 
	\begin{tabular}{c c c c}
		\hline\hline 
		Layer & kernel size & $N_{\rm input}$ & $N_{\rm output}$\\
		\hline
		Conv2d	&  3$\times$3 & 1$\times$18$\times$18 & 64$\times$16$\times$16\\
		Conv2d & 3$\times$3 & 64$\times$16$\times$16 & 64$\times$14$\times$14 \\
		Max pool & 2$\times$2 & 64$\times$14$\times$14 & 64$\times$13$\times$13\\
		Conv2d	&  3$\times$3 & 64$\times$13$\times$13 & 128$\times$11$\times$11\\
		Conv2d	&  3$\times$3 & 128$\times$11$\times$11 & 128$\times$9$\times$9\\
		Max pool&  2$\times$2 & 128$\times$9$\times$9 & 128$\times$8$\times$8\\
		Conv2d	&  3$\times$3 & 128$\times$8$\times$8 & 256$\times$6$\times$6\\
		Conv2d	&  3$\times$3 & 256$\times$6$\times$6 & 256$\times$4$\times$4\\
		Conv2d	&  3$\times$3 & 256$\times$4$\times$4 & 256$\times$2$\times$2\\
		Max pool&  2$\times$2 & 256$\times$2$\times$2 & 256$\times$1$\times$1\\
		linear & -- & 256+N$^{a}$ & 2000 \\
		linear & -- & 2000 & 1000 \\
		linear & -- & 1000 & 500 \\
		linear & -- & 500 & 1\\
		\hline
	\end{tabular}\\
    $^a$ N is equal to the number of input filters (i.e., 4 or 9)
\end{table}
The CNN has as input either of the sets of magnitudes of the DLNN (i.e. only \Euclid filters or \Euclid and ancillary filters), but it also includes the simulated $H_{\scriptscriptstyle\rm E}$-band images. The architecture of this second network is summarised in Table \ref{tab:CNN}. It consists of a series of convolutional layers applied to the $H_{\scriptscriptstyle\rm E}$-band images, whose outputs are then combined with the magnitudes and processed through a set of linear layers. The convolutional layers are key for identifying features and shapes inside each $H_{\scriptscriptstyle\rm E}$-band image. We chose a deep network with $3\times3$ kernels, instead of a network with a less layers but larger kernels, as this architecture make the decision function more discriminative and reduces the number of free parameters \citep{Simonyan15}. \par
After every two convolutional layers we applied a max-pooling layer, which is used for down-sizing the images which reduces the number of parameters of the network. As for the DLNN, the linear layers are interspersed with ReLu functions. The CNN is trained separately with the two sub-samples with different  S/N cuts in the $H_{\scriptscriptstyle\rm E}$-band images. \par
As a preliminary approach, we decided to derive, in both the CNN and DLNN runs, the redshift, stellar mass, and SFR independently. In this way we can investigate the challenges of the derivation of the three properties separately. We leave the combined analysis to a future work, but as reported in Appendix \ref{sec:MS}, we do not find evidence of galaxies with unrealistic combinations of physical properties, e.g. very low mass galaxies at very high-$z$. \par
\subsubsection*{Re-scaling}
We re-scaled all input parameters, i.e. stellar mass, SFR, redshift, and magnitudes, in order to have values between 0 and 1. This is performed by subtracting from each parameter its minimum value and dividing it by the difference between its maximum and minimum values. This is performed in logarithmic scale for the stellar mass and SFR and in linear scale for the redshift. We do not consider magnitudes below  ${\rm S/N}<3$ in the re-scaling, as we assigned a value of -1 to all of them (Sec. \ref{sec:fluxes}). A similar re-scaling is also applied to each simulated $H_{\scriptscriptstyle\rm E}$-band  image in order to have pixel values between 0 and 1. The same re-scaling is applied to the entire sample, but it is calculated using only the galaxies considered for the training (see later). This re-scaling is an important step in machine learning as the inputs and outputs may differ over orders of magnitude and, therefore, the largest one may dominate the training process. It is necessary to keep in mind, when comparing the CNN to the DLNN in the next sections, that because of the re-scaling performed separately for each galaxy, each $H_{\scriptscriptstyle\rm E}$-band image has lost information about the overall galaxy flux and mainly contains the information on features and shapes, which is what we aim to train on. \par
\subsubsection*{Hyper-parameters}
We also divide the full sample into batches of 200 objects, which are used serially to update the training process, to increase stochasticity, and reduce the problem of local minima.  In both the CNN and DLNN runs we implement an Adam algorithm \citep{Kingma2015}, which is an optimisation function based on stochastic gradient descent, to optimise the hyper-parameters of the networks. To evaluate the difference between the data and the predictions, for each update of the network we derived a loss function based on the mean squared error (MSE), i.e. $l(x,y)=\sum_{n=0}^{N}(x_{n}-y_{n})^{2}/N$, where $x$ are the predicted values, $y$ the target ones, and $N$ is the total number of galaxy in input. \par
\subsubsection*{The training, the validation and the test samples}
We randomly split all the samples in three sub-samples, of equal numbers, and we then apply augmentation only to the training and validation samples (see Figure \ref{fig:distributions} and Section \ref{sec:aug}).  Each network is trained with the first sub-sample (training sample), while the second sub-sample (validation sample) is used to estimate in an independent way the loss function and stop the training when it converges to avoid over-fitting (i.e. over learning features specific of the training sample), which may happen if we only analyse the loss function derived with the sub-sample used for training. The third sub-sample (test sample) is never seen by the networks, and is only used to evaluate the results. This sub-sample is not augmented so that the final statistics corresponds to a realistic galaxy sample. The split is performed once for all networks and is done before augmentation to avoid having the same object present in the training and in the test sample. \par
\subsubsection*{Networks combination methods}
All networks are run ten times using different randoms seeds. This number was chosen as a compromise between computational time and stochasticity. We then combine these runs with three different approaches. We report the results for the following: 
\begin{itemize}
    \item The best network, defined as the network with the smallest outlier fraction considering the full sample for the redshift estimation, then the sub-sample with a flat stellar mass distribution for the stellar mass derivation, and then the sub-sample above ${\rm SFR}>1 \Msolar {\rm yr}^{-1}$ for the SFR measurements.
    \item The median of the outputs of the ten networks.
    \item A Meta-learner \citep[][Humphrey et al. in prep.]{Wolpert1992} that is an additional machine learning network used as a linear discriminant among the different runs. This allows us to take into account the fact that some runs may have identified features peculiar to a subset of data. This Meta-learner consists of a DLNN with the architecture shown in Table \ref{tab:NN}, but it uses the results of the ten runs, instead of the magnitudes, as inputs.
\end{itemize}

\section{Galaxy properties derived with machine learning}\label{sec:propderiv}

In this section we report and discuss the results for the redshift, stellar mass, and SFR estimates based on machine learning methods. We highlight that in order to train the machine learning algorithms we need to have a sample with known output values. In this work we rely on simulated data and with real data we could rely on spectroscopic redshifts and SFRs derived from a combination of different tracers (e.g. ultraviolet and infrared stellar continuum). However, there is not an equivalent method to derive the true stellar mass of galaxies and we need to rely on the SED fitting applied to a sub-sample of galaxies with plenty of ancillary data. The power of machine learning is the capability of deriving the properties with a better accuracy.\par
For comparison, we also report the results derived with the same SED fitting procedure used to retrieve mock magnitudes, but using both the four \Euclid filters and the nine \Euclid and ancillary filters, as inputs. This of course corresponds to an ideal situation, as the same code and the same set of templates are used to retrieve the mock magnitudes and to estimate physical properties. It is however necessary to take into account that other SED fitting codes may perform differently, not only because of different SED libraries, but also because of the use of priors, which are instead not used here. This test is anyway useful for a direct comparison with the machine learning algorithms considered.\par

\subsection{Computational performance}
One of the main advantages of machine learning algorithms is the time necessary to derive the desired results. In general, the time necessary to apply a SED fitting procedure, regardless of the considered code, depends on the number of templates considered. For example, with the setup considered in this work, i.e., 14 SED templates (see Section \ref{sec:fluxes}), 12 dust extinction values, 23 age values, and with redshift steps of 0.05 up to $z=6$, \textit{LePhare} takes 0.23 seconds per object\footnote{\label{pc} This is using a machine with 12 central processing units of 3.20 GHz and a 16 GB random access memory.}, requiring more than 4 hours for $\sim63\times10^{3}$ objects. Conversely, the DLNN training requires around 20 to 40 minutes, depending on the sample size (e.g. $\sim24$ or $63\times 10^3$ objects), while the evaluation requires less then a minute in total for the same 63$\times 10^{3}$ objects and using the same machine. The CNN, using a graphics processing unit, requires a longer time for training, up to 13 hours, and for evaluating ($\sim 15$ minutes), given the larger complexity of the set of inputs. There is in any case a huge improvement on time cost moving from SED fitting codes to DLNN or CNN, as machine learning networks, once the training is performed, require only to apply a set of linear transformations, or convolution for the CNN, to calculate the output values, while a SED fitting procedure requires more complex steps, e.g. chi-square derivation for each combination of SED template and object. \par

\subsection{Redshift derivation}\label{sec:redshift}

\begin{table}
	\caption{Statistics of the redshift derivation. Columns are: 1) considered algorithms, 2) numbers of input filters $N_{\rm in}$, 3)  S/N cuts, 4) method used to combine the ten runs of each network, 5) fraction of outliers, 6) bias, 7) NMAD, and 8) MSE. For the combination method, we include the results of the best run, the median among the ten runs, and the results for the Meta-learner applied to the ten runs. The first four lines correspond to results derived with SED fitting.}
	\label{tab:fout_z} 
	\centering 
	\resizebox{0.47\textwidth}{!}{
	\begin{tabular}{c c c c c c c c}
		\hline\hline 
		Algorithm & $N_{\rm in}$ &  S/N & Combination & $f_{\rm out}$ & $\langle\Delta z\rangle$ & NMAD & MSE\\
		(1) & (2) & (3) & (4) & (5) & (6) & (7) & (8) \\
		\hline
        SED  & 4 & 3 & & 0.604 & 0.090 & 0.327 & 0.442\\
        SED  & 4 & 10 & & 0.596 & 0.105 & 0.315 & 0.280\\
        SED  & 9 & 3 & & 0.127 & $-$0.002 & 0.045 & 0.081\\
        SED  & 9 & 10 & & 0.040 & $-$0.003 & 0.029 & 0.028\\
		\hline
		DLNN	 &  4 &  3 & best & 0.099 & 0.011 & 0.052 & 0.014\\
		 	     &    &    & median & 0.103 & 0.011 & 0.050 & 0.014\\ %
		    	 &    &    & Meta-learner & 0.088 & 0.005 & 0.050 & 0.014\\
		DLNN	 &  4 & 10 & best & 0.076 & 0.005 &  0.050 & 0.010\\
		    	 &    &    & median & 0.081 & 0.003 &  0.050 & 0.010\\ %
		    	 &    &    & Meta-learner & 0.068 & 0.004 &  0.048 & 0.010\\
		CNN	&  4 & 3  & best & 0.133 & 0.015 & 0.073 & 0.017\\
		   	&    &    & median & 0.138 & 0.009 & 0.071 & 0.017\\ %
		   	&    &    & Meta-learner & 0.119 & 0.008 & 0.064 & 0.015 \\
		CNN	&  4 & 10 & best & 0.133 & $-$0.012 & 0.077 & 0.014\\
		 	&    &    & median & 0.144 & $-$0.001 & 0.081 & 0.015\\ %
		 	&    &    & Meta-learner & 0.117 & 0.013 & 0.071 & 0.013\\
		DLNN	 &  9 & 3 & best &  0.001 & $-$0.002 &  0.008 & 0.000 \\
		   	     &    &   & median &  0.002 & $-$0.001 & 0.010 & 0.001\\ %
		 	     &    &   & Meta-learner &  0.001 & 0.001 &  0.006 & 0.000\\
		DLNN	 &  9 & 10 & best &  0.002 & 0.001 & 0.013 & 0.001\\
		    	 &    &    & median &  0.002 & 0.001  & 0.014 & 0.001\\ %
		    	 &    &    & Meta-learner &  0.002 & 0.000 & 0.010 & 0.000\\
		CNN	&  9 & 3  & best &  0.002 & 0.005 & 0.028 & 0.001\\
		 	&    &    & median &  0.003 & $-$0.001 & 0.022 & 0.001\\ %
		 	&    &    & Meta-learner &  0.002 & $-$0.003 & 0.017 & 0.001\\
		CNN	&  9 & 10 & best  & 0.003 & $-$0.009 & 0.030 & 0.001\\
		 	&    &    & median  & 0.005 & 0.000 & 0.027 & 0.001\\ %
		 	&    &    & Meta-learner  & 0.002 & $-$0.003 & 0.023 & 0.001\\
		\hline
	\end{tabular}}
\end{table}

\begin{figure*}
	\centering
	\includegraphics[width=0.23\linewidth, keepaspectratio]{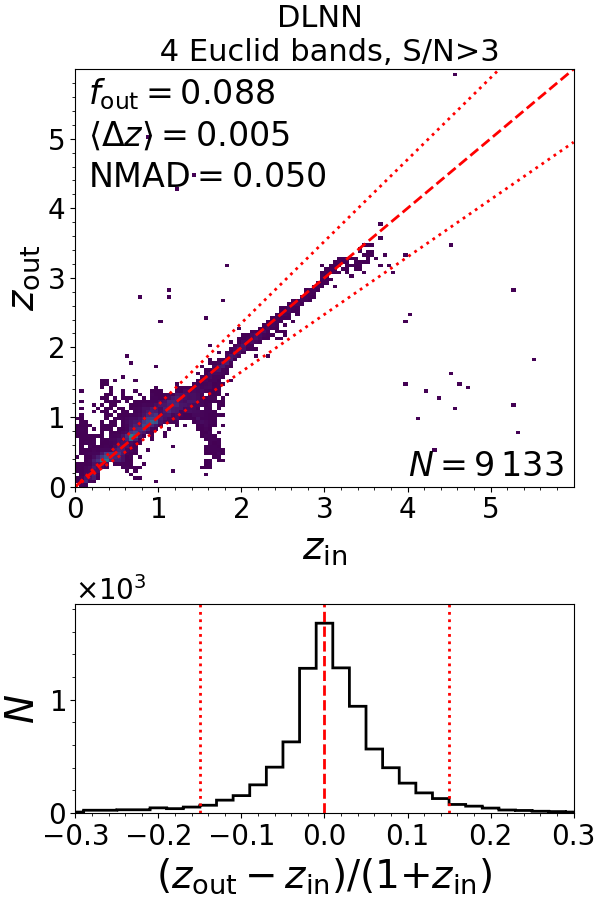}
	\includegraphics[width=0.23\linewidth, keepaspectratio]{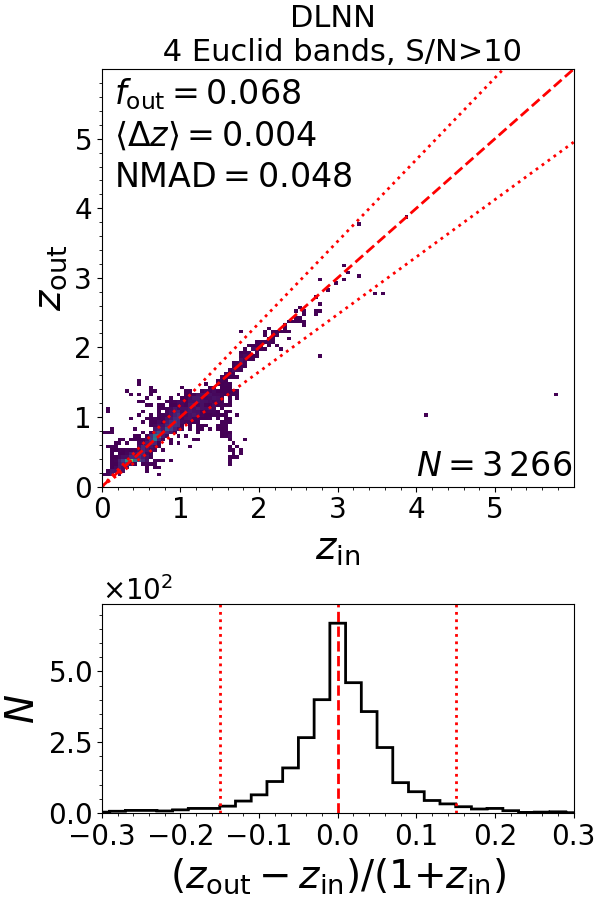}
	\includegraphics[width=0.23\linewidth, keepaspectratio]{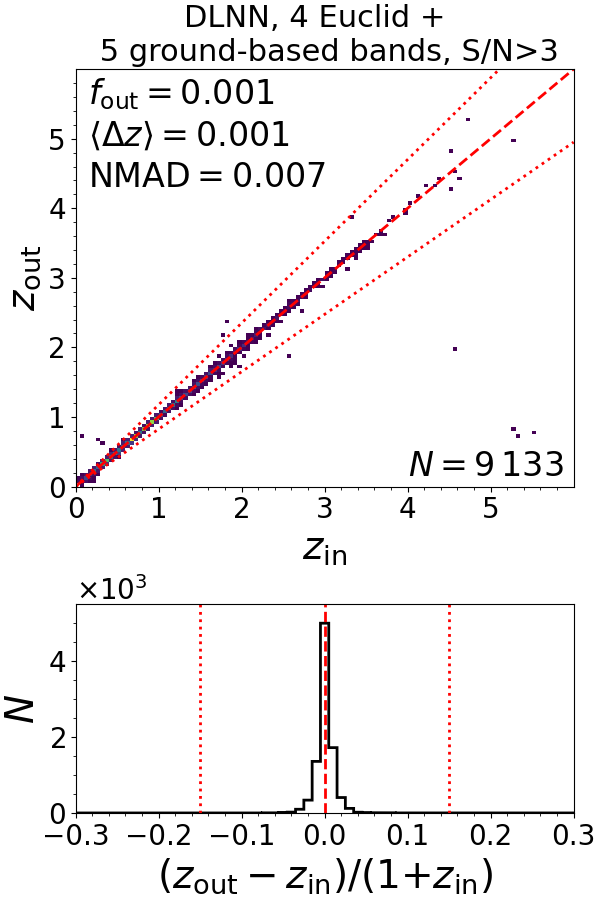}
	\includegraphics[width=0.23\linewidth, keepaspectratio]{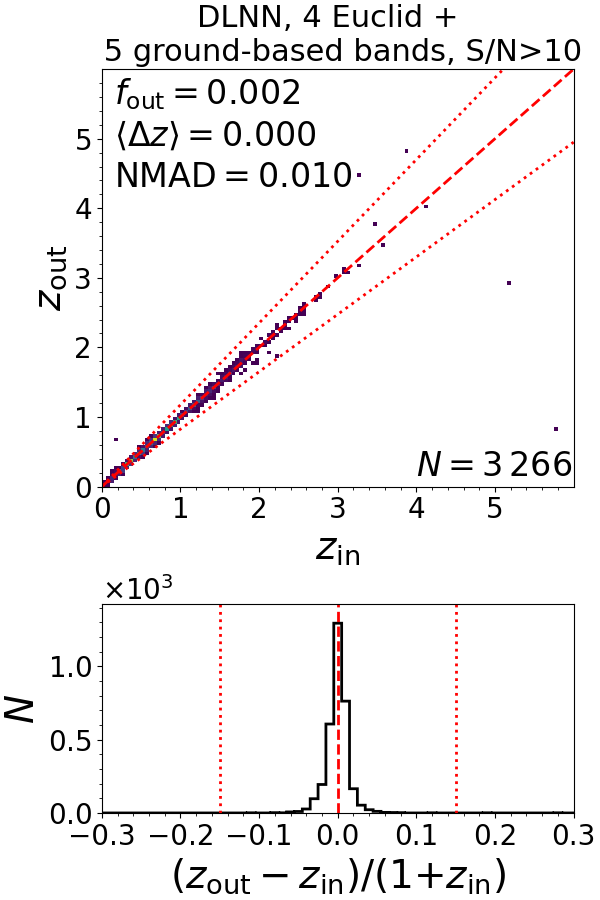}
	\caption{\textit{Top:} Comparison between the recovered redshift and the input one for the DLNN methods. Points are coloured depending on the number of galaxies with the same combination of input and output redshift, following a linear scale from blue to yellow corresponding to 1 and 450 (200) galaxies with ${\rm S/N}>3$ (${\rm S/N}>10$). The red dashed line is the identity and the red dotted lines indicate the outlier limits, i.e. $|\Delta z|=0.15\,(1+z_{\rm in})$. On the top left of each panel we report the fraction of outliers, the bias, and the NMAD. On the bottom right we report the number of objects in the test sample. \textit{Bottom:} distribution of the absolute normalised redshift difference. The red vertical dashed line shows a null difference and the red dotted lines correspond to values of 0.15 and $-$0.15. \textit{From left to right}: redshift recovered using DLNN with four \Euclid filters considering objects with  ${\rm S/N}>3$ and with  ${\rm S/N}>10$, redshift recovered using DLNN with nine input filters considering objects with  ${\rm S/N}>3$ and with  ${\rm S/N}>10$. The ten runs of each network are combined using a Meta-learner. }
	\label{fig:ML_NN_z}
\end{figure*}

\begin{figure*}
	\centering
	\includegraphics[width=0.23\linewidth, keepaspectratio]{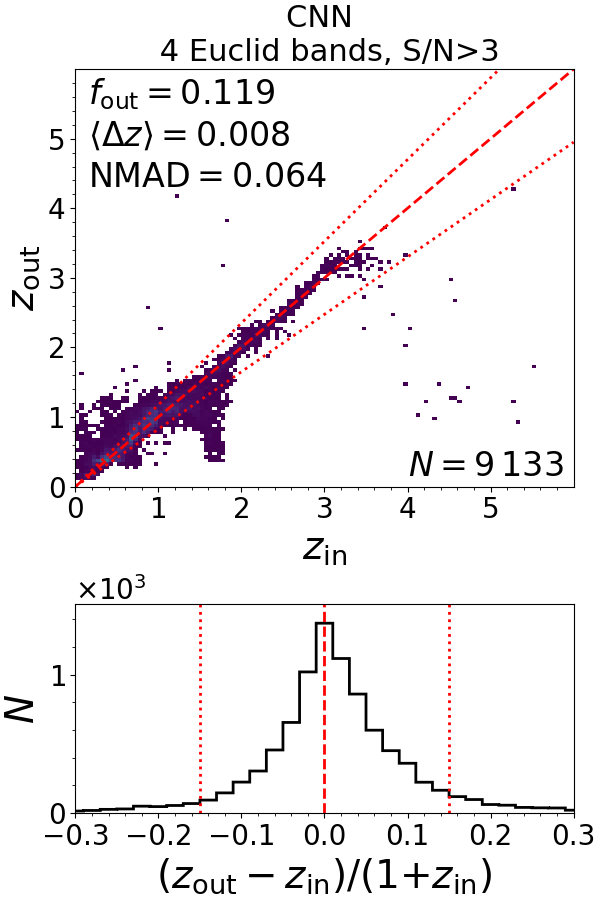}
	\includegraphics[width=0.23\linewidth, keepaspectratio]{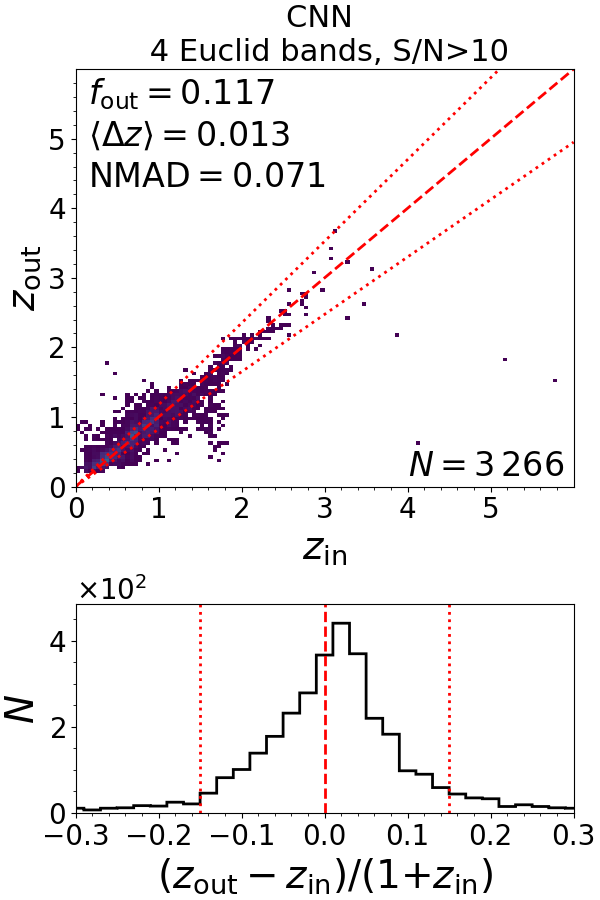}
	\includegraphics[width=0.23\linewidth, keepaspectratio]{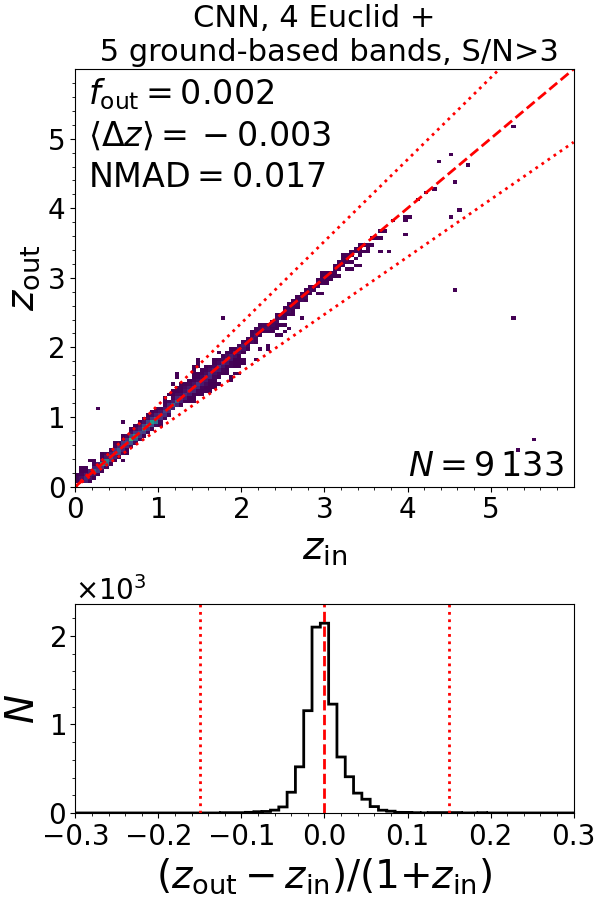}
	\includegraphics[width=0.23\linewidth, keepaspectratio]{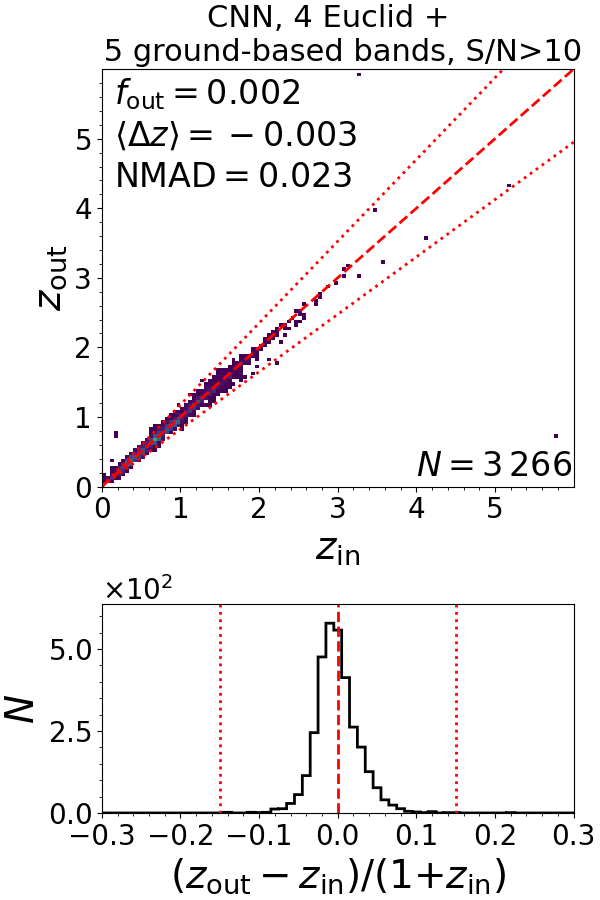}
	\caption{Same as Figure \ref{fig:ML_NN_z}, but for the runs using the CNN.} 
	\label{fig:ML_CNN_z}
\end{figure*}

\begin{figure}
    \centering
    \includegraphics[width=0.75\linewidth, keepaspectratio]{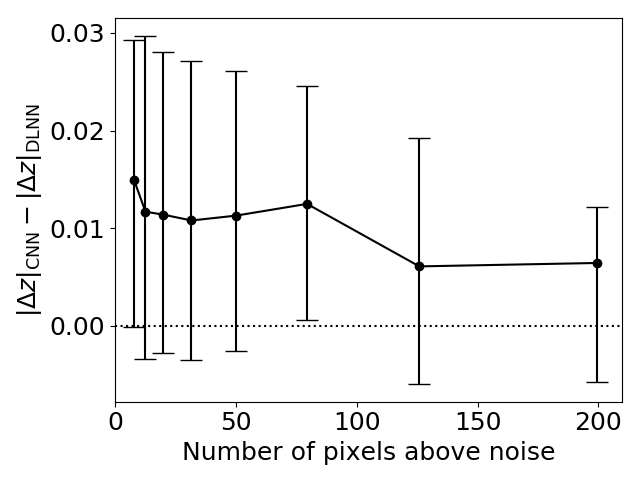}
    \caption{Difference between the recovered redshift in the CNN and DLNN with respect to the number of pixels that are three times above the noise level. Solid circles show the median difference, while the error bars show the central 25$\%$ of the distribution.}
    \label{fig:Size_z}
\end{figure}

In Table \ref{tab:fout_z} we report the fraction of outliers ($f_{\rm out}$), defined as objects with $|z_{\rm out}-z_{\rm in}|>0.15\,(1+z_{\rm in})$ as commonly defined in the literature \citep[e.g.,][]{Ilbert2010,Laigle2016}, the bias 
\begin{equation}
    \langle\Delta z\rangle={\rm median}[(z_{\rm out}-z_{\rm in})/(1+z_{\rm in})],
\end{equation}
and the normalised median absolute deviation\footnote{This is equivalent to the standard deviation for a normal distribution.}
\begin{equation}
{\rm NMAD}=1.48\,[|z_{\rm out}-z_{\rm in}|/(1+z_{\rm in})],
\end{equation}
of the recovered redshifts for all networks. In the same Table we list results for the best and the median of the ten runs of each network, as well as the results derived considering the Meta-learner. The latter are also shown in Figures \ref{fig:ML_NN_z} and \ref{fig:ML_CNN_z}, for the four (i.e. two sets of inputs and two  S/N cuts) DLNN and the four CNN runs, respectively. \par

First, when focusing on each network to compare the ten different runs, it is evident that the Meta-learner (see Section \ref{sec:ML}) gives in general better results than both the best of the ten runs and the median of them. The fraction of outliers of the Meta-learner is always the smallest, even if for some networks with nine input filters the other two approaches give comparable results. The improvement in the fraction of outliers goes up to $\Delta\,f_{\rm out}=0.016$ (0.027), when comparing with the best (median) of the ten runs with four input filters. When we considered the networks with nine input filters, both for the DLNN and the CNN, the fraction of outliers are very small in all cases and the difference is, at maximum, $\Delta f_{\rm out}=0.003$. In addition, the NMAD of the Meta-learner is always the smallest, showing that this approach not only generally decreases the fraction of outliers, but also improves the overall redshift accuracy. As an additional test, we analyse the impact of including at the end of the network a dropout layer, which randomly set to zero some of the elements of the inputs during training with probability 0.5 using samples from a Bernoulli distribution. This method, whose results are not shown here, even if it can identify different trends present in the data, as the Meta learner, does not have the advantage of using the results from multiple runs. Indeed, it performs worse than the Meta learner (e.g. $f_{\rm out}=0.129$, $\langle\Delta z\rangle=0.005$ and NMAD$=0.066$ for the DLNN with four input filters and images with ${\rm S/N}>3$) even when doubling the nodes of the last hidden layer (i.e. $f_{\rm out}=0.111$, $\langle\Delta z\rangle=-0.005$ and NMAD$=0.058$). 
\par
We now compare the results of the DLNN and the CNN methods. First, if we run the CNN without including any additional flux, but only the $H_{\scriptscriptstyle\rm E}$-band images, the fraction of outliers, averaging the ten runs, is quite large: i.e., $f_{\rm out}=0.601$ when limiting the sample to images with ${\rm S/N}>3$. Second, the CNN does not show an improvement with respect to the DLNN with any combination of S/N cuts or the number of input filters. Even in the cases where the fraction of outliers remains similar, which happens in networks with nine input filters, the NMAD increases. The inclusion of the $H_{\scriptscriptstyle\rm E}$-band image adds information about the size of the objects, which could in principle improve the redshift estimation, but this is probably limited by the $H_{\scriptscriptstyle\rm E}$-band spatial resolution (\ang{;;0.3}, $\sim2.5$ kpc at $z=1.5$). Indeed, as can be seen in Figure \ref{fig:Size_z}, as the number of pixels above the noise level increases there is an increase of objects for which the CNN gives better results than the DLNN. In the future, the inclusion of images in multiple filters could be tested to allow the CNN to identify features in the SED, like the 4000 \AA\, break, and to take advantage of the higher angular resolution of the $I_{\scriptscriptstyle\rm E}$ filter (\ang{;;0.1}, $\sim0.8$ kpc at $z=1.5$). 
\par
We now focus on the results of the DLNN runs combined using the Meta-learner, which gives the best redshift estimation. By comparing the two samples of galaxies with  ${\rm S/N}>10$ and  ${\rm S/N}>3$, the redshift estimation is improved only when four filters are considered as input.
This shows that, in the case with nine input filters, the improvement in data quality given by selecting only  ${\rm S/N}>10$ is shadowed by a decrease of the number of objects in the training sample (see Section \ref{sec:mass} for further discussion). The inclusion of the additional five ground-based filters, i.e. $u$, $g$, $r$, $i$, and $z$, decreases the fraction of outliers from 0.066-0.088 to 0.001-0.002, depending on the  S/N limit. The bias is always very small, below 0.001, while the NMAD decreases from $\sim0.05$ to $<0.01$, when changing from four to nine input magnitudes. \par
When four input filters are considered as input, there are galaxies at $z\sim1.7$ for which the redshift is underestimated. In particular, in this redshift range the 4000 \AA\, break is inside the $I_{\scriptscriptstyle\rm E}$ filter, so galaxies generally have a red  $I_{\scriptscriptstyle\rm E}$-$I_{\scriptscriptstyle\rm Y}$ colour. These outliers are intermediate-mass ($\langle\ M_{*}\rangle=10^{9.8}\,\rm M_{\odot}$) star-forming galaxies that, given they relative high specific star-formation rates, have $I_{\scriptscriptstyle\rm E}$-$I_{\scriptscriptstyle\rm Y}$ colours similar to galaxies at lower redshifts. For these galaxies, LSST filters are probably necessary to add information blue-ward the 4000 \AA\, break. The importance of the optical filters is highlighted also by the sensitivity analysis reported in Appendix \ref{sec:Sensitivity}. 
\par

\subsubsection{Comparison with the \Euclid photometric-redshift challenge}
The \Euclid photometric-redshift challenge presented in \citet{Euclidz2020} compared the photometric redshift estimation derived using thirteen different methods, nine of which are based on machine-learning techniques. The considered machine learning networks are based on the nearest neighbour (i.e. Directional Neighborhood Fitting by \citealt{Devicente2016}; frankenz and the Nearest-Neighbor Photometric Redshift by \citealt{Tanaka2018}), boosted decision trees, random forest \citep{Pedregosa2011}, Gaussian processes, and neural networks (Machine-learning Estimation Tool for Accurate PHOtometric Redshifts by \citealt{Cavuoti2017,Amaro2019}; ANNz \citealt{Collister2004}). We refer to each specific papers and the work by \citet{Euclidz2020} for all the details about these methods. \par
A precise comparison between our work and their results needs to be considered with caution, given the differences in the considered input samples and filters, but it can still be used to put our work into contest. In particular, their work uses in input magnitudes in eight optical-to-near-IR filters (no $u$ band) derived from observations available in the COSMOS field. Their analysis is restricted to galaxies with available spectroscopic redshifts and the derived one-point statistics, such as outlier fraction and NMAD, are calculated weighting the spectroscopic sample in order to match the colour-space of the parent photometric catalogue. Therefore, on one hand their sample may be prone to biases due to the spectroscopic selection, but, on the other hand, our input sample may be missing some galaxy populations not included in the considered SED templates. \par
Taking these differences in mind, results obtained with their machine learning algorithms correspond to a fraction of outliers and NMAD varying from 0.031 to 0.326 and from 0.053 to 0.114, respectively. Both quantities are smaller than the ones derived in this work considering nine input filters, but at least some of them are better than our results derived with only \Euclid filters. 

\subsubsection{Comparison with the \textit{LePhare} SED fitting}
Results obtained with the CNN or DLNN all outperformed results from the SED fitting, using their same set of input magnitudes, even in the ideal case where both the code and the SED templates are the same, and are thus used to create the mock magnitudes. The fraction of outliers with the SED fitting corresponds to 0.604 and 0.127, when considering the sample with  ${\rm S/N}>3$ and four and nine filters as input, respectively. For comparison, \citet{Euclidz2020} used eight input filters (no $u$ filter), the same SED fitting code, but a different input sample not corresponding to the templates used to generate photometric magnitudes (see previous section), finding an outlier fraction of 0.134 and a NMAD of 0.056. The outlier fractions we find range between four and 100 times more than the fraction of outliers derived with any CNN or DLNN runs. The improvement of the machine learning algorithms over SED fitting has also been shown by Humphrey et al. (in prep) when selecting passive galaxies. These authors argued that this is due to the machine learning networks capability to optimally weight the different input data points, while the SED fitting methods generally use a more direct weighting method, based on the  S/N. More details on the SED fitting results are reported in Appendix \ref{sec:SEDfit}. \par
Overall, among the different cases tested here, the best network for redshift estimation consists of the DLNN with nine input filters and  ${\rm S/N}>3$, combined using a Meta-learner.

\subsection{Stellar masses}\label{sec:mass}

\begin{figure*}
	\centering
	\includegraphics[width=0.23\linewidth, keepaspectratio]{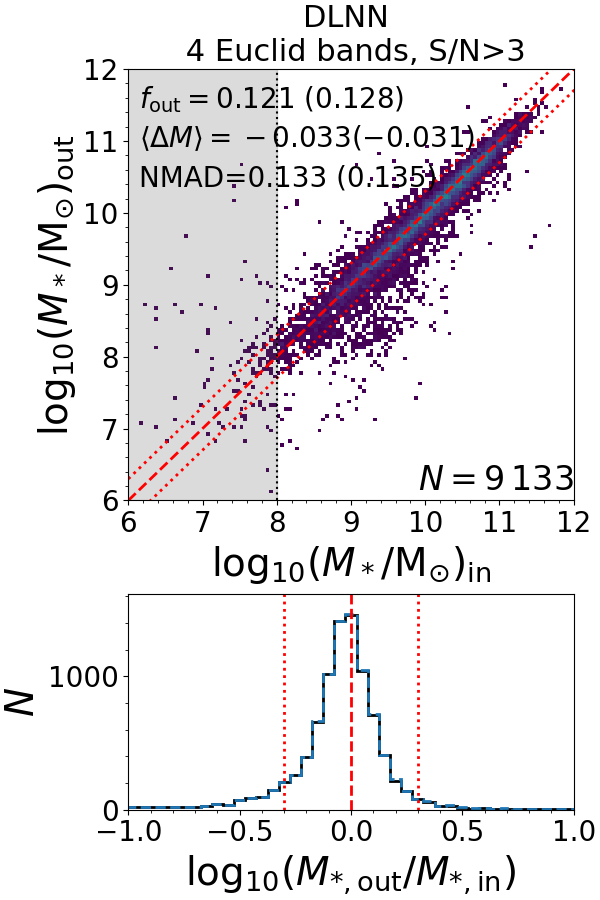}
	\includegraphics[width=0.23\linewidth, keepaspectratio]{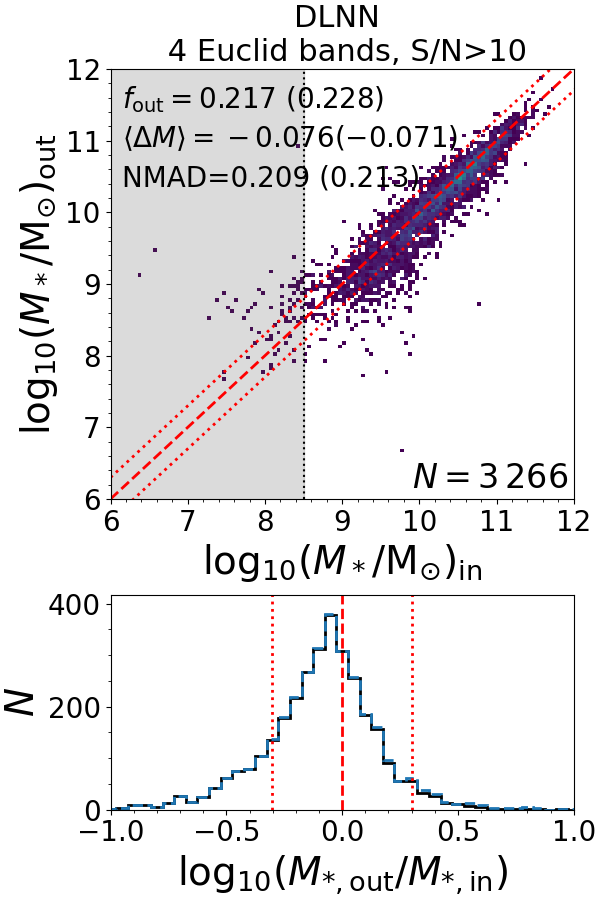}
	\includegraphics[width=0.23\linewidth, keepaspectratio]{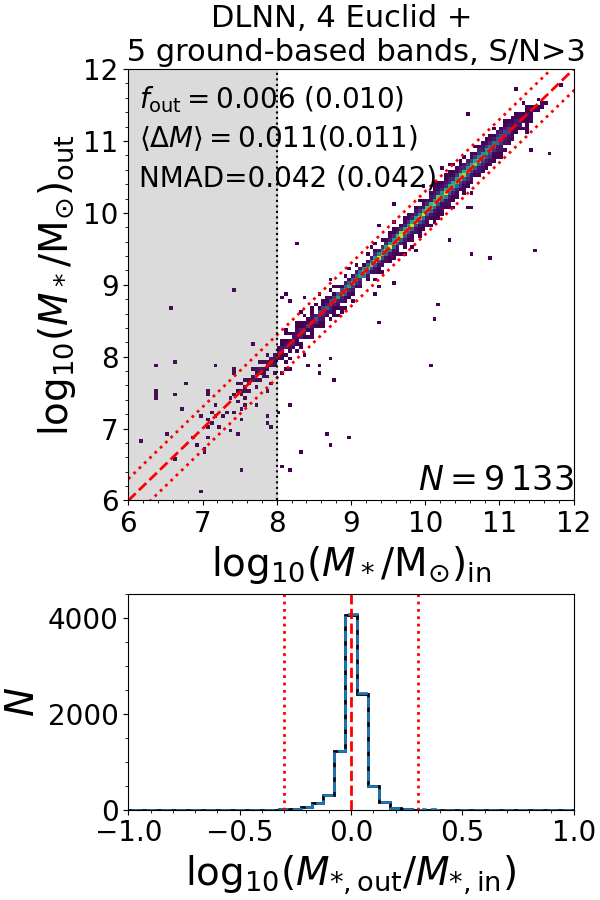}
	\includegraphics[width=0.23\linewidth, keepaspectratio]{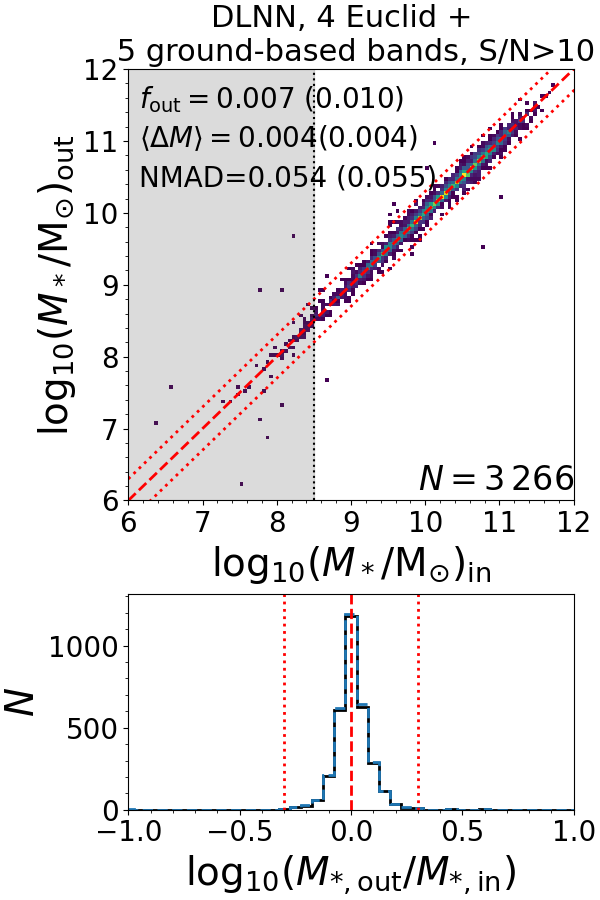}
	\caption{\textit{Top:} Comparison between the recovered stellar mass and the input one for the DLNN methods. Points are coloured depending on the number of galaxies with the same combination of input and output stellar mass, following a linear scale from blue to yellow corresponding to 1 and 100 (40) galaxies with ${\rm S/N}>3$ (${\rm S/N}>10$). The grey shaded area indicate the stellar mass range in which the input stellar mass distribution is not flat but underrepresented in the training sample, i.e. $M_{*}<10^{8.5} \Msolar$ for  ${\rm S/N}>3$ and $M_{*}<10^{8} \Msolar$ for  ${\rm S/N}>10$. The red dashed line is the identity and the red dotted lines indicate output stellar mass equal to twice or half the input one, which corresponds to the definition of an outlier. On the top left of each panel we report the fraction of outliers, bias, and NMAD of the sample with $M_{*}>10^{8.5} \Msolar$ or $M_{*}>10^{8} \Msolar$, depending on the  S/N cut. In parentheses we reported the same values for the full sample. On the bottom right we report the number of objects in the test sample. \textit{Bottom:} distribution of the difference between the output and input stellar mass, for the full sample (\textit{blue dashed line}) and for galaxies with $M_{*}>10^{8.5} \Msolar$ or $M_{*}>10^{8} \Msolar$ (\textit{black solid line}), depending on the  S/N cut. The red vertical dashed line shows a null difference and the red dotted lines indicate output stellar mass equal to twice or half the input one. \textit{From left to right}: stellar mass recovered using DLNN with four \Euclid filters considering objects with  ${\rm S/N}>3$ and with  ${\rm S/N}>10$, stellar mass recovered using DLNN with four \Euclid filters and five ancillary bands considering objects with  ${\rm S/N}>3$ and with  ${\rm S/N}>10$. The ten runs of each network are combined using a Meta-learner.}
	\label{fig:ML_NN_M}
\end{figure*}

\begin{figure*}
	\centering
	\includegraphics[width=0.23\linewidth, keepaspectratio]{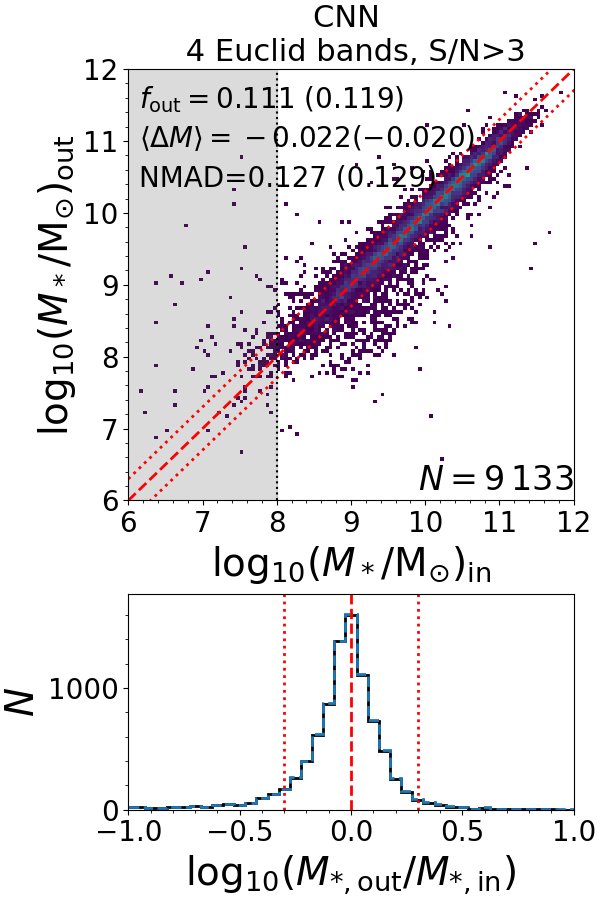}
	\includegraphics[width=0.23\linewidth, keepaspectratio]{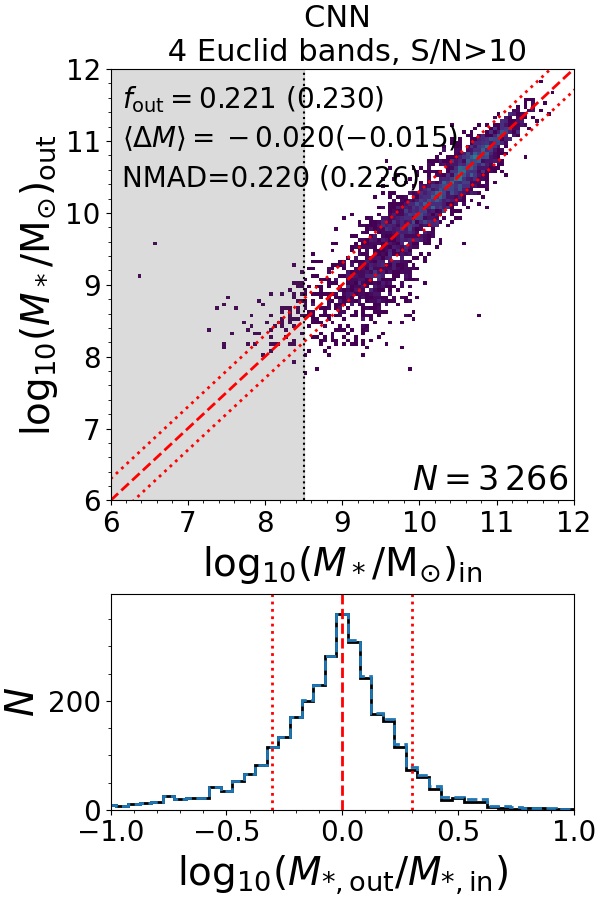}
	\includegraphics[width=0.23\linewidth, keepaspectratio]{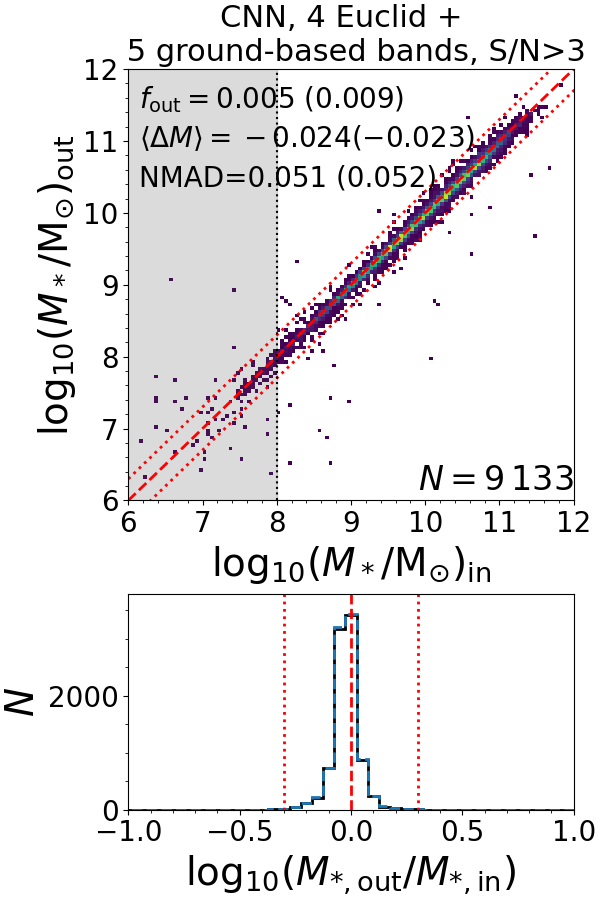}
	\includegraphics[width=0.23\linewidth, keepaspectratio]{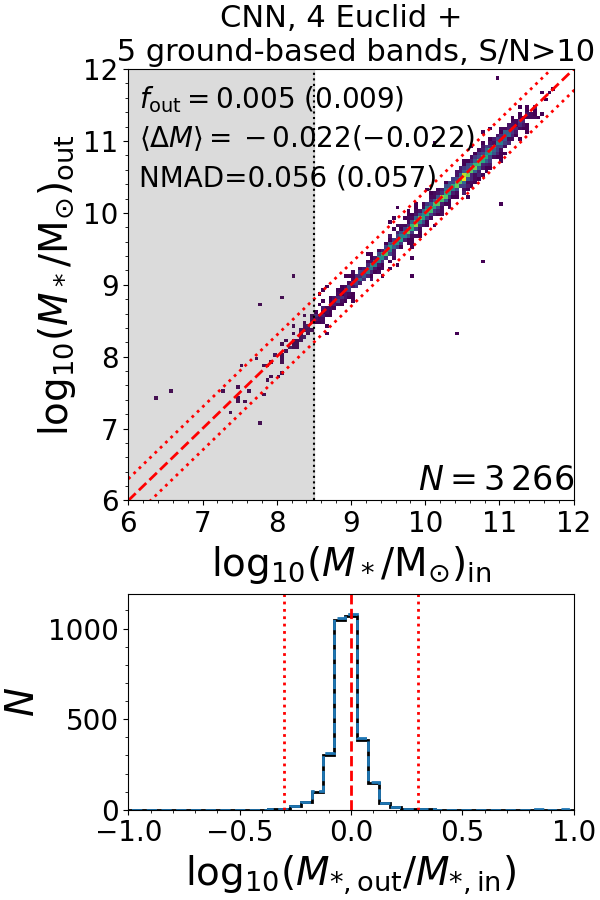}
	\caption{Same as Figure \ref{fig:ML_NN_M}, but for the runs using the CNN. }
	\label{fig:ML_CNN_M}
\end{figure*}
\begin{table*}
	\caption{Same as Table \ref{tab:fout_z}, but for the stellar mass. The results correspond to the mass range where the sample distribution is flat, i.e. $\logten(M\,\Msolar^{-1})>8$ for  ${\rm S/N}>3$ and 8.5 for  ${\rm S/N}>10$, while values in parentheses correspond to the full sample. The first five lines correspond to results derived with a constant M/L$_{H}$ ratio and with a SED fitting.} 
	\centering 
	\begin{tabular}{c c c c c c c c}
		\hline\hline 
		Algorithm & $N_{\rm in}$ &  S/N & Combination &  $f_{\rm out}$ & $\langle\Delta M_{*}\rangle$ & NMAD & MSE\\
		(1) & (2) & (3) & (4) & (5) & (6) & (7) & (8) \\
		\hline
		$M/L_{H}=$0.62 & 1 & 3 & & 0.298(0.300) & 0.000$^{a}$(0.003) &  0.300(0.302) & 0.084(0.085)\\
        SED  & 4 & 3 & & 0.403(0.412) & 0.134(0.140) & 0.341(0.348) & 0.215(0.268)\\
        SED  & 4 & 10 & & 0.432(0.436) & 0.193(0.196) & 0.375(0.378) & 0.181(0.217)\\
        SED  & 9 & 3 & & 0.128(0.135) & 0.001(0.002) & 0.120(0.121) & 0.112(0.130)\\
        SED  & 9 & 10 & & 0.048(0.051) & 0.012(0.012) & 0.094(0.095) & 0.040(0.051)\\
		\hline
		DLNN	&  4 & 3& best         &  0.132(0.139) & $-$0.037($-$0.036) & 0.146(0.148) & 0.073(0.089)\\ %
			    &    &   & median      &  0.123(0.129) & $-$0.025($-$0.024) & 0.129(0.131) & 0.067(0.082)\\ %
			    &    &   & Meta-learner &  0.121(0.128) & $-$0.033($-$0.031) & 0.133(0.135) & 0.068(0.085)\\ %
		DLNN	&  4 & 10& best         &  0.217(0.228) & $-$0.034($-$0.031) & 0.208(0.212) & 0.079(0.098)\\ %
			    &    &   & median      &  0.223(0.231) & $-$0.057($-$0.052) & 0.221(0.223) & 0.082(0.099)\\ %
			    &    &   & Meta-learner &  0.217(0.228) & $-$0.076($-$0.071) & 0.209(0.213) & 0.093(0.103) \\ %
		CNN	    &  4 & 3 & best         &  0.128(0.136) & $-$0.034($-$0.032) & 0.141(0.144) & 0.079(0.097)\\
		        &    &   & median      &  0.131(0.139) & $-$0.024($-$0.022)  & 0.134(0.136) & 0.068(0.083)\\ %
		        &    &   & Meta-learner &  0.111(0.119) & $-$0.022($-$0.020)  & 0.127(0.129) & 0.062(0.079)\\ %
		CNN	    &  4 & 10& best         &  0.252(0.262) & $-$0.086($-$0.082)  & 0.235(0.240) & 0.093(0.112)\\ %
		        &    &   & median      &  0.263(0.273) & $-$0.087($-$0.082) & 0.239(0.242) & 0.099(0.117)\\ %
		        &    &   & Meta-learner &  0.221(0.230) & $-$0.020($-$0.015)  & 0.220(0.226) & 0.088(0.103)\\ %
		DLNN	&  9 & 3 & best         &  0.005(0.008)   & $-$0.017($-$0.017) & 0.054(0.054) & 0.008(0.013) \\ %
		        &    &   & median      &  0.005(0.008)   & 0.003(0.003) & 0.041(0.041) & 0.008(0.014) \\ %
		        &    &   & Meta-learner &  0.006(0.010)   & 0.011(0.011) & 0.042(0.042) & 0.009(0.014)\\ %
		DLNN    &  9 & 10& best         &  0.007(0.012)   & $-$0.001(0.000)   & 0.066(0.066) & 0.013(0.022)\\ %
		        &    &   & median      &  0.009(0.012)   & $-$0.019($-$0.019)  & 0.068(0.070) & 0.009(0.016)\\ %
		        &    &   & Meta-learner &  0.007(0.010)   &  0.004(0.004)  & 0.054(0.054) & 0.007(0.015)\\ %
		CNN	    &  9 & 3 & best         &  0.006(0.011)   & 0.006(0.006) & 0.050(0.050) & 0.008(0.015)\\ %
		        &    &   & median      &  0.006(0.010)   & $-$0.001($-$0.001)  & 0.045(0.045) & 0.007(0.012)\\%
		        &    &   & Meta-learner &  0.005(0.009)   & $-$0.024($-$0.023) & 0.051(0.051) & 0.009(0.015)\\ %
		CNN	    &  9 & 10& best         &  0.006(0.010)   & $-$0.013($-$0.013)  & 0.057(0.058) & 0.008(0.015)\\ %
		        &    &   & median      &  0.023(0.025)   & $-$0.030($-$0.030) & 0.081(0.082) & 0.019(0.026)\\ %
		        &    &   & Meta-learner &  0.005(0.009)   & $-$0.022($-$0.022)  & 0.056(0.057) & 0.008(0.015)\\ %
		\hline
	\end{tabular}\label{tab:fout_M}\\
	$^{a}$ the bias is null by construction, as the used $M/L_{H}$ is equal to the median value of the sample.
	
\end{table*}

The results for the stellar mass retrieval with machine learning are summarised in Table \ref{tab:fout_M} for both the CNN and the DLNN methods, considering all the different inputs, both  S/N cuts and the different methods to combine the ten runs of each network. We remind the reader that the redshift is not among the inputs when deriving the stellar mass, as the two quantities are derived with separate networks.  We estimate for the entire sample the fraction of outliers, arbitrary defined as galaxies for which the stellar mass is overestimated or underestimated by a factor of two ($\sim0.3 \rm dex$). In addition, we estimate for each method the bias 
\begin{equation}
    \langle\Delta M_{*}\rangle={\rm median}[\logten(M_{*,{\rm out}}/M_{*,{\rm in}})]    
\end{equation}
and the normalised median absolute deviation of the recovered stellar mass 
\begin{equation}
    {\rm NMAD}=1.48\,{\rm median}[|\logten(M_{*,{\rm out}}/M_{*,{\rm in}})|].
\end{equation}
Figures \ref{fig:ML_NN_M} and \ref{fig:ML_CNN_M} show the results for the DLNN and CNN runs, after combining the results using a Meta-learner.\par

As for the redshift, we first focus on the three methods to combine the ten runs of each network. In general, the differences among the methods are less evident than for the redshift, with the three methods alternating on what gives the best results. However, the fraction of outliers derived with the Meta-learner is the smallest, except for the DLNN with nine filters as input and  ${\rm S/N}>3$, with a difference in the fraction of outliers $\Delta f_{\rm out}\leq0.042$ with respect to the best and the median of the ten runs. Given the improvement, even if small, offered by the Meta-learner, we will focus on the results obtained with this method in the rest of this section. \par

We now compare the results of the DLNN, which includes only integrated magnitudes, and the CNN, which contains both integrated magnitudes and $H_{\scriptscriptstyle\rm E}$-band images. We remind the reader that the $H_{\scriptscriptstyle\rm E}$-band images include information about the features and shapes, but not the overall $H_{\scriptscriptstyle\rm E}$-band magnitude. The use of only $H_{\scriptscriptstyle\rm E}$-band images, without any integrated flux, is not sufficient to estimate the stellar mass, as it results, for example, in a large outlier fraction $f_{\rm out}=0.668$ when averaging the results of the ten runs of the sample limited to images with ${\rm S/N}>3$ and $\logten(M\,\Msolar^{-1})>8$. Using the $H_{\scriptscriptstyle\rm E}$-band images together with the integrated fluxes reduces instead the outlier fraction in the stellar mass with respect to the results obtained with the DLNN (see Table \ref{tab:fout_M}). An exception is the case with only four input filters and the sample limited to images with ${\rm S/N}>10$. The improvement in the fraction of outliers using the CNN is generally $\Delta f_{\rm out}<0.014$, but it is present even when the fraction is already very small. This happens for example in the networks with nine input filters, for which the fractions of outliers in the DLNN are below 0.007, but the inclusion of the $H_{\scriptscriptstyle\rm E}$-band images produces an improvement by $\Delta f_{\rm out}=0.001$--0.002. \par

The $H_{\scriptscriptstyle\rm E}$-band filter traces light from a relatively old stellar population, at least at low redshift, so we expect it to be a good tracer of the stellar mass and drive the improvement when adding the $H_{\scriptscriptstyle\rm E}$-band images. To verify this point, we analyse the fraction of outliers which are evolved galaxies, also called quiescent, (i.e. number of galaxies that are outlier and quiescent divided by the total number of quiescent) and the fraction of outliers which are galaxies currently forming stars and, therefore, including a younger stellar population (i.e. number of galaxies that are outlier and star-forming divided by the total number of star-forming galaxies). The first population is defined as galaxies with input specific star-formation rates $\logten[{\rm sSFR}/({\rm yr}^{-1})]<-10.5$, while the second has $\logten[{\rm sSFR}/({\rm yr}^{-1})]\geq-10.5$.
 
\par
\begin{figure}
    \centering
    \includegraphics[width=0.7\linewidth, keepaspectratio]{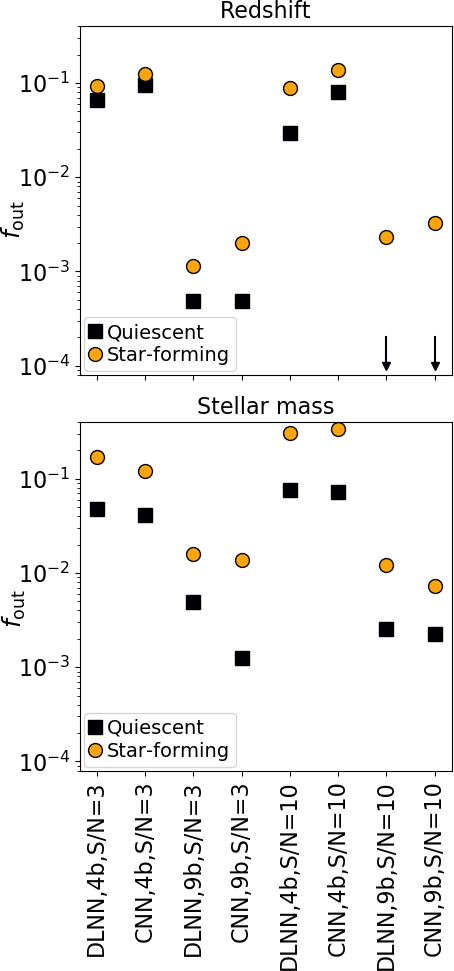}
    \caption{Fraction of outliers which are quiescent (i.e. $\logten[{\rm SFR}/(\Msolar\,{\rm yr}^{-1})]<-10.5$, \textit{black squares}) and star-forming galaxies (i.e. $\logten[{\rm SFR}/(\Msolar\,{\rm yr}^{-1})]>-10.5$, \textit{yellow circles}) for different networks, as derived for the redshift (\textit{top}) and the stellar mass (\textit{bottom}). Arrows in the top panel correspond to $f_{\rm out}=0$.}
    \label{fig:fout_QSvsSF}
\end{figure}

The comparison between star-forming and quiescent outlier galaxies is shown for the stellar mass and for the redshift (Figure \ref{fig:fout_QSvsSF}). The inclusion of $H_{\scriptscriptstyle\rm E}$-band images results on an improvement on the measurements of the stellar mass, but not of the redshift (see Section \ref{sec:redshift}). While for redshift the outlier fraction of evolved galaxies generally increases for CNN with respect to DLNN, the opposite happens for the stellar mass. Moreover, there is no improvement in the stellar mass measurement of star-forming galaxies between DLNN and CNN when there are four input filters and  ${\rm S/N}>10$. There is instead an improvement, even if small, in the mass outlier fraction for evolved galaxies, even if they are less than 30$\%$ of star-forming galaxies in the training sample. This explains the improvement on the stellar mass measures introduced by the CNN. \par

\begin{figure}
    \centering
    \includegraphics[width=0.8\linewidth, keepaspectratio]{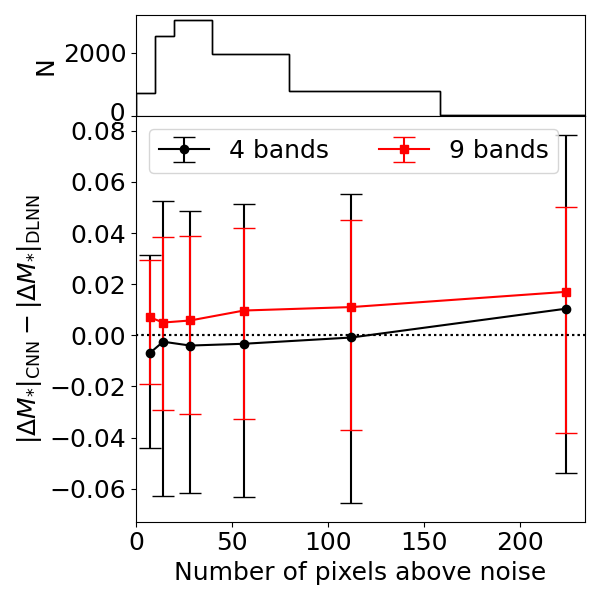}
    \caption{In the main panel we shown the difference between the absolute stellar mass errors in the CNN and DLNN with respect to the number of pixels that are three times above the noise level. Solid symbols show the median errors, while the error bars show the central 68$\%$ ($\pm1\sigma$) of the distribution. The difference is shown for the networks when using four (\textit{black circles}) and nine input filters (\textit{red squares}) and for the sample with images having ${\rm S/N}>3$. In the top panel we report the distribution of the objects as a function of the number of pixels that are three times above the noise level.}
    \label{fig:DM_size}
\end{figure}

In addition, we investigate the impact of galaxy size in the $H_{\scriptscriptstyle\rm E}$-band images on the stellar mass derivation, by examining the number of pixels that are above three times each image's noise level, or  ${\rm S/N}>3$. It is necessary to consider that a compact and unresolved structure is information that the network is using, therefore the introduction of the $H_{\scriptscriptstyle\rm E}$-band images may also improve the stellar mass derivation of unresolved galaxies. Indeed, comparing DLNN and CNN with  ${\rm S/N}>3$, galaxies for which the stellar mass measurement improved adding the $H_{\scriptscriptstyle\rm E}$-band images have on average 35 pixels that are three times above the noise level. On the other hand, galaxies for which the $H_{\scriptscriptstyle\rm E}$-images worsen the stellar mass derivation have on average 36 pixels above the noise. Such a small difference indicates that the stellar mass estimation is not affected by the galaxy size. This is further visible in Figure \ref{fig:DM_size}, where we analyse the improvement in the stellar mass measurament as a function of the number of pixels that are three times above the noise level. The median difference is always quite small, and is almost constant with the number of pixels above the noise level, except for the largest galaxies, which are probably only partially included inside the cutout image (i.e., $18\times18$ pixel). We will analyse these extended galaxies in a future work focused on local galaxies. In the same figure it is also shown that when four filters are used as input the stellar mass improves when adding the $H_{\scriptscriptstyle\rm E}$-band images for more than 50$\%$ of galaxies at each size bin, except for the largest galaxies, justifying the additional computational effort of including images. The improvement is below $\Delta \logten(M_{*}/\rm M_{\odot})=0.08$ for most (68$\%$) of the galaxies.  \par

\begin{figure}
    \centering
    \includegraphics[width=0.8\linewidth, keepaspectratio]{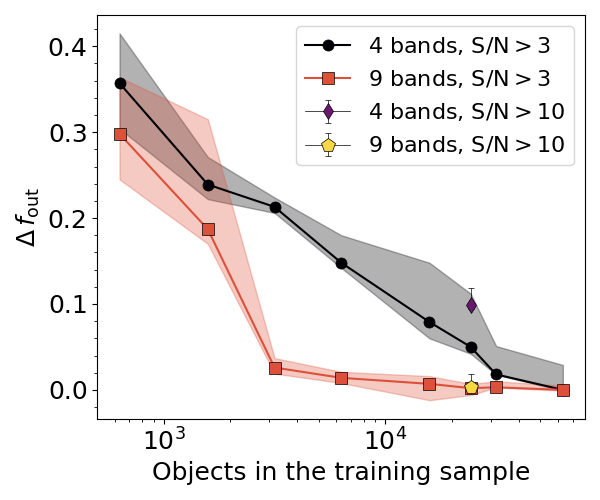}
    \caption{Difference in the outlier fraction of the stellar mass with decreasing size of the training sample. The difference is derived for the DLNN with four input magnitudes and  ${\rm S/N}>3$ (\textit{black circles}) and for the DLNN with nine input magnitudes and  ${\rm S/N}>3$ (\textit{red squares}), considering as zero point the fraction of outlier of the complete sample. For comparison, we report also the difference in the outlier fraction of the sample with  ${\rm S/N}>10$ with respect to the sample with  ${\rm S/N}>3$, considering the DLNN with four (\textit{purple diamond}) and nine input magnitudes (\textit{yellow pentagon}). Fractions are derived averaging the results of the ten runs and the shaded areas show the standard variation within the ten runs.}
    \label{fig:size_test}
\end{figure}

Limiting the sample to those objects with a $H_{\scriptscriptstyle\rm E}$-band image at  ${\rm S/N}>10$ produces different results, depending on the number of input filters. Indeed, when only the four \Euclid filters are considered as inputs, the fraction of outliers increases from $f_{\rm out}\sim0.107$--0.128 to $f_{\rm out}\sim0.217$--0.230 for samples with  ${\rm S/N}>10$. On the contrary, the fraction of outliers remains stable when nine filters are used as input, while the bias improves by 0.002--0.007, depending on the considered network architecture. \par
To investigate the cause of the different impacts of limiting the sample to  ${\rm S/N}>10$, we explore how the size of the training sample impacts the resulting fraction of outliers (Figure \ref{fig:size_test}), using the DLNN as an example and varying the size of the training sample down to 1$\%$ of the complete one. This test is performed by randomly removing galaxies from the training sample after augmentation, therefore the stellar mass distribution, with some limitation once the sample size is very small, should be similar to the one of the training sample (see in Figure \ref{fig:distributions}, top central panel). \par First, the fractions of outliers in both the ${\rm S/N}>10$ and the ${\rm S/N}>3$ samples with nine filters as inputs are consistent within the errors with the fractions of samples of similar size, but with  ${\rm S/N}>3$. Second, the outlier fraction increases with decreasing sample size. However, this decrease, when nine magnitudes are used as input, becomes relevant (e.g., $\Delta f_{\rm out}>0.02$) at smaller sample sizes (i.e., $<3\times10^{3}$ objects) than in the case of four input magnitudes (i.e., $<3\times10^{4}$ objects). Therefore, when only four filters are used as input, there is not enough information available, and it is preferable to have a larger, even if more noisy, sample. On the contrary, when more information is available, i.e. nine input filters, it is possible to focus on quality over quantity. \par
We now focus on the improvement of the stellar mass retrieval given by the inclusion of the $u$, $g$, $r$, $i$, and $z$ ground-based filters, focusing again on the results obtained with the Meta-learner. The improvement is evident by looking at the fraction of outliers that varies between $f_{\rm out}=0.107$--0.230 when only \Euclid filters are used as input, while it never exceeds $f_{\rm out}=0.010$ when all nine filters are used as input. The presence of long wavelength filters, such as the $H_{\scriptscriptstyle\rm E}$-band at least up to $z\sim1.5$, is indeed fundamental for obtaining a reliable stellar mass, as evident by the relative good stellar mass estimation when only four \Euclid filters are used as input. However, the inclusion of shorter wavelength filters probably helps anchor the overall SED template to estimate very accurately the stellar mass. As for the redshift, the importance of the optical filters is highlighted also by the sensitivity analysis reported in Appendix \ref{sec:Sensitivity}. \par

\begin{figure}
    \centering
    \includegraphics[width=0.9\linewidth, keepaspectratio]{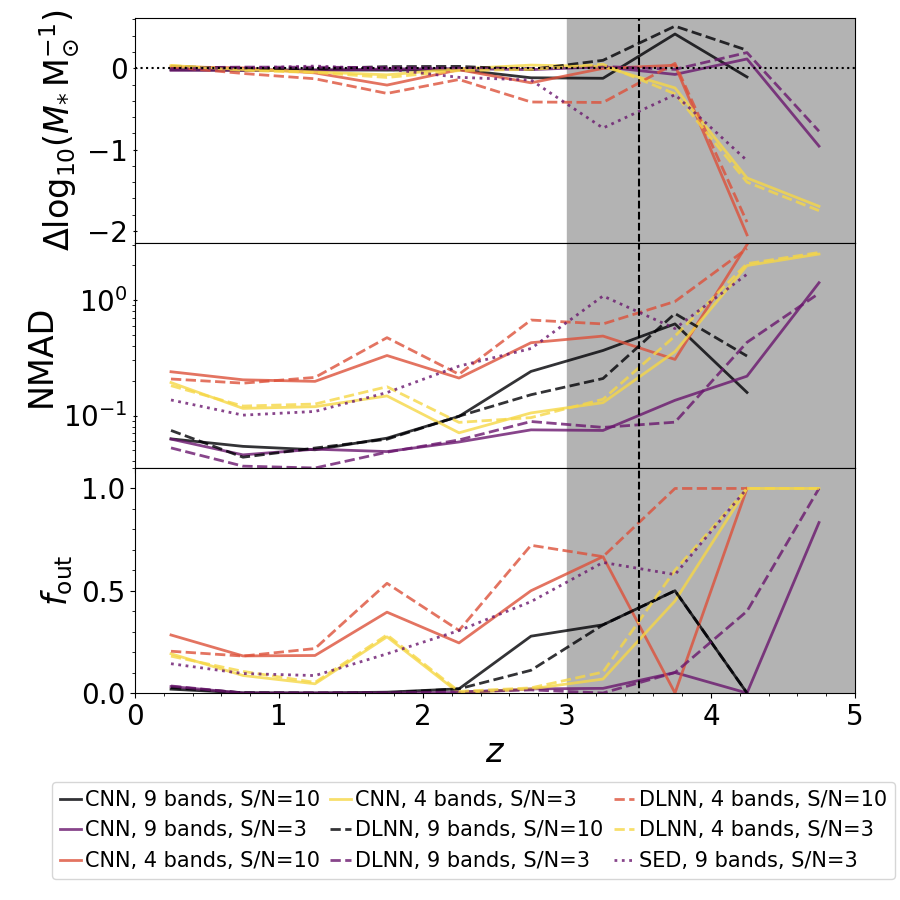}
    \caption{Redshift variation of the bias (\textit{top}), NMAD (\textit{centre}), and outlier fraction (\textit{bottom}) of the recovered stellar mass. 
    Different symbols indicate different algorithm architectures and different inputs (see legend). The grey area (black vertical dashed line) shows the redshift bins with less than 100 objects for the training samples which have  ${\rm S/N}>10$ ( ${\rm S/N}>3$).}
    \label{fig:M_z}
\end{figure}

\begin{figure}
    \centering
    \includegraphics[width=0.9\linewidth, keepaspectratio]{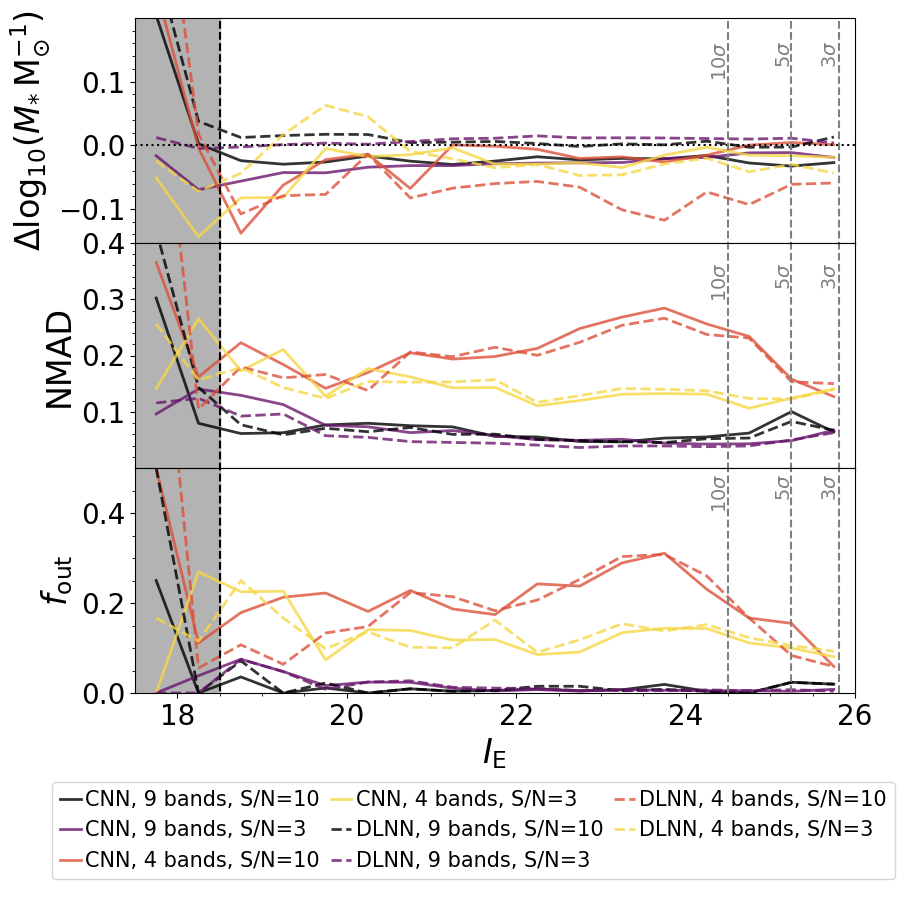}
    \caption{Same as Figure \ref{fig:M_z}, but focusing on the variation with $I_{\scriptscriptstyle\rm E}$ magnitude.}
    \label{fig:M_VIS}
\end{figure}

\subsubsection{Variation with redshift and $I_{\scriptscriptstyle\rm E}$ magnitude}
In Figures \ref{fig:M_z} and \ref{fig:M_VIS} we show the variation of the stellar mass measures with redshift and $I_{\scriptscriptstyle\rm E}$ magnitude, respectively. There is a clear trend with redshift, as the fraction of outliers is below 0.28 and the NMAD is below 0.06 at $z<1.5$. This trend with redshift is at least partially driven by the small number of galaxies available at $z>3$ for the training, i.e. galaxies at lower redshift are more numerous and, therefore, they dominate the training process of the network. A similar effect is seen when looking at the variation with the $I_{\scriptscriptstyle\rm E}$ magnitude, as the fraction of outliers and the NMAD value increases at the brightest magnitudes, i.e. $I_{\scriptscriptstyle\rm E}<18$, that are poorly represented in the sample. In the future, a larger sample with a flat multi-dimensional (i.e. $z$, stellar mass, SFR, magnitude) distribution may help improve the measurement of these objects. \par
The deterioration of the stellar mass measurement with redshift may also be explained with the $H_{\scriptscriptstyle\rm E}$-band, which is the filter at the longest wavelength among the analysed ones, tracing shorter wavelengths at larger redshift and, therefore, are less sensitive to the light emitted by the old stellar populations which make up most of the stellar mass. The stellar mass measurement at magnitude $I_{\scriptscriptstyle\rm E}>18$ is instead generally constant, except for the networks with four input filters and at  ${\rm S/N}>10$, for which it becomes worse between $I_{\scriptscriptstyle\rm E}=22$ and 24. The decrease visible at fainter magnitudes is probably spurious and driven by the limited number of galaxy in the sample ($<100$ at $I_{\scriptscriptstyle\rm E}>25$) caused by the S/N cut.\par

\subsubsection{Comparison with the \textit{LePhare} SED fitting}
Finally, to put these results into context, we derived the stellar mass directly from $H_{\scriptscriptstyle\rm E}$-band magnitudes assuming a single mass-to-light ratio, which is a simplistic but direct method, and using a SED fitting procedure. For the first case, we considered an ideal situation where we calculate the $H_{\scriptscriptstyle\rm E}$-band luminosity from the true redshift and we assumed a mass-to-light ratio equal to the median value (i.e. M/$L_{H}\sim0.6$), obtained by comparing the $H_{\scriptscriptstyle\rm E}$-band luminosity directly with the true stellar mass. For the full sample with  ${\rm S/N}>3$ and $\logten(M_{*}\,\Msolar^{-1})>8$, we obtained a fraction of outliers of 0.298, $\langle\Delta \logten(M_{*}\,\Msolar^{-1})\rangle=0$, by construction, and ${\rm NMAD}=0.3$ (Table \ref{tab:fout_M}), which is overall a worse result than that obtained with both the DLNN and CNN methods for the same sample. \par
With the SED fitting (more details in Appendix \ref{sec:SEDfit}), when considering only the four \Euclid filters as input, results are even worse than with a constant $M/L_{H}$ ratio,  which is however derived considering the true median $M/L_{H}$. These results however improve when nine filters are used as input, but these still perform less well than the best CNN or DLNN results. Indeed, with the SED fitting we obtain $f_{\rm out}=0.128$--0.048 compared with $f_{\rm out}\leq0.02$, considering all combinations methods, or $f_{\rm out}\leq0.007$ focusing on results derived with the Meta-learner. In the SED fitting, redshift is kept free and the improvement when adding the $u$, $g$, $r$, $i$, and $z$ filters is also driven by the improvement in the redshift estimation (see Section \ref{sec:redshift}).\par
Finally, in Figure \ref{fig:M_z} we also report the variation of bias, NMAD and outlier fraction with redshift for one of the best SED estimates, i.e. nine filters used as input and ${\rm S/N}>3$. Surprisingly, ML results are more precise and accurate than SED fitting ones at all redshfits, even in the situation when the training sample is limited in number. This may also be linked to a difficult redshift estimation (see Appendix \ref{sec:SEDfit}).
\subsection{Star formation rate derivation}\label{sec:sfr}

\begin{figure*}
	\centering
	\includegraphics[width=0.23\linewidth, keepaspectratio]{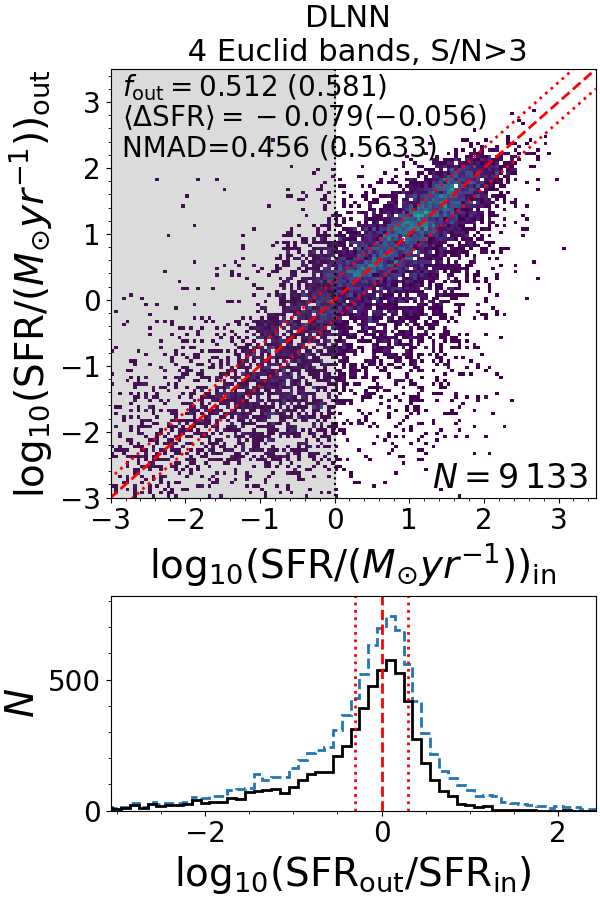}
	\includegraphics[width=0.23\linewidth, keepaspectratio]{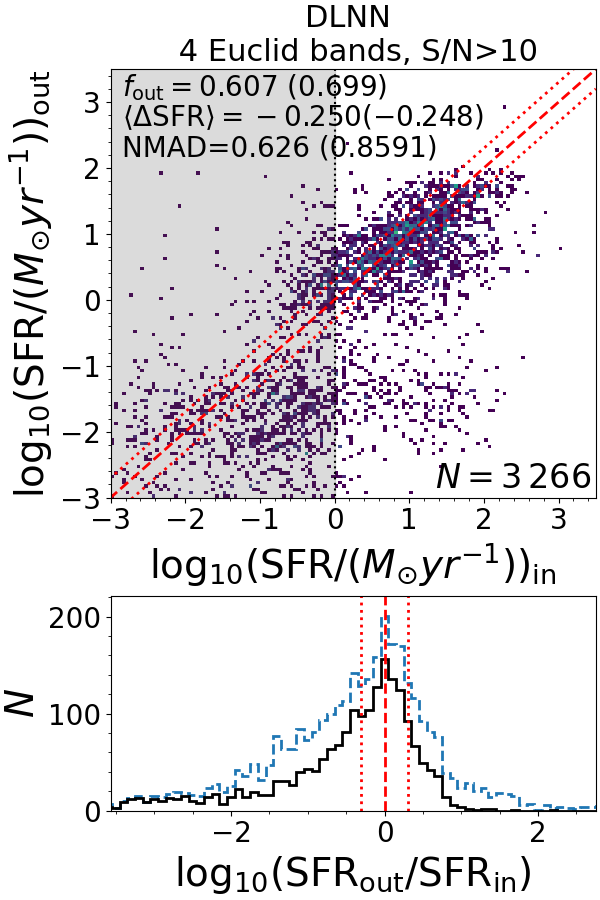}
	\includegraphics[width=0.23\linewidth, keepaspectratio]{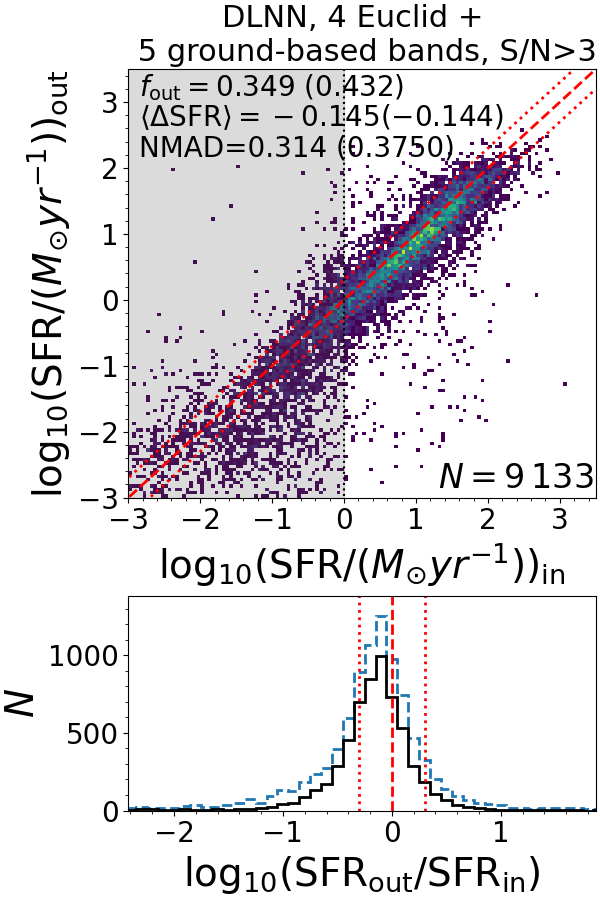}
	\includegraphics[width=0.23\linewidth, keepaspectratio]{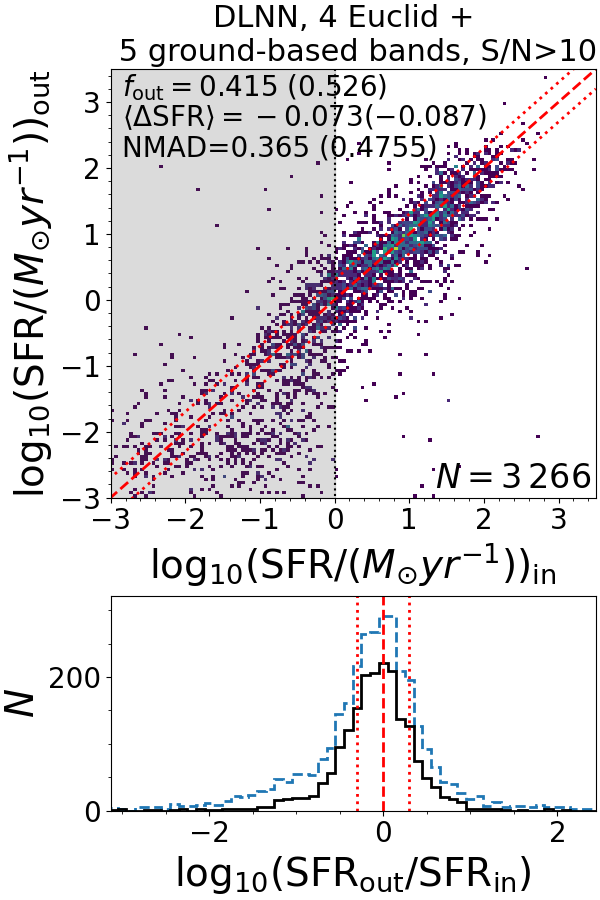}
	\caption{\textit{Top:} Comparison between the recovered SFR and the input one for the DLNN methods. Points are coloured depending on the number of galaxies with the same combination of input and output SFR, following a linear scale from blue to yellow corresponding to 1 and 25 (10) galaxies with ${\rm S/N}>3$ (${\rm S/N}>10$). The grey shaded area indicates ${\rm SFR}<1\, \Msolar\, {\rm yr}^{-1}$. The red dashed line is the identity and the red dotted lines indicate output SFR equal to twice or half the input SFR. On the top left of each panel we report the fraction of outliers, bias, and NMAD of the sample with ${\rm SFR}>1\,\Msolar\,{\rm yr}^{-1}$ and, in parentheses, the values for the full sample. On the bottom right we report the number of objects in the test sample. \textit{Bottom:} distribution of the difference between the output and input SFR, for the full sample (\textit{blue dashed line}) and for galaxies with ${\rm SFR}>1\, \Msolar\, {\rm yr}^{-1}$ (\textit{black solid line}). The red vertical dashed line shows a null difference and the red dotted lines indicate output SFR equal to twice or half the input SFR. \textit{From left to right}: SFR recovered using DLNN with four \Euclid filters considering objects with  ${\rm S/N}>3$, SFR recovered using DLNN with four \Euclid filters and five ancillary bands considering objects with  ${\rm S/N}>3$, SFR recovered using DLNN with four \Euclid filters considering objects with  ${\rm S/N}>10$, SFR recovered using DLNN with four \Euclid filters and five ancillary bands considering objects with  ${\rm S/N}>10$. The ten runs of each network are combined using a Meta-learner.}
	\label{fig:ML_NN_SFR}
\end{figure*}

\begin{figure*}
	\centering
	\includegraphics[width=0.23\linewidth, keepaspectratio]{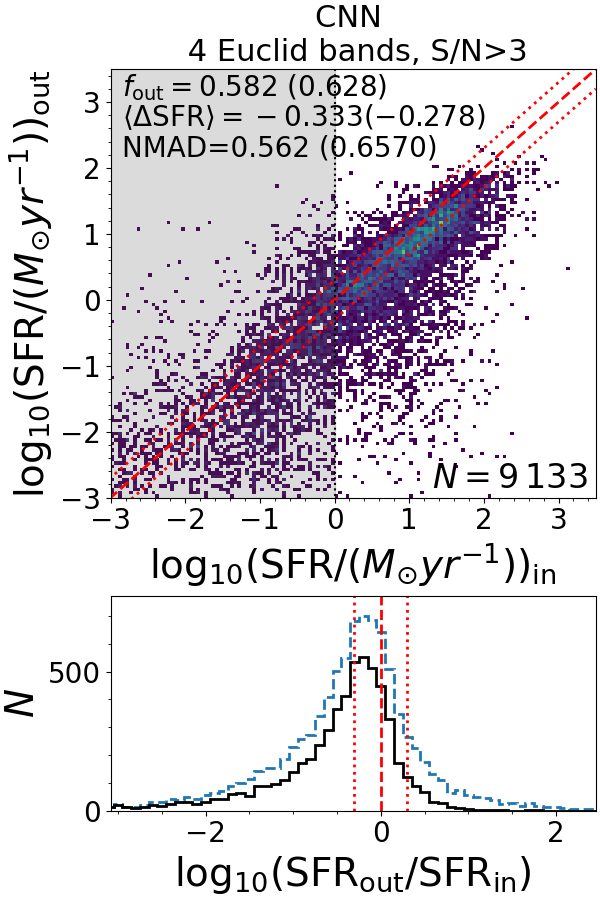}
	\includegraphics[width=0.23\linewidth, keepaspectratio]{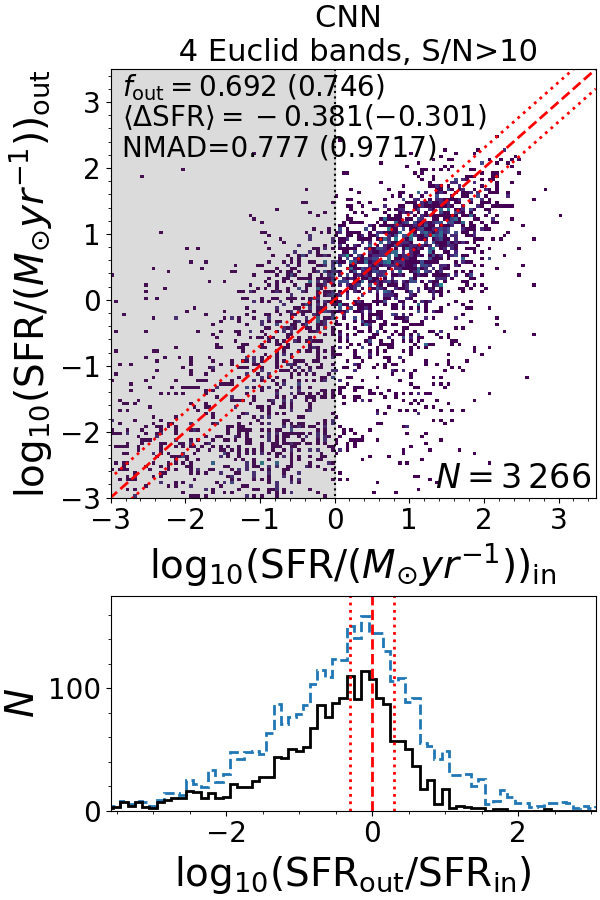}
	\includegraphics[width=0.23\linewidth, keepaspectratio]{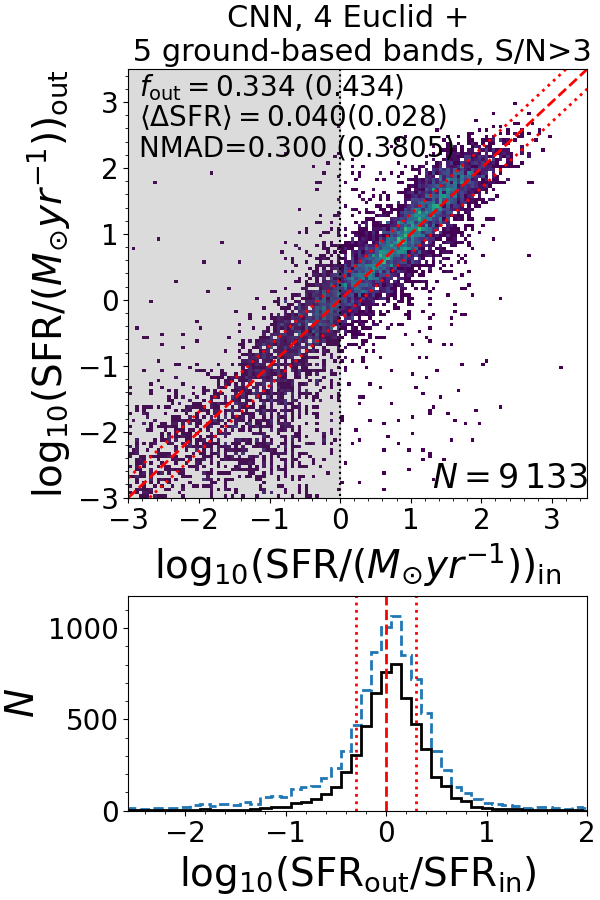}
	\includegraphics[width=0.23\linewidth, keepaspectratio]{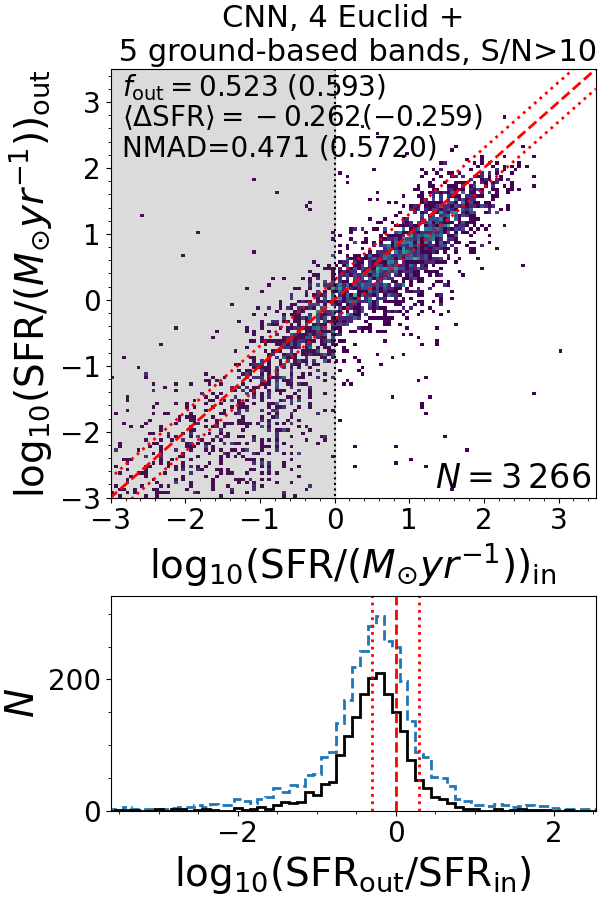}
	\caption{Same as Figure \ref{fig:ML_NN_SFR}, but for the runs using the CNN. }
	\label{fig:ML_CNN_SFR}
\end{figure*}

\begin{table*}
	\caption{Same as Table \ref{tab:fout_z}, but for the SFR. The results correspond to all galaxies with ${\rm SFR}>1\,\Msolar\,{\rm yr}^{-1}$, while values in parentheses correspond to the full sample. The first four lines correspond to results derived with SED fitting.} 
	\centering 
	\begin{tabular}{c c c c c c c c}
		\hline\hline 
		Algorithm & $N_{\rm in}$ &  S/N & Combination &  $f_{\rm out}$ & $\langle\Delta$SFR$\rangle$ & NMAD & MSE\\
		(1) & (2) & (3) & (4) & (5) & (6) & (7) & (8) \\
		\hline
        SED  & 4 & 3 & & 0.752(0.814) & 0.151(0.500) & 0.997(1.412) & 6.331(7.758)\\
        SED  & 4 & 10 & & 0.784(0.856) & 0.286(0.807) & 1.142(1.890) & 5.036(10.887)\\
        SED  & 9 & 3 & & 0.560(0.622) & $-$0.115($-$0.065) & 0.521(0.637) & 1.577(5.680)\\
        SED  & 9 & 10 & & 0.511(0.629) & $-$0.053($-$0.028) & 0.459(0.669) & 1.313(9.075)\\
		\hline
		DLNN	&  4 & 3 & best         &  0.515(0.587) & $-$0.155($-$0.136) & 0.467(0.592) & 0.810(0.956)\\ %
		        &    &   & median      &  0.530(0.598) & $-$0.157($-$0.125) & 0.484(0.605) & 0.791(0.920)\\ %
		        &    &   & Meta-learner &  0.512(0.581) & $-$0.079($-$0.056)  & 0.456(0.563) & 0.697(0.904)\\ %
		DLNN	&  4 & 10& best         &  0.677(0.737) & $-$0.183($-$0.143) & 0.715(0.931) & 1.207(1.463)\\ %
			    &    &   & median      &  0.708(0.756) & $-$0.300($-$0.233) & 0.840(0.989) & 1.462(1.559)\\ %
			    &    &   & Meta-learner &  0.607(0.699) & $-$0.250($-$0.248) & 0.626(0.859) & 1.263(1.479)\\ %
		CNN	    &  4 & 3 & best         &  0.553(0.619) & $-$0.162($-$0.157)  & 0.516(0.634) & 0.956(1.067) \\ %
		        &    &   & median      &  0.559(0.625) & $-$0.185($-$0.141) & 0.528(0.656) & 0.841(0.986)\\ %
		        &    &   & Meta-learner &  0.582(0.628) & $-$0.333($-$0.278) & 0.562(0.657) & 0.803(0.976)\\ %
		CNN	    &  4 & 10& best         &  0.682(0.749) & $-$0.355(0.255)  & 0.775(0.997) & 1.137(1.503)\\ %
		        &    &   & median      &  0.715(0.761) & $-$0.459($-$0.334) & 0.891(1.065) & 1.273(1.473)\\ %
		        &    &   & Meta-learner &  0.692(0.746) & $-$0.381($-$0.301)  & 0.777(0.972) & 1.180(1.533)\\ %
		DLNN	&  9 & 3 & best         &  0.310(0.411) & $-$0.023($-$0.021)  & 0.280(0.350) & 0.235(0.453)\\ %
		        &    &   & median      &  0.319(0.419) & $-$0.033($-$0.0046) & 0.292(0.363) & 0.246(0.461)\\ %
		        &    &   & Meta-learner &  0.349(0.432) & $-$0.145($-$0.144)   & 0.314(0.375) & 0.249(0.478)\\ %
		DLNN    &  9 & 10& best         &  0.419(0.545) & 0.028(0.016)  & 0.374(0.495) & 0.369(0.730)\\ %
		        &    &   & median      &  0.453(0.563) & $-$0.039($-$0.072)  & 0.393(0.529) & 0.367(0.752)\\ %
		        &    &   & Meta-learner &  0.415(0.526) & $-$0.073(0.087)  & 0.365(0.475) & 0.369(0.780)\\ %
		CNN	    &  9 & 3 & best         &  0.325(0.440) & $-$0.046(0.016)  & 0.293(0.383) & 0.265(0.546)\\ %
		        &    &   & median      &  0.342(0.446) & $-$0.056($-$0.065) & 0.304(0.390) & 0.249(0.493)\\ %
		        &    &   & Meta-learner &  0.334(0.434) & 0.040(0.028) & 0.300(0.380) & 0.245(0.524)\\ %
		CNN	    &  9 & 10& best         &  0.443(0.547) & $-$0.111($-$0.112)  & 0.386(0.505) & 0.395(0.822)\\ %
		        &    &   & median      &  0.472(0.588) & $-$0.047($-$0.067)  & 0.417(0.562) & 0.378(0.795)\\ %
		        &    &   & Meta-learner &  0.523(0.593) & $-$0.262($-$0.259)  & 0.471(0.572) & 0.412(0.888)\\ %
		\hline
	\end{tabular}
	\label{tab:fout_SFR}
\end{table*}

In this section we report the results for the SFR retrieval with machine learning, which are summarised in Table \ref{tab:fout_SFR} for all networks, and are plotted in Figures \ref{fig:ML_NN_SFR} and \ref{fig:ML_CNN_SFR}, for the DLNN and CNN runs combined with the Meta-learner. Neither the redshift nor the stellar mass are among the inputs when deriving the SFR. As for the stellar mass, outliers are arbitrary defined as galaxies with SFR which are incorrect by, at least, a factor of two ($\sim0.3 \rm dex$). The bias is defined as 

\begin{equation}
 \langle\Delta {\rm SFR}\rangle={\rm median}[\logten({\rm SFR}_{\rm out}/{\rm SFR}_{\rm in})]   
\end{equation}
and the normalised median absolute deviation for the SFR corresponds to 
\begin{equation}
{\rm NMAD}=1.48\,{\rm median}[|\logten({\rm SFR}_{\rm out}/{\rm SFR}_{\rm in})|].    
\end{equation}
\par

 We find that the SFR is much more challenging to estimate than the stellar mass, as is evident by looking at the fraction of outliers, which ranges from 0.310 to 0.715, and the NMAD, which is always above 0.28, i.e. $\sim32\%$ of the sample have a SFR wrong by at least 0.28 dex. The results obtained with the three methods to combine the ten runs of each network are generally similar, with the Meta-learner and the best of the ten runs giving slightly better results than the median of ten runs. This is probably due to the large variation between the different runs, whose output SFRs have a mean standard deviation between 0.16 to 0.43. In the rest of this section we focus on results obtained with the Meta-learner, for consistency with the redshift and stellar mass measures. \par

The DLNN gives in general a more precise value of the SFR than the CNN. This improvement is mainly driven by a reduction in the outlier fraction, down to $\Delta\,f_{\rm out}=0.11$, but also by a small decrease of the NMAD (i.e. $\Delta\,{\rm NMAD}\leq0.15$). One exception is the case with  ${\rm S/N}>3$ and using nine input filters, for which the CNN results slightly improves (i.e., $\Delta\,f_{\rm out}=0.015$) over the DLNN ones. Not surprisingly, the use of only $H_{\scriptscriptstyle\rm E}$-band images alone is not sufficient to estimate the SFR, as it results, for example, in an outlier fraction of $f_{\rm out}=0.887$ when averaging the output of the ten runs of the sample limited to images with ${\rm S/N}>3$ and ${\rm SFR}>1\,\Msolar\,{\rm yr}^{-1}$. In the future, the inclusion of images at wavelength shorter than the $H_{\scriptscriptstyle\rm E}$ band, which are more sensitive to the SFR and which will likely improve the predictions of this physical property, can be tested. \par

On one hand, for the SFR estimation it is not useful to limit the sample to galaxies with a high  S/N, as the fraction of outliers increases by 0.07--0.16 when comparing the results of the samples with  ${\rm S/N}>3$ and  ${\rm S/N}>10$, similarly to what has been seen for the stellar mass (Section \ref{sec:mass}). On the other hand, it is evident that the inclusion of filters at short wavelengths, like the $u$, $g$, $r$, $i$, and $z$ ground-based filters, improves the SFR estimates, lowering the outlier fraction by $\Delta\,f_{\rm out}=0.15$--0.25. The importance of the optical filters is also evident by performing a sensitivity analysis of the input features (Appendix \ref{sec:Sensitivity}). This is not surprising considering that ultra-violet wavelengths are better tracers of the SFR than near-IR ones \citep{Pforr2012,Pforr2013}. However, even with nine input filters, the SFR measures remain more challenging than measuring the stellar mass or the redshift with the setup we use. \par

\subsubsection{Variation with redshift and $I_{\scriptscriptstyle\rm E}$ magnitude}
\begin{figure}
    \centering
    \includegraphics[width=0.9\linewidth, keepaspectratio]{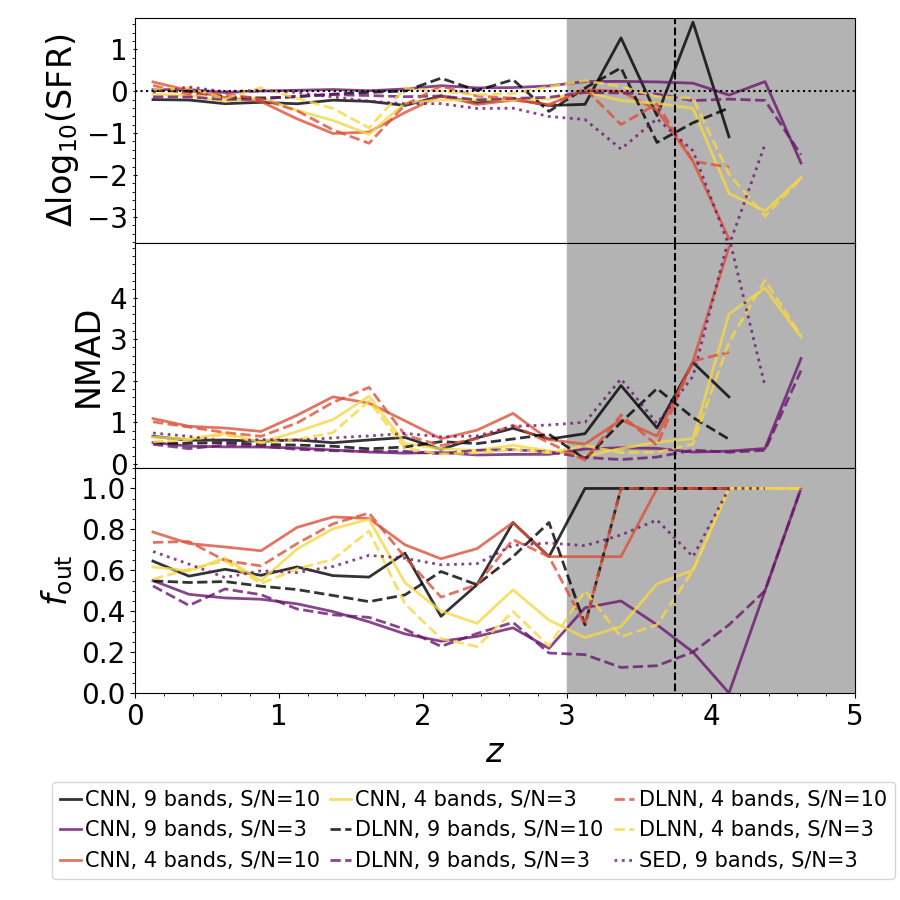}
    \caption{Redshift variation of the bias (\textit{top}), NMAD (\textit{centre}), and outlier fraction (\textit{bottom}) of the recovered SFR. 
    Solid and dashed lines indicate the statistics of CNN and DLNN, respectively, with different colors depending on the set of inputs (see legend). The grey area (black vertical dashed line) shows the redshift bins with less than 100 objects in the training samples which have  ${\rm S/N}>10$ ( ${\rm S/N}>3$).}
    \label{fig:SFR_z}
\end{figure}

\begin{figure}
    \centering
    \includegraphics[width=0.9\linewidth, keepaspectratio]{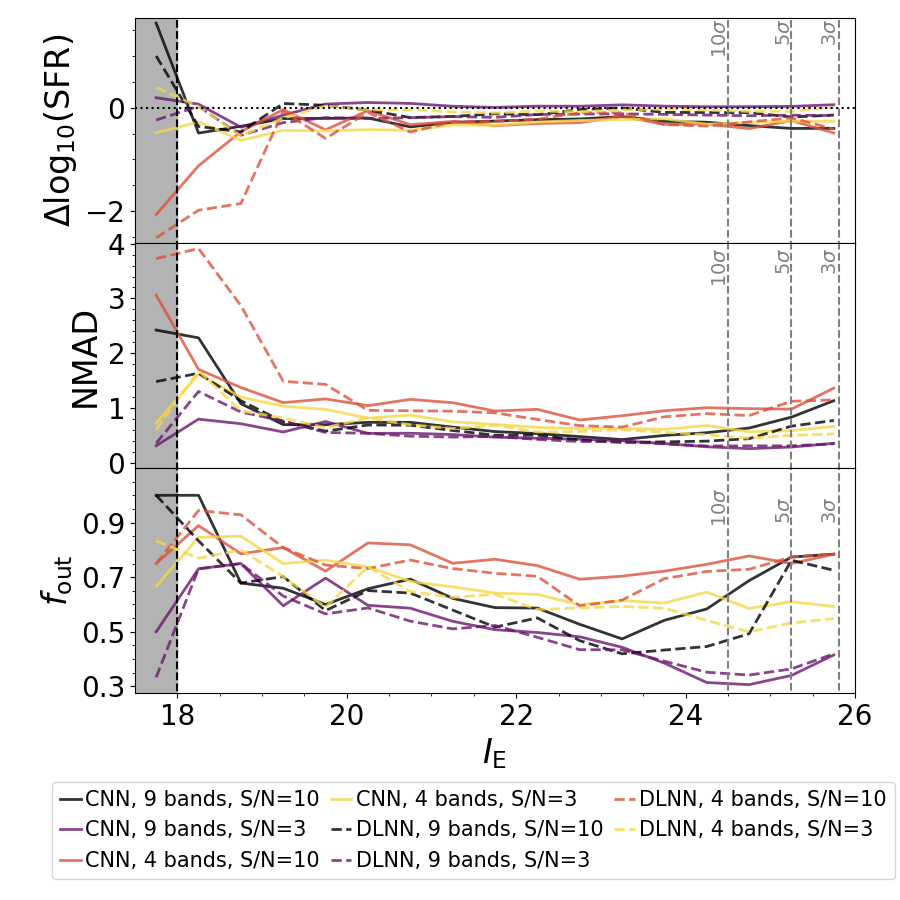}
    \caption{Same as Figure \ref{fig:SFR_z}, but focusing on the variation with the $I_{\scriptscriptstyle\rm E}$ magnitude.}
    \label{fig:SFR_VIS}
\end{figure}

In Figures \ref{fig:SFR_z} and \ref{fig:SFR_VIS}, we show the variation with redshift and $I_{\scriptscriptstyle\rm E}$ magnitude of the fraction of outliers, the bias, and the NMAD of the recovered SFR. As for the stellar mass, there is a rapid deterioration of the SFR estimation as soon as the number of objects available for training is relative small, i.e. $z>4$ and $I_{\scriptscriptstyle\rm E}<18$. In addition, for some of the most accurate networks, i.e. CNN and DLNN with nine input filters and  ${\rm S/N}>3$ (purple solid and dashed lines line in Figure \ref{fig:SFR_z}), the fraction of outliers decreases at increasing redshift, ranging from $f_{\rm out}=0.52$--0.55 at $z=0.125$ to $f_{\rm out}\sim0.13$--0.33 at $z=3.6$. For these networks the outlier fraction also decreases toward fainter $I_{\scriptscriptstyle\rm E}$ magnitudes. We can speculate on different effects driving these dependencies. First, at increasing redshifts our filters trace shorter rest-frame wavelengths, which are more sensitive to SFR. Second, the average SFR of star-forming galaxies increases with redshift \citep[i.e.][]{Noeske2007,Brinchmann2004,Bisigello2018}, making the SFR easier to estimate for the networks. In the future the SFR estimation at low redshift can be further analysed with a sample more focused on low-z galaxies than the one analysed in this work. The dependence of the outlier fraction with $I_{\scriptscriptstyle\rm E}$ magnitude can be linked with the dependence with redshift, as high-redshift galaxies are expected to be fainter than lower-redshift ones. \par

\subsubsection{Comparison with the \textit{LePhare} SED fitting}
Finally, as done for redshift and stellar mass, we compare the results obtained with the CNN and DLNN methods with the results derived with a SED fitting procedure, for which we give more details in Appendix \ref{sec:SEDfit}. Our machine learning algorithms perform better than the SED fitting, but the improvement is not as pronounced as it is for the redshift and the stellar mass, with a difference $\Delta f_{\rm out}<0.250$. It is however worth noticing that the fraction of outliers derived with the DLNN with  ${\rm S/N}>3$ and four filters as input ($f_{\rm out}=0.512$) is lower than the fraction of outliers derived with the SED fitting method for the same sample, but when using nine input filters ($f_{\rm out}=0.560$).\par
Finally, in Figure \ref{fig:SFR_z} we investigate the variation of bias, NMAD, and outlier fraction with redshift, in the case of SED fitting applied to galaxies with ${\rm S/N}>3$ and nine input filters. As for the stellar mass, the SFR estimation derived with SED fitting is always worst than the one derived with DLNN or CNN at any given redshift. This however may be due to a wrong redshift estimation (see Appendix \ref{sec:SEDfit}). 

\section{Summary}\label{sec:end}

This paper is a general exploration of using machine learning to determine and measure the most basic properties of galaxies, particularly those at higher redshifts. This will be a critical process for the next generation of galaxy surveys as \Euclid, \textit{Rubin}/LSST, and the \textit{Roman} Space Telescope. We investigate this problem in several ways, including different machine learning methods and by using as input different forms of data. We use information from the \Euclid Space Telescope as a baseline for understanding how these estimates can be done on other telescopes with similar data. We thus investigate how well machine learning does in retrieving three main features of galaxies - redshifts, stellar masses, and star-formation rates. This work presents only point estimates for all these quantities and the inclusion of probability distribution functions or statistical errors will be investigated on a future work. \par
Our main results are the following:
\begin{itemize}
\item Our machine learning algorithms performs better than traditional methods. In particular, to estimate the stellar mass we consider a simple but direct method consisting on a constant $M/L_{H}$, which is derived from the true median mass-to-light ratio of the sample. As a second method we test a SED fitting procedure, using the same code and templates considered to derive the mock magnitudes, for redshift, stellar mass, and SFR. The redshift and stellar mass machine learning runs outperformed the other methods, while the improvement in the SFR estimation is more limited. It is however necessary to keep in mind that ML networks, on the contrary of SED fitting procedure, are limited to the parameter space of the sample used for training.
\item We verify that it is preferable to combine the results of different runs using a Meta-learner, i.e. an additional DLNN which uses the results of the other networks as inputs. The Meta-learner outperforms the median of the results and the best among the different runs for the redshift predictions, with an improvement in the outlier fraction even up to $\Delta\,f_{\rm out}=0.029$.
\item The inclusion of $H_{\scriptscriptstyle\rm E}$-band images, in addition to the integrated magnitudes, is particularly useful for the stellar mass estimation, due to the fact that the $H_{\scriptscriptstyle\rm E}$-band filter traces the light from relatively old stellar populations, at least at low redshift, and it is therefore a good tracer of stellar mass through structure. The inclusion of images in this filter has a small impact on the redshift estimation, while it mainly introduces noise in the SFR derivation. In the future, the impact of images on the SFR and redshift retrieval may be further tested by including images at shorter wavelengths, which are more sensitive to on-going star formation, and SED features useful for redshift estimation, such as the 4000 \AA, or images with a smaller angular resolution than the ones tested.\
\item Limiting the input sample only to galaxies with  ${\rm S/N}>10$ in the $H_{\scriptscriptstyle\rm E}$-band improves the results only in a few cases. This selection improves the quality of the input data, but, at the same time, it reduces the number of galaxies in the training sample. 
\item We compare results obtained using only the four \Euclid filters and complementing them with additional five LSST-like filters. The improvement is evident in all cases, with the fraction of outliers decreasing by $\Delta\,f_{\rm out}=0.15$--0.25 for the SFR estimation, while it decreases to below $f_{\rm out}\leq0.020$ and 0.005 for the stellar mass and the redshift, respectively. These results indicate the necessity of an eventual coordinated effort from \textit{Rubin}/LSST and \Euclid to improve the measurement of physical properties such as redshift, stellar mass, and SFR. \par

\end{itemize}

\section*{Acknowledgements}
LB and CC acknowledge the support of the STFC Cosmic Vision funding. LB acknowledges the financial support of Agenzia Spaziale Italiana (ASI) under the research contract 2018-31-HH.0. SvM acknowledges funding from the European Research Council through the award of the Consolidator Grant ID 681627-BUILDUP. HH is supported by a Heisenberg grant of the Deutsche Forschungsgemeinschaft (Hi 1495/5-1) as well as an ERC Consolidator Grant (No. 770935). MB acknowledges financial contributions from the agreement ASI/INAF 2018-23-HH.0, Euclid ESA mission – Phase D.
\AckEC
In this work we made use of the Numpy \citep{Numpy} package for Python.

\section*{Data availability}
Data included in this paper will be available on request.



\bibliographystyle{mnras}
\bibliography{mybib} 

\vspace{10pt}
\vspace{10pt}
\noindent

$^{1}$ Dipartimento di Fisica e Astronomia "G.Galilei", Universit\'a di Padova, Via Marzolo 8, I-35131 Padova, Italy\\
$^{2}$ INAF-Osservatorio di Astrofisica e Scienza dello Spazio di Bologna, Via Piero Gobetti 93/3, I-40129 Bologna, Italy\\
$^{3}$ School of Physics and Astronomy, University of Nottingham, University Park, Nottingham NG7 2RD, UK\\
$^{4}$ Jodrell Bank Centre for Astrophysics, Department of Physics and Astronomy, University of Manchester, Oxford Road, Manchester M13 9PL, UK\\
$^{5}$ Sterrenkundig Observatorium, Universiteit Gent, Krijgslaan 281 S9, B-9000 Gent, Belgium\\
$^{6}$ INFN section of Naples, Via Cinthia 6, I-80126, Napoli, Italy\\
$^{7}$ INAF-Osservatorio Astronomico di Capodimonte, Via Moiariello 16, I-80131 Napoli, Italy\\
$^{8}$ Department of Physics "E. Pancini", University Federico II, Via Cinthia 6, I-80126, Napoli, Italy\\
$^{9}$ Instituto de Astrof\'isica e Ci\^encias do Espa\c{c}o, Universidade do Porto, CAUP, Rua das Estrelas, PT4150-762 Porto, Portugal\\
$^{10}$ INAF-Osservatorio Astrofisico di Arcetri, Largo E. Fermi 5, I-50125, Firenze, Italy\\
$^{11}$ Institute of Cosmology and Gravitation, University of Portsmouth, Portsmouth PO1 3FX, UK\\
$^{12}$ Istituto Nazionale di Astrofisica (INAF) - Osservatorio di Astrofisica e Scienza dello Spazio (OAS), Via Gobetti 93/3, I-40127 Bologna, Italy\\
$^{13}$ Kapteyn Astronomical Institute, University of Groningen, PO Box 800, 9700 AV Groningen, The Netherlands\\
$^{14}$ Universit\'e Paris-Saclay, CNRS, Institut d'astrophysique spatiale, 91405, Orsay, France\\
$^{15}$ Dipartimento di Fisica e Astronomia, Universit\'a di Bologna, Via Gobetti 93/2, I-40129 Bologna, Italy\\
$^{16}$ INFN-Sezione di Bologna, Viale Berti Pichat 6/2, I-40127 Bologna, Italy\\
$^{17}$ Universit\"ats-Sternwarte M\"unchen, Fakult\"at f\"ur Physik, Ludwig-Maximilians-Universit\"at M\"unchen, Scheinerstrasse 1, 81679 M\"unchen, Germany\\
$^{18}$ Max Planck Institute for Extraterrestrial Physics, Giessenbachstr. 1, D-85748 Garching, Germany\\
$^{19}$ INAF-Osservatorio Astrofisico di Torino, Via Osservatorio 20, I-10025 Pino Torinese (TO), Italy\\
$^{20}$ Dipartimento di Fisica, Universit\'a degli studi di Genova, and INFN-Sezione di Genova, via Dodecaneso 33, I-16146, Genova, Italy\\
$^{21}$ INFN-Sezione di Roma Tre, Via della Vasca Navale 84, I-00146, Roma, Italy\\
$^{22}$ Dipartimento di Fisica, Universit\'a degli Studi di Torino, Via P. Giuria 1, I-10125 Torino, Italy\\
$^{23}$ INFN-Sezione di Torino, Via P. Giuria 1, I-10125 Torino, Italy\\
$^{24}$ INAF-IASF Milano, Via Alfonso Corti 12, I-20133 Milano, Italy\\
$^{25}$ Institut de F\'{i}sica d'Altes Energies (IFAE), The Barcelona Institute of Science and Technology, Campus UAB, 08193 Bellaterra (Barcelona), Spain\\
$^{26}$ Port d'Informaci\'{o} Cient\'{i}fica, Campus UAB, C. Albareda s/n, 08193 Bellaterra (Barcelona), Spain\\
$^{27}$ Institut d'Estudis Espacials de Catalunya (IEEC), Carrer Gran Capit\'a 2-4, 08034 Barcelona, Spain\\
$^{28}$ Institute of Space Sciences (ICE, CSIC), Campus UAB, Carrer de Can Magrans, s/n, 08193 Barcelona, Spain\\
$^{29}$ INAF-Osservatorio Astronomico di Roma, Via Frascati 33, I-00078 Monteporzio Catone, Italy\\
$^{30}$ Dipartimento di Fisica e Astronomia "Augusto Righi" - Alma Mater Studiorum Universit\'a di Bologna, Viale Berti Pichat 6/2, I-40127 Bologna, Italy\\
$^{31}$ Institute for Astronomy, University of Edinburgh, Royal Observatory, Blackford Hill, Edinburgh EH9 3HJ, UK\\
$^{32}$ ESAC/ESA, Camino Bajo del Castillo, s/n., Urb. Villafranca del Castillo, 28692 Villanueva de la Ca\~nada, Madrid, Spain\\
$^{33}$ European Space Agency/ESRIN, Largo Galileo Galilei 1, 00044 Frascati, Roma, Italy\\
$^{34}$ Univ Lyon, Univ Claude Bernard Lyon 1, CNRS/IN2P3, IP2I Lyon, UMR 5822, F-69622, Villeurbanne, France\\
$^{35}$ Observatoire de Sauverny, Ecole Polytechnique F\'ed\'erale de Lau- sanne, CH-1290 Versoix, Switzerland\\
$^{36}$ Mullard Space Science Laboratory, University College London, Holmbury St Mary, Dorking, Surrey RH5 6NT, UK\\
$^{37}$ Departamento de F\'isica, Faculdade de Ci\^encias, Universidade de Lisboa, Edif\'icio C8, Campo Grande, PT1749-016 Lisboa, Portugal\\
$^{38}$ Instituto de Astrof\'isica e Ci\^encias do Espa\c{c}o, Faculdade de Ci\^encias, Universidade de Lisboa, Campo Grande, PT-1749-016 Lisboa, Portugal\\
$^{39}$ Department of Astronomy, University of Geneva, ch. d\'Ecogia 16, CH-1290 Versoix, Switzerland\\
$^{40}$ Department of Physics, Oxford University, Keble Road, Oxford OX1 3RH, UK\\
$^{41}$ INFN-Padova, Via Marzolo 8, I-35131 Padova, Italy\\
$^{42}$ Universit\'e Paris-Saclay, Universit\'e Paris Cit\'e, CEA, CNRS, Astrophysique, Instrumentation et Mod\'elisation Paris-Saclay, 91191 Gif-sur-Yvette, France\\
$^{43}$ INAF-Osservatorio Astronomico di Trieste, Via G. B. Tiepolo 11, I-34143 Trieste, Italy\\
$^{44}$ Aix-Marseille Univ, CNRS/IN2P3, CPPM, Marseille, France\\
$^{45}$ Istituto Nazionale di Fisica Nucleare, Sezione di Bologna, Via Irnerio 46, I-40126 Bologna, Italy\\
$^{46}$ INAF-Osservatorio Astronomico di Padova, Via dell'Osservatorio 5, I-35122 Padova, Italy\\
$^{47}$ Dipartimento di Fisica "Aldo Pontremoli", Universit\'a degli Studi di Milano, Via Celoria 16, I-20133 Milano, Italy\\
$^{48}$ INAF-Osservatorio Astronomico di Brera, Via Brera 28, I-20122 Milano, Italy\\
$^{49}$ INFN-Sezione di Milano, Via Celoria 16, I-20133 Milano, Italy\\
$^{50}$ Institute of Theoretical Astrophysics, University of Oslo, P.O. Box 1029 Blindern, N-0315 Oslo, Norway\\
$^{51}$ Jet Propulsion Laboratory, California Institute of Technology, 4800 Oak Grove Drive, Pasadena, CA, 91109, USA\\
$^{52}$ von Hoerner \& Sulger GmbH, Schlo{\ss}Platz 8, D-68723 Schwetzingen, Germany\\
$^{53}$ Technical University of Denmark, Elektrovej 327, 2800 Kgs. Lyngby, Denmark\\
$^{54}$ Max-Planck-Institut f\"ur Astronomie, K\"onigstuhl 17, D-69117 Heidelberg, Germany\\
$^{55}$ Universit\'e de Gen\`eve, D\'epartement de Physique Th\'eorique and Centre for Astroparticle Physics, 24 quai Ernest-Ansermet, CH-1211 Gen\`eve 4, Switzerland\\
$^{56}$ Department of Physics and Helsinki Institute of Physics, Gustaf H\"allstr\"omin katu 2, 00014 University of Helsinki, Finland\\
$^{57}$ NOVA optical infrared instrumentation group at ASTRON, Oude Hoogeveensedijk 4, 7991PD, Dwingeloo, The Netherlands\\
$^{58}$ Argelander-Institut f\"ur Astronomie, Universit\"at Bonn, Auf dem H\"ugel 71, 53121 Bonn, Germany\\
$^{59}$ Dipartimento di Fisica e Astronomia "Augusto Righi" - Alma Mater Studiorum Universit\`{a} di Bologna, via Piero Gobetti 93/2, I-40129 Bologna, Italy\\
$^{60}$ Department of Physics, Institute for Computational Cosmology, Durham University, South Road, DH1 3LE, UK\\
$^{61}$ Universit\'e C\^{o}te d'Azur, Observatoire de la C\^{o}te d'Azur, CNRS, Laboratoire Lagrange, Bd de l'Observatoire, CS 34229, 06304 Nice cedex 4, France\\
$^{62}$ Institute of Physics, Laboratory of Astrophysics, Ecole Polytechnique F\'{e}d\'{e}rale de Lausanne (EPFL), Observatoire de Sauverny, 1290 Versoix, Switzerland\\
$^{63}$ European Space Agency/ESTEC, Keplerlaan 1, 2201 AZ Noordwijk, The Netherlands\\
$^{64}$ Department of Physics and Astronomy, University of Aarhus, Ny Munkegade 120, DK-8000 Aarhus C, Denmark\\
$^{65}$ Space Science Data Center, Italian Space Agency, via del Politecnico snc, 00133 Roma, Italy\\
$^{66}$ Centre National d'Etudes Spatiales, Toulouse, France\\
$^{67}$ Institute of Space Science, Bucharest, Ro-077125, Romania\\
$^{68}$  Universit\'e Paris Cit\'e, CNRS, Astroparticule et Cosmologie, F-75013 Paris, France\\
$^{69}$ Departamento de F\'isica, FCFM, Universidad de Chile, Blanco Encalada 2008, Santiago, Chile\\
$^{70}$ Centro de Investigaciones Energ\'eticas, Medioambientales y Tecnol\'ogicas (CIEMAT), Avenida Complutense 40, 28040 Madrid, Spain\\
$^{71}$ Instituto de Astrof\'isica e Ci\^encias do Espa\c{c}o, Faculdade de Ci\^encias, Universidade de Lisboa, Tapada da Ajuda, PT-1349-018 Lisboa, Portugal\\
$^{72}$ Universidad Polit\'ecnica de Cartagena, Departamento de Electr\'onica y Tecnolog\'ia de Computadoras, 30202 Cartagena, Spain\\
$^{73}$ Infrared Processing and Analysis Center, California Institute of Technology, Pasadena, CA 91125, USA\\
$^{74}$ Instituto de Astrof\'isica de Canarias, Calle V\'ia L\'actea s/n, E-38204, San Crist\'obal de La Laguna, Tenerife, Spain\\
$^{75}$ Institut d'Astrophysique de Paris, UMR 7095, CNRS, and Sorbonne Universit\'e, 98 bis boulevard Arago, 75014 Paris, France\\
$^{76}$ AIM, CEA, CNRS, Universit\'{e} Paris-Saclay, Universit\'{e} de Paris, F-91191 Gif-sur-Yvette, France\\
$^{77}$ NASA Ames Research Center, Moffett Field, CA 94035, USA\\
$^{78}$ INFN-Bologna, Via Irnerio 46, I-40126 Bologna, Italy\\
$^{79}$ IFPU, Institute for Fundamental Physics of the Universe, via Beirut 2, 34151 Trieste, Italy\\
$^{80}$ Dipartimento di Fisica e Scienze della Terra, Universit\'a degli Studi di Ferrara, Via Giuseppe Saragat 1, I-44122 Ferrara, Italy\\
$^{81}$ INAF, Istituto di Radioastronomia, Via Piero Gobetti 101, I-40129 Bologna, Italy\\
$^{82}$ Institut de Recherche en Astrophysique et Plan\'etologie (IRAP), Universit\'e de Toulouse, CNRS, UPS, CNES, 14 Av. Edouard Belin, F-31400 Toulouse, France\\
$^{83}$ Institute for Theoretical Particle Physics and Cosmology (TTK), RWTH Aachen University, D-52056 Aachen, Germany\\
$^{84}$ Department of Physics \& Astronomy, University of California Irvine, Irvine CA 92697, USA\\
$^{85}$ University of Lyon, UCB Lyon 1, CNRS/IN2P3, IUF, IP2I Lyon, France\\
$^{86}$ Aix-Marseille Univ, CNRS, CNES, LAM, Marseille, France\\
$^{87}$ INFN-Sezione di Genova, Via Dodecaneso 33, I-16146, Genova, Italy\\
$^{88}$ INAF-Istituto di Astrofisica e Planetologia Spaziali, via del Fosso del Cavaliere, 100, I-00100 Roma, Italy\\
$^{89}$ School of Physics, HH Wills Physics Laboratory, University of Bristol, Tyndall Avenue, Bristol, BS8 1TL, UK\\
$^{90}$ Instituto de F\'isica Te\'orica UAM-CSIC, Campus de Cantoblanco, E-28049 Madrid, Spain\\
$^{91}$ Department of Physics, P.O. Box 64, 00014 University of Helsinki, Finland\\
$^{92}$ Ruhr University Bochum, Faculty of Physics and Astronomy, Astronomical Institute (AIRUB), German Centre for Cosmological Lensing (GCCL), 44780 Bochum, Germany\\
$^{93}$ Department of Physics, Lancaster University, Lancaster, LA1 4YB, UK\\
$^{94}$ Instituto de Astrof\'isica de Canarias (IAC); Departamento de Astrof\'isica, Universidad de La Laguna (ULL), E-38200, La Laguna, Tenerife, Spain\\
$^{95}$ Universit\'e de Paris, F-75013, Paris, France, LERMA, Observatoire de Paris, PSL Research University, CNRS, Sorbonne Universit\'e, F-75014 Paris, France\\
$^{96}$ Department of Physics and Astronomy, University College London, Gower Street, London WC1E 6BT, UK\\
$^{97}$ Astrophysics Group, Blackett Laboratory, Imperial College London, London SW7 2AZ, UK\\
$^{98}$ Univ. Grenoble Alpes, CNRS, Grenoble INP, LPSC-IN2P3, 53, Avenue des Martyrs, 38000, Grenoble, France\\
$^{99}$ Centre de Calcul de l'IN2P3, 21 avenue Pierre de Coubertin F-69627 Villeurbanne Cedex, France\\
$^{100}$ University of Applied Sciences and Arts of Northwestern Switzerland, School of Engineering, 5210 Windisch, Switzerland\\
$^{101}$ Dipartimento di Fisica - Sezione di Astronomia, Universit\'a di Trieste, Via Tiepolo 11, I-34131 Trieste, Italy\\
$^{102}$ INFN, Sezione di Trieste, Via Valerio 2, I-34127 Trieste TS, Italy\\
$^{103}$ Department of Mathematics and Physics E. De Giorgi, University of Salento, Via per Arnesano, CP-I93, I-73100, Lecce, Italy\\
$^{104}$ INFN, Sezione di Lecce, Via per Arnesano, CP-193, I-73100, Lecce, Italy\\
$^{105}$ Institute for Computational Science, University of Zurich, Winterthurerstrasse 190, 8057 Zurich, Switzerland\\
$^{106}$ Higgs Centre for Theoretical Physics, School of Physics and Astronomy, The University of Edinburgh, Edinburgh EH9 3FD, UK\\
$^{107}$ Institut d'Astrophysique de Paris, 98bis Boulevard Arago, F-75014, Paris, France\\
$^{108}$ Institut f\"ur Theoretische Physik, University of Heidelberg, Philosophenweg 16, 69120 Heidelberg, Germany\\
$^{109}$ Universit\'e St Joseph; Faculty of Sciences, Beirut, Lebanon\\
$^{110}$ Department of Astrophysical Sciences, Peyton Hall, Princeton University, Princeton, NJ 08544, USA\\
$^{111}$ Helsinki Institute of Physics, Gustaf H{\"a}llstr{\"o}min katu 2, University of Helsinki, Helsinki, Finland\\
$^{112}$ SISSA, International School for Advanced Studies, Via Bonomea 265, I-34136 Trieste TS, Italy

\appendix
\section{Physical parameter combination}\label{sec:MS}
In this section we verify if the independent derivation of the three physical properties, i.e. redshift, stellar mass, and SFR, bring to some unrealistic galaxies. Some examples of these objects are low-mass galaxies at high-$z$ or low-mass quiescent galaxies, as both galaxy populations would be undetectable given the considered observational limits. \par
\begin{figure}
    \centering
    \includegraphics[width=0.75\linewidth, keepaspectratio]{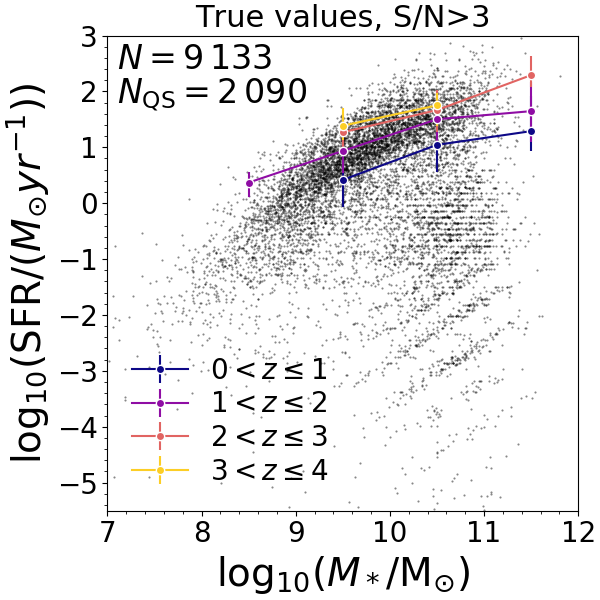}
    \includegraphics[width=0.75\linewidth, keepaspectratio]{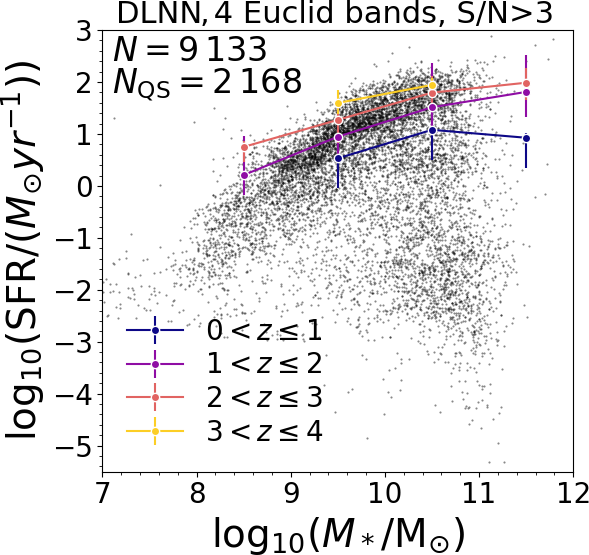}
    \caption{$M_{*}-$SFR plane for all galaxies with ${\rm S/N}>3$ considering the true values for redshift, stellar-mass, and SFR (\textit{top}) and the values derived using a DLNN with four \Euclid filters (\textit{bottom}). Coloured lines indicate the median SFR of the star-forming galaxies (i.e. $\logten[{\rm SFR}/(\Msolar\,{\rm yr}^{-1})]\geq-10.5$) at $z=1$ to 4. On the top left of each panel we report the total number of object in the sample and the number of quiescent galaxies (i.e. $\logten[{\rm SFR}/(\Msolar\,{\rm yr}^{-1})]>-10.5$).}
    \label{fig:MS}
\end{figure}

In Figure \ref{fig:MS} we report the $M_{*}-$SFR plane of galaxies with ${\rm S/N}>3$ using the true values of the physical properties. In the same Figure we show, as example, the $M_{*}-$SFR plane using the physical properties derived with the DLNN with only \Euclid filters for galaxies with ${\rm S/N}>3$. In addition, as a proxy for the star-formation main-sequence \citep[MS, e.g.][]{Noeske2007}, we derive the median SFR value at different redshift, after selecting only star-forming galaxies, approximated as galaxies with  $\logten[{\rm SFR}/(\Msolar\,{\rm yr}^{-1})]\geq-10.5$.\par
We can see that the derived SFR, stellar mass, and redshift are not completely unrelated, even if they have been derived independently. Indeed, the MS is present and its normalisation increases with redshift, as observed in the literature \citep[e.g.][]{Speagle2014,Schreiber2016} and as visible when considering the true values for the physical parameters. In addition, the number of quiescent galaxies ($\logten[{\rm SFR}/(\Msolar\,{\rm yr}^{-1})]<-10.5$) derived using DLNN or CNN varies between the 95-114$\%$ respect to their number in the input samples, indicating that the sSFR is recovered well enough for their identification.\par
Therefore, given the presence of the MS and the little variation on the recovered number of quiescent galaxies, we can conclude that there are no significant evidence of galaxies with nonphysical combinations of SFR, stellar mass, and redshift, even if these quantities are derived independently. We leave to future works the investigation of possible improvement when deriving the mentioned physical properties simultaneously with a single network as well as the full analysis of the derived MS and sSFR.

\section{Sensitivity analysis}\label{sec:Sensitivity}
To understand the importance of the different inputs considered in this work, we performed a sensitivity analysis \citep{Guyon2003} of the parameter space used for redshift, stellar mass, and SFR estimations. For this test we considered the mock catalogue with ${\rm S/N}>3$ and nine input bands, i.e. the four \Euclid filters and the $u$, $g$, $r$, $i$, and $z$ ground-based ones. \par
In particular, we used the Light Gradient Boosting Model \citep[LGBM;][]{LGBM} as the base regression model to perform different feature space analyses, using the knowledge base made by the training and testing sources and the related true targets. We optimised the model hyper-parameters through an automatic grid search for each of the regression use cases (redshift, stellar mass, and SFR), we also introduced 
six additional random features in the parameter space, made by a simple white noise. We then derived the informative contribution given by the input features by alternating four different methods, including:
\begin{itemize}
    \item a feature importance calculation based on a standard tree-based method \citep[i.e., XGBoost;][]{XGBoost};
    \item a Recursive Feature Elimination \citep[RFE;][]{Chen2008} that recursively fits a supervised algorithm considering a smaller sets of features. The excluded features are the ones that are considered less important according to the magnitude of some weights (e.g., the coefficients of linear models, or the feature importances for tree-based models).
    \item Boruta \citep{Boruta2010}, which is a wrapper-based technique for feature selection. In particular, we iteratively fitted a supervised algorithm (a tree-based model) on an extended version of the tabular data. The extended version, in each iteration, is composed of the original data with a horizontally attached shuffled copy of the columns. In each iteration, we maintained only the features that have a higher importance than the best of the shuffled features and are better than the expected random chance (using a binomial distribution).
    \item the SHAP method \citep[SHapley Additive exPlanations;][]{Lipovetsky2001AnalysisOR,Strumbelj2013,Lundberg2017}, which was demonstrated to be effective on mitigating the effects in the selection of high-frequency or high-cardinality features. Taken from cooperative game theory, this method allows us to derive the contributions given by the presence, or absence, of the different input features.

\end{itemize}
We repeated the feature analysis for different training and testing by splitting seeds to mitigate randomness in data selection. We mixed the aforementioned methods through five different combinations, i.e. standard tree-based importance alone, RFE and standard tree-based importance, RFE and SHAP importance, Boruta and standard tree-based importance, and Boruta and SHAP importance. \par
\begin{figure}
    \centering
    \includegraphics[trim={1cm 0 0 0},clip,width=0.75\linewidth, keepaspectratio]{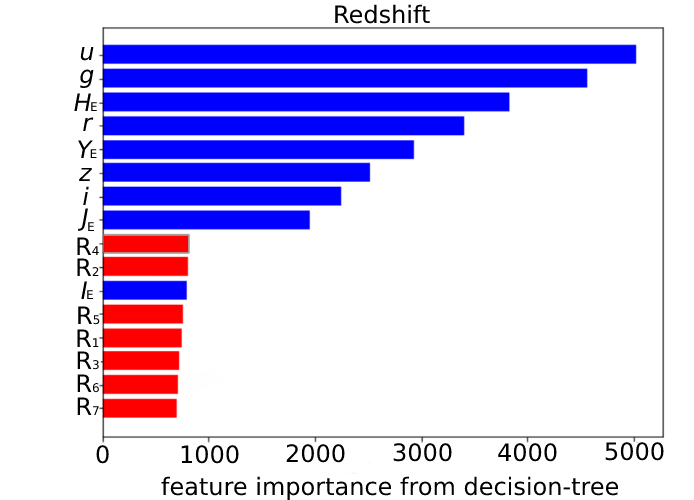}
    \includegraphics[trim={1cm 0 0 0},clip,width=0.75\linewidth, keepaspectratio]{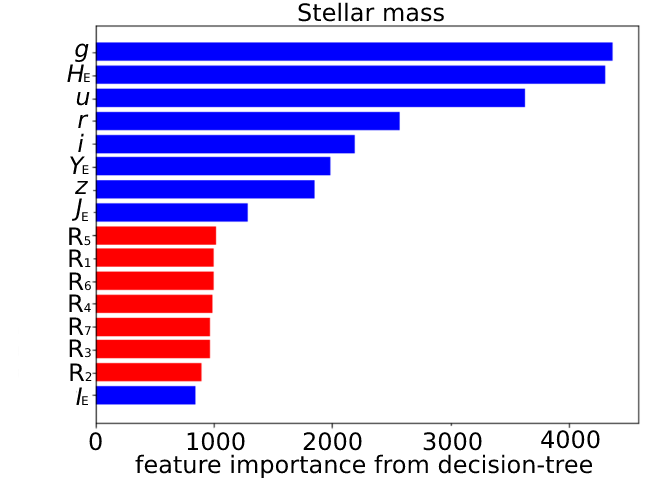}
    \includegraphics[trim={1cm 0 0 0},clip,width=0.75\linewidth, keepaspectratio]{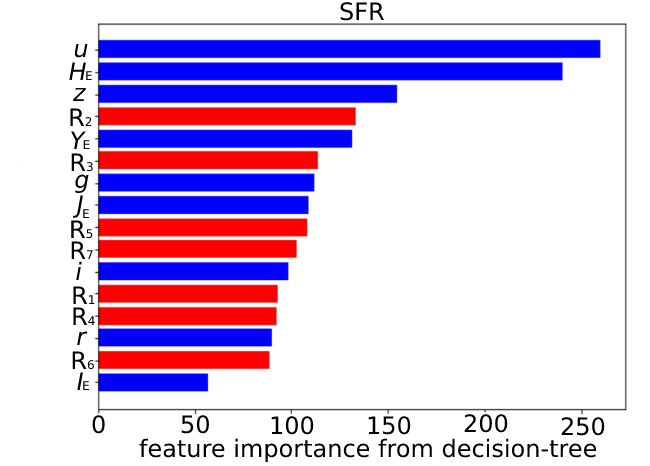}
    \caption{The feature importance calculation based on a standard tree-based method for redshift (\textit{top}), stellar mass (\textit{centre}), and SFR (\textit{bottom}). Red bars indicate random features, while the blue ones are the nine considered filters.}
    \label{fig:feat_imp}
\end{figure}

 
In Figure \ref{fig:feat_imp} we show the results for the feature importance calculation based on the standard tree-based method as example, but the other bring to similar results. Overall, for all redshift, stellar mass, and SFR the optical bands have a high importance, while the $I_{\scriptscriptstyle\rm E}$ band is the less relevant. This may be caused by the wide wavelength range of the $I_{\scriptscriptstyle\rm E}$ filter that is already covered, but with higher spectral resolution, by the $r$, $i$, and $z$ bands. It is however necessary to consider that the importance of the $I_{\scriptscriptstyle\rm E}$ filter may increase for galaxies, a minority in our mock catalogue, that are too faint to be observed in the single $r$, $i$, and $z$ filters.

\color{black}

\section{SED fitting results}\label{sec:SEDfit}
In this section we give more details on results derived with the SED fitting described in Section \ref{sec:propderiv}. \par
\begin{figure*}
    \centering
    \includegraphics[width=0.23\linewidth, keepaspectratio]{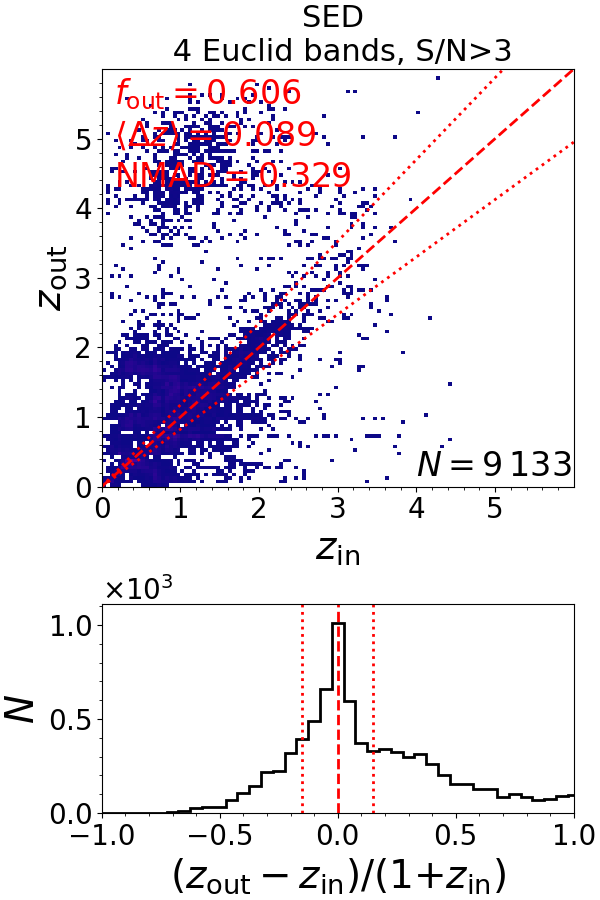}
    \includegraphics[width=0.23\linewidth, keepaspectratio]{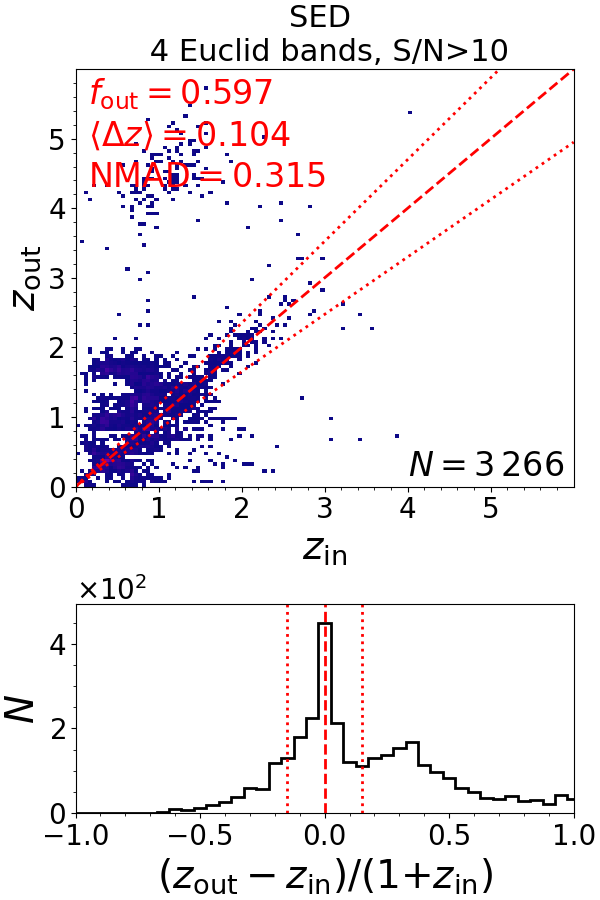}
    \includegraphics[width=0.23\linewidth, keepaspectratio]{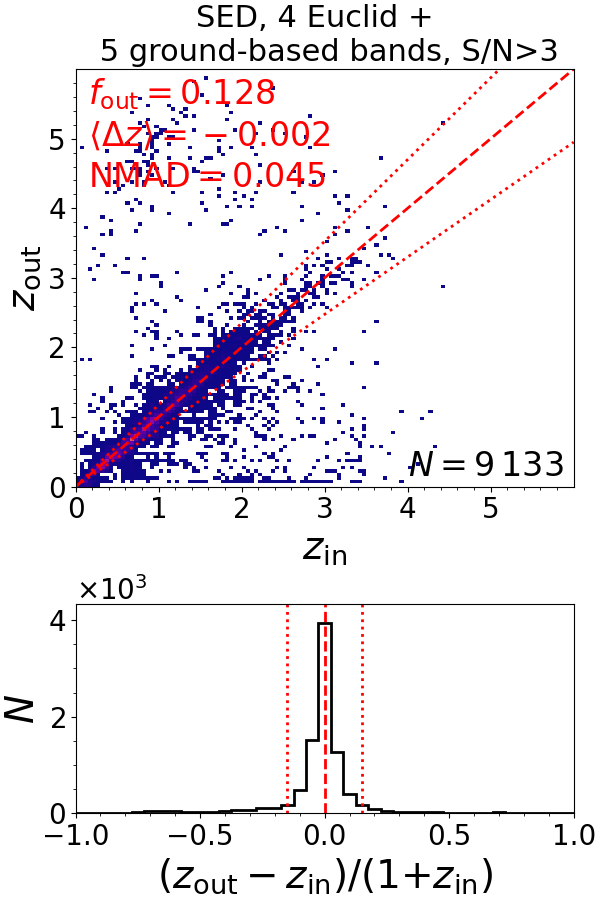}
    \includegraphics[width=0.23\linewidth, keepaspectratio]{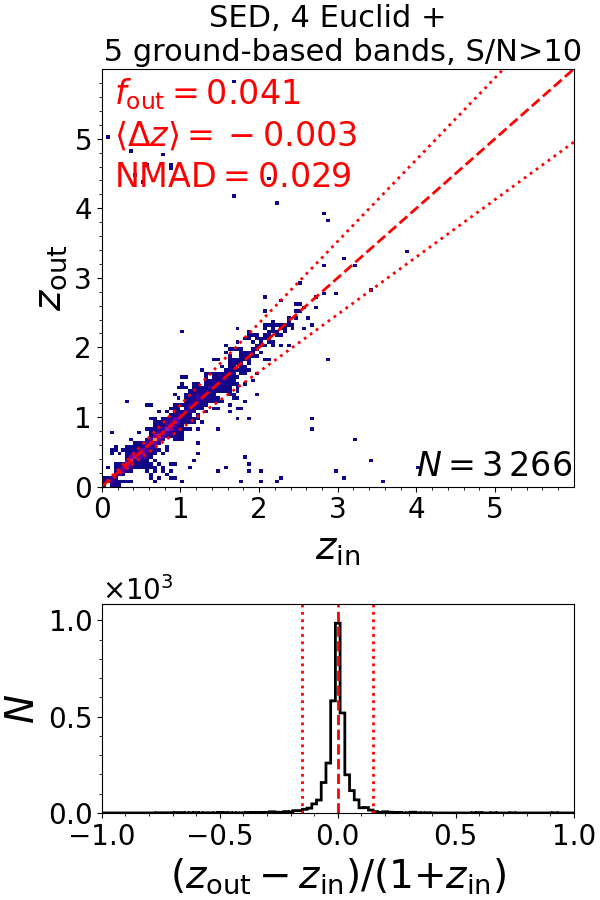}
    \caption{Same as Figure \ref{fig:ML_NN_z}, but for the SED fitting.}
    \label{fig:z_SED}
\end{figure*}

In particular, we report in Figure \ref{fig:z_SED} the results for the redshift. Some of the most extreme outliers, when only the \Euclid filters are as input, are galaxies between $z=1$ and 2 that are wrongly classified as galaxies at $z=4-5$. This is due to 4000$\AA$-break that is wrongly identify by the the SED fitting code as the Lyman-break at 912$\AA$, due to the limited wavelength coverage. However, when nine filters are as input, the improvement is not limited to these extreme outliers, but the general dispersion at $z<2$ is reduced. \par
\begin{figure*}
    \centering
    \includegraphics[width=0.23\linewidth, keepaspectratio]{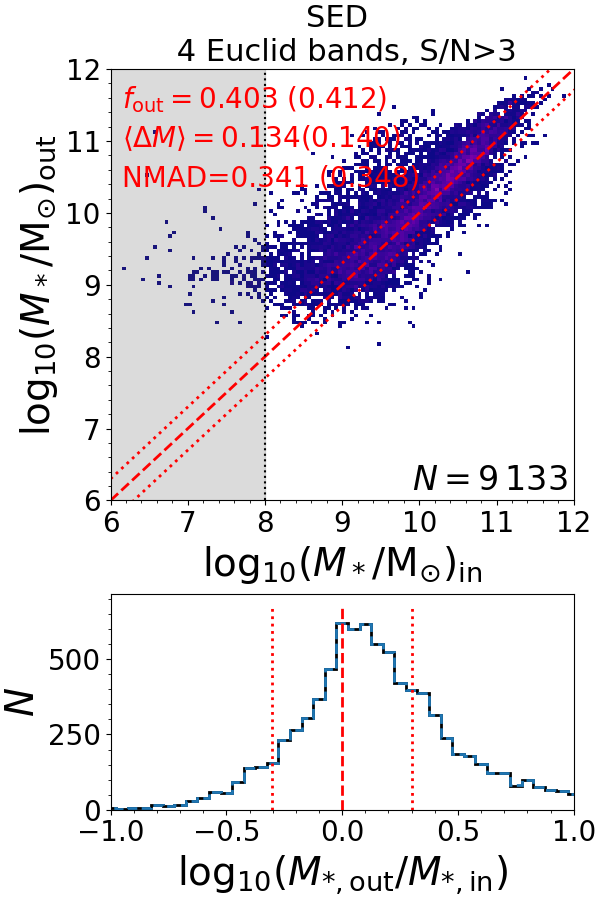}
    \includegraphics[width=0.23\linewidth, keepaspectratio]{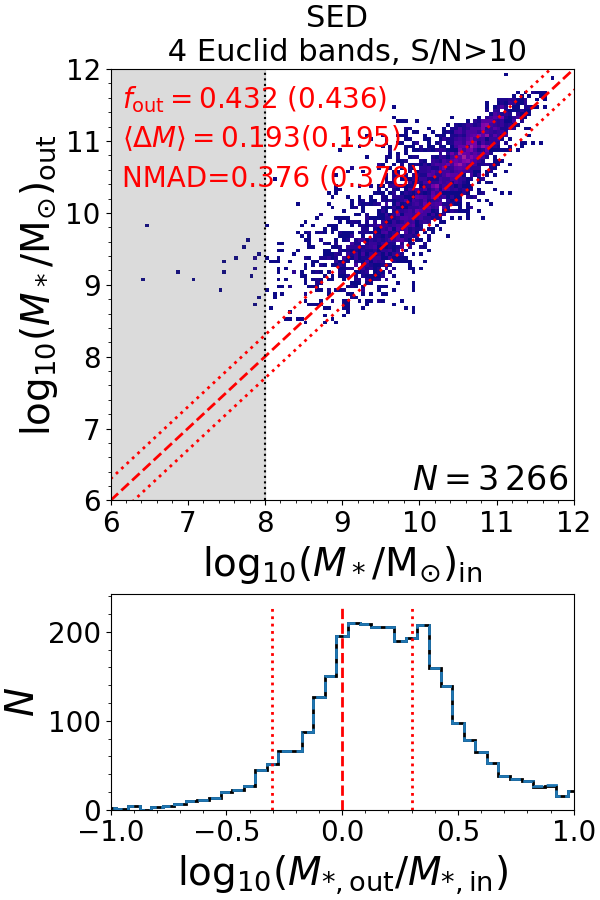}
    \includegraphics[width=0.23\linewidth, keepaspectratio]{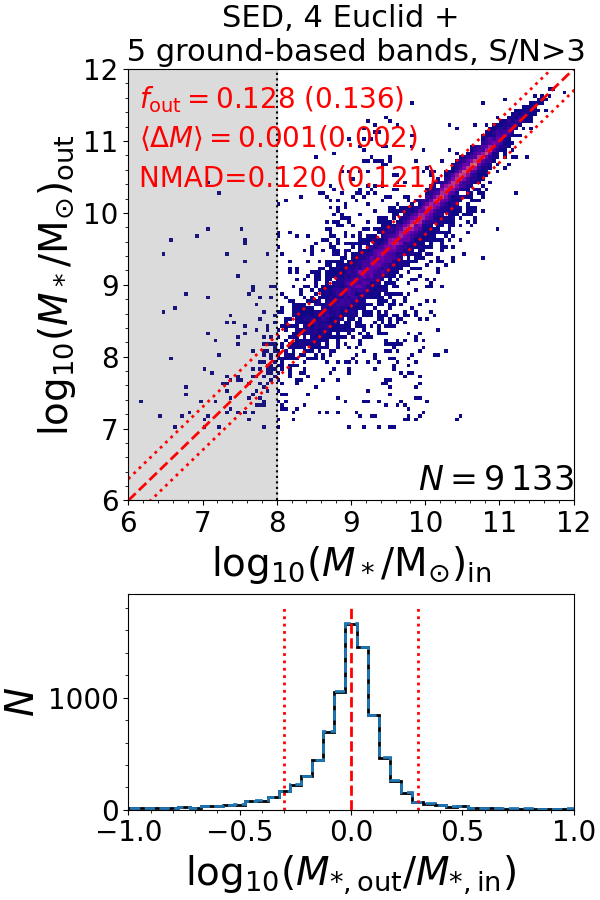}
    \includegraphics[width=0.23\linewidth, keepaspectratio]{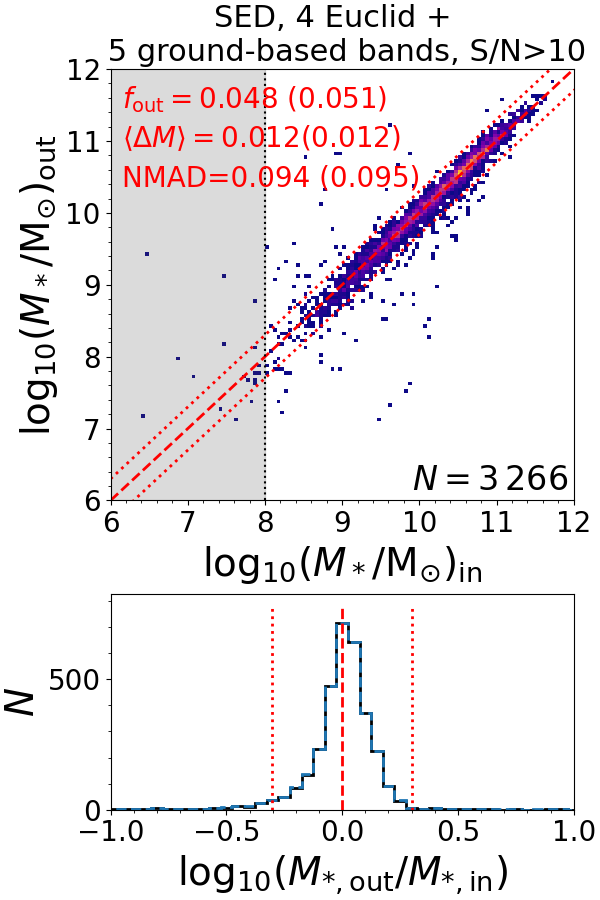}
    \caption{Same as Figure \ref{fig:ML_NN_M}, but for the SED fitting.}
    \label{fig:M_SED}
\end{figure*}

In Figure \ref{fig:M_SED} we show the results for the stellar mass. As the stellar mass is derived in the same run as the redshift, a wrong redshift estimation will influence the stellar mass derivation. This explain the spread visible in the stellar mass derived when only \Euclid filters are considered as input. \par
\begin{figure*}
    \centering
    \includegraphics[width=0.23\linewidth, keepaspectratio]{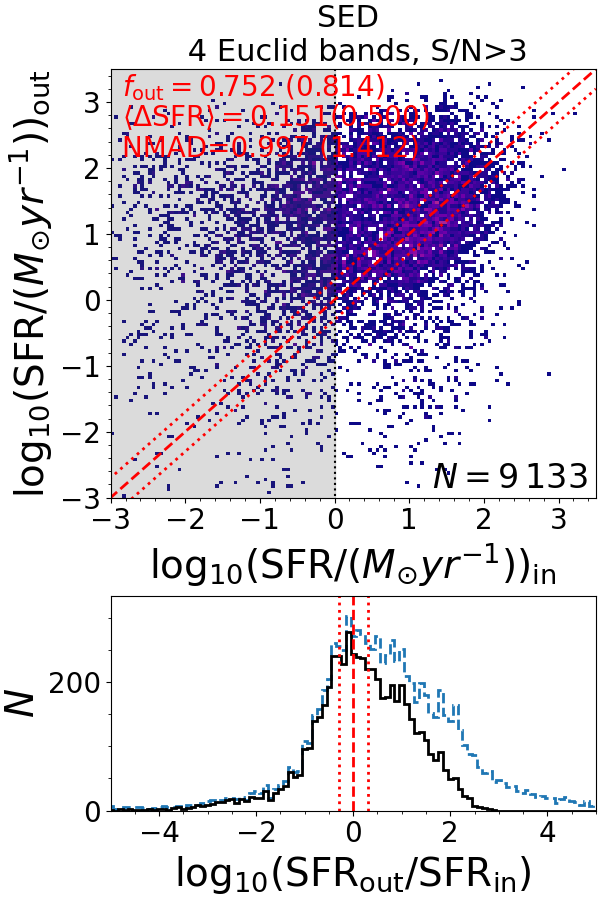}
    \includegraphics[width=0.23\linewidth, keepaspectratio]{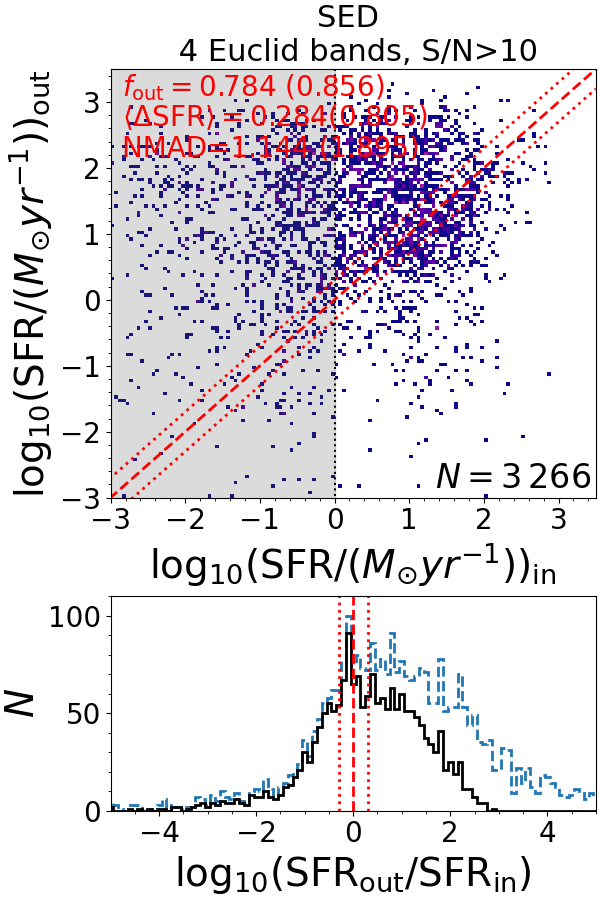}
    \includegraphics[width=0.23\linewidth, keepaspectratio]{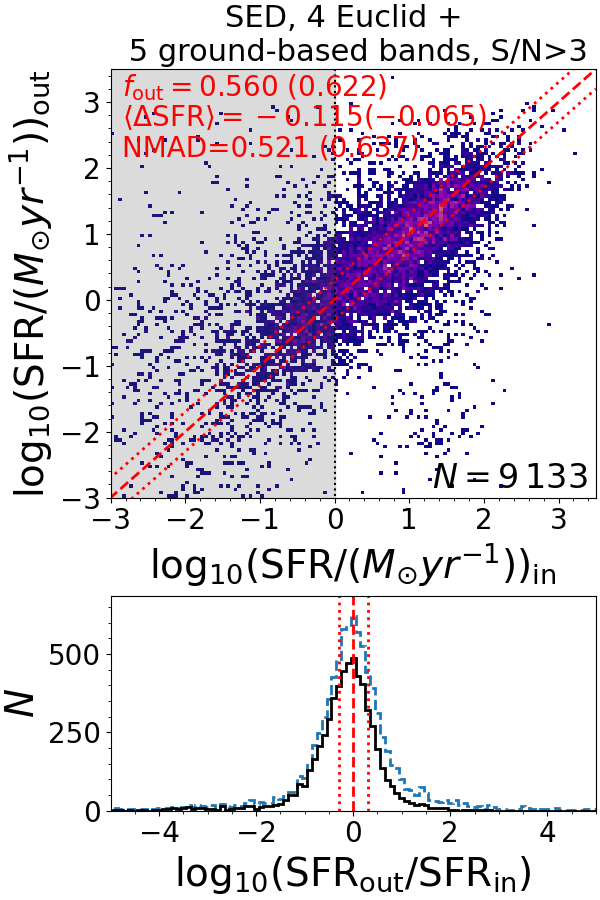}
    \includegraphics[width=0.23\linewidth, keepaspectratio]{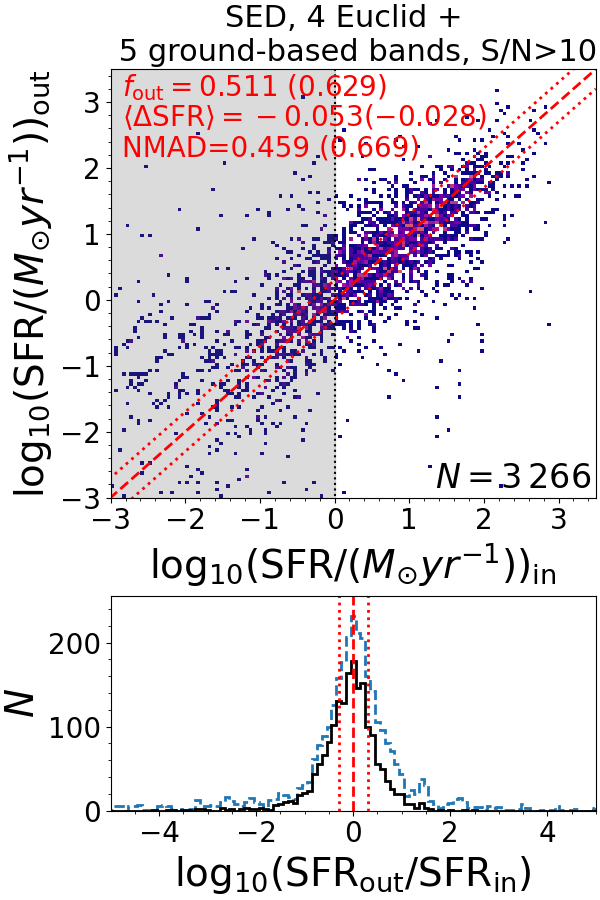}
    \caption{Same as Figure \ref{fig:ML_NN_SFR}, but for the SED fitting.}
    \label{fig:SFR_SED}
\end{figure*}

Finally, in Figure \ref{fig:SFR_SED} we report the results for the SFR, as derived with SED fitting. The estimation largely improves when considering nine filters as inputs instead of four filters. This may be due to two reasons: i) a improvement in the redshift estimation, as mentioned before for the stellar mass; ii) the presence of the $u$-band filter which is more sensitive to SFR than the $I_{\scriptscriptstyle\rm E}$ band, being at shorter wavelengths.\par

\bsp	
\label{lastpage}
\end{document}